\documentclass[ a4paper, 10pt]{article}
\pdfoutput=1

\usepackage{amsmath,amsfonts,amscd,amssymb, amsthm,mathrsfs, ascmac}
\usepackage{latexsym}
\usepackage[english]{babel}
\usepackage[utf8]{inputenc}
\usepackage{graphicx}
\usepackage{bbm}
\usepackage{enumitem}
\usepackage{nicefrac}
\usepackage{bm}
\usepackage[svgnames,psnames]{xcolor}
\usepackage[colorlinks,citecolor=DarkGreen,linkcolor=FireBrick,linktocpage,unicode]{hyperref}
\usepackage{braket}
\usepackage{xparse}
\usepackage{tikz-cd}
\usetikzlibrary{positioning}

\usepackage{comment}
\usepackage{authblk}

\usetikzlibrary{intersections, calc, arrows}

\newtheorem{thm}{Theorem}[section]
\newtheorem{Def}[thm]{Definition}
\newtheorem{Lemma}[thm]{Lemma}
\newtheorem{Prop}[thm]{Proposition}
\newtheorem{Coro}[thm]{Corollary}

\theoremstyle{definition}
\newtheorem*{Proof}{Proof}
\newtheorem{rem}[thm]{Remark}
\renewcommand{\labelenumi}{(\roman{enumi})}\makeatletter
\def\appendix{%
\def\theequation{\Alph{section}\arabic{equation}}%
\setcounter{equation}{0}%
\def\thesection{\Alph{section}}%
\@addtoreset{equation}{section}%
\setcounter{section}{0}%
\def\@seccntformat##1{%
\@nameuse{prefix@##1}%
\@nameuse{the##1}%
\@nameuse{postfix@##1}\quad}%
\def\prefix@section{Appendix~}%
\def\postfix@section{:}%
} 
\makeatletter
  \renewcommand{\theequation}{%
  \thesection.\arabic{equation}}
  \@addtoreset{equation}{section}
\makeatother
\title{
{\bf Electron-phonon interaction  in  Kondo  lattice systems}
}
\author[1]{Tadahiro Miyao}
\author[1]{Hayato Tominaga}

\affil[1]{Department of Mathematics, Hokkaido University

Sapporo 060-0810, Japan}

\date{}
\newcounter{lastenumi}
\def\i<#1>{\langle #1 \rangle}
\def\l<#1>{\left\langle #1 \right\rangle}
\def\b<#1>{\big\langle #1 \big\rangle}
\def\tr{\mathrm{Tr}}
\def\wrt{\ \text{w.r.t.}\ }
\newcommand{\no}{\nonumber \\}
\newcommand{\bs}{\boldsymbol}
\newcommand{\bq}{{\boldsymbol q}}
\newcommand{\SN}{\boldsymbol{\mathcal{S}}_N}
\newcommand{\bome}{{\boldsymbol \omega}}
\newcommand{\bphi}{\boldsymbol q}
\newcommand{\bomega}{\boldsymbol \omega}
\newcommand{\bfphi}{{\boldsymbol \vphi}}
\newcommand{\vphi}{\varphi}
\newcommand{\Hph}{\mathcal{H}_{\rm ph}}

\newcommand{\bd}{\beta}

\def\up{\uparrow}
\def\down{\downarrow}
\def\bsigma{{\boldsymbol\sigma}}
\allowdisplaybreaks
\usepackage[top=22truemm,bottom=22truemm,left=14truemm,right=14truemm]{geometry}
\begin{document}

\maketitle
\begin{abstract}
We study ground state properties  of the Kondo lattice model with an electron-phonon interaction.
The ground state is proved to be unique; in addition, the total spin of the ground state is determined  according to the lattice structure. To prove the assertions, an extension of the method of spin reflection positivity is given in terms of order preserving operator inequalities.
\end{abstract}

\section{Introduction}

\subsection{Background}
The Kondo lattice model (KLM)  describes the interaction between localized spins and  band conduction electrons.
In particular, the half-filled KLM can be regarded as a model for the Kondo insulator.
Because the KLM has a  wide  variety of applications, it has been actively studied, see, e.g.,  \cite{Capponi2001, Peters2007, Santos2002, TSU1997} and references therein.
Although there are a large number of literatures  concerning theoretical analysis of the KLM,  
only few rigorous results are currently known:
Yanagisawa and Shimoi showed  the ground state of the KLM with an extra
on-site Coulomb repulsion  is singlet if the strength of the Coulomb repulsion, $U$,  is large \cite{Yanagisawa1995}; in \cite{Tsunetsugu1997},
Tsunetsugu provided a proof for $U=0$; properties of the spin-spin correlations in the ground state were examined by Shen \cite{Shen1996}.  
\medskip

The subtle interplay of electrons and phonons induces various physical phenomena.
For example, the Holstein-Hubbard model, a prototype model for the electron-phonon coupling,  describes antiferromagnetic,   superconducting and charge-density-wave orders.
Despite the importance of  electron-phonon interactions, there are only few studies examining effects of  electron-phonon interactions in the KLM.
The aim of the present paper is to examine rigorously  the ground state properties of the half-filled KLM with the electron-phonon interaction. 
We prove the uniqueness of the ground state of the model and provide an expression for its total spin, see Theorem \ref{MainThm}.
\medskip

A main tool for the proof is the spin-reflection positivity invented by Lieb  \cite{Lieb1989}. The concept of the reflection positivity originates from the axiomatic 
quantum field theory \cite{Osterwalder1973, Osterwalder1975}. In his  seminal  paper \cite{Lieb1989}, Lieb applied the idea of the reflection positivity  to  the spin space of electrons and  studied 
the magnetic properties of the ground states for the Hubbard model.  
Yanagisawa and Shimoi first applied the method of  the spin reflection positivity to the KLM \cite{Yanagisawa1995}. Further applications of the method to the KLM were discussed  by several authors \cite{Shen1996,Tsunetsugu1997}. 
Freericks and Lieb were  the first to  extend the spin reflection positivity to  electron-phonon interacting systems \cite{Freericks1995}. 
Miyao further generalized the method of the spin reflection positivity in terms of  order operator inequalities and provided a larger   variety of applications including the electron-phonon interacting systems \cite{Miyao2012, Miyao2016, Miyao2018, Miyao2019}.  For reviews on the spin-reflection positivity, see,  e.g., \cite{Shen1998, Tasaki2020,Tian2004}.  For recent developments, see  \cite{yoshida2020rigorous} and references therein.
In the present paper, we apply the method of  the spin reflection positivity  to the KLM with the electron-phonon interaction by  properly extending Miyao\rq{}s idea.
\medskip

\subsection{Main results}\label{MainSubs}

Let us consider the Kondo lattice model with  an electron-phonon interaction:
\begin{align}
{\bf  H}&=-\sum_{x,y\in\Lambda}\sum_{\sigma=\up,\down}t_{x,y}c_{x\sigma}^*c_{y\sigma}+\sum_{x\in\Lambda,u\in\Omega}J_{x,u}{\boldsymbol s}_x\cdot{\boldsymbol S}_u \no
&\quad+\sum_{x,y\in\Lambda}U_{x,y}(n_x^c-1)(n_y^c-1)+\sum_{x,y\in\Lambda}g_{x,y}n_x^c(b_y^*+b_y)+\omega_0\sum_{x\in\Lambda}b_x^*b_x.
\end{align}
We denote by $\Lambda$ a lattice of the conduction electrons, and by $\Omega$ a set of sites on which the localized electrons are located. 
The operator ${\bf H}$ acts on $\mathcal{H}_{\rm c} \otimes \mathcal{H}_{\rm f} \otimes \mathcal{H}_{\rm ph}$, where
 \begin{align}
\mathcal H_{\rm c}&=\mathcal F_{\rm F}(\ell^2(\Lambda))\otimes\mathcal F_{\rm F}(\ell^2(\Lambda)),\\
\mathcal H_{\rm f}&=\mathcal F_{\rm F}(\ell^2(\Omega))\otimes\mathcal F_{\rm F}(\ell^2(\Omega)),\\
\mathcal H_{\rm ph}&=L^2(\mathbb{R}^{|\Lambda|}).
\end{align}
Here, $\mathcal F_{\rm F}(\ell^2(\Lambda))$ and $\mathcal F_{\rm F}(\ell^2(\Omega))$ are the fermionic Fock space over $\ell^2(\Lambda)$ and $\ell^2(\Omega)$,  respectively; More precisely, 
$\mathcal{F}_{\rm F}(\mathcal{X})=\bigoplus_{n=0}^{\dim \mathcal{X}} \bigwedge^n \mathcal{X}$,  where $\bigwedge^n \mathcal{X}$ is the $n$-fold  antisymmetric tensor product of $\mathcal{X}$ with 
$\bigwedge^0 \mathcal{X}=\mathbb{C}$.  
\medskip

$c_{x\sigma}$ 
 denotes the annihilation operator of the conduction electrons, 
and $f_{u\sigma}$ denotes the annihilation operator of the localized spins.
These operators satisfy the standard anticommutation relations:
\begin{align}
\{c_{x\sigma},c_{x'\sigma'}^*\}=\delta_{x,x'}\delta_{\sigma,\sigma'},\quad\{c_{x\sigma},c_{x'\sigma'}\}=0,\quad\{f_{u\sigma},f_{u'\sigma'}^*\}=\delta_{u,u'}\delta_{\sigma,\sigma'},\quad\{f_{u\sigma},f_{u'\sigma'}\}=0, \\
\{c_{x\sigma}, f_{u\sigma\rq{}}\}=\{c_{x\sigma}, f_{u\sigma\rq{}}^*\}=0,
\end{align}
where $\delta_{a, b}$ is the Kronecker delta.\footnote{
One may think that   Hilbert space of the   electrons should be $\mathcal{F}_{\rm F}(\ell^2((\Lambda\sqcup \Omega) \times \{\uparrow, \downarrow\}))$, where $\Lambda\sqcup \Omega$ indicates the discriminated union of $\Lambda$ and $\Omega$.
In the above, we have used the identification:  $\mathcal{F}_{\rm F}(\ell^2((\Lambda\sqcup \Omega) \times \{\uparrow, \downarrow\}))=\mathcal{H}_{\rm c}\otimes \mathcal{H}_{\rm f}$. 
Note that this representation  is   very useful  in the following sections.
For readers\rq{} convenience, we briefly explain this identification below:    Let $\mathcal{X}$ and $\mathcal{Y}$ be Hilbert spaces.
 For $\mathcal{Z}=\mathcal{X}, \mathcal{Y}, \mathcal{X}\oplus \mathcal{Y}$, 
we denote by $A_{\mathcal{Z}}(f)$ the annihilation operator on $\mathcal{F}_{\rm F}(\mathcal{Z})$.
Similarly, the Fock vacuum in $\mathcal{F}_{\rm F}(\mathcal{Z})$ is denoted by $\Omega_{\mathcal{Z}}$.
For each $f\in \mathcal{X}$ and $g\in \mathcal{Y}$, we set 
$B(f, g)=A_{\mathcal{X}}(f)\otimes 1+(-1)^{N_{\mathcal{X}}}\otimes A_{\mathcal{Y}}(g)$, where
$N_{\mathcal{X}}$ is the number operator on $\mathcal{F}_{\rm F}(\mathcal{X})$.  We readily confirm that 
 the family of operators $\{B(f, g)| f\in \mathcal{X}, g\in \mathcal{Y}\}$ satisfies the same anticommutation relations as $\{A_{\mathcal{X}\oplus \mathcal{Y}}(f\oplus g)| f\in \mathcal{X}, g\in \mathcal{Y}\}$, e.g., $\{B(f, g), B(f\rq{},  g\rq{})^*\}=\langle f\oplus g|f\rq{}\oplus g\rq{}\rangle$. In addition, it holds that $B(f, g)\Omega_{\mathcal{X}}\otimes \Omega_{\mathcal{Y}}=0$.
Therefore, we can construct  a natural unitary operator, $\tau$,  from   $\mathcal{F}_{\rm F}(\mathcal{X}\oplus \mathcal{Y})$
onto $\mathcal{F}_{\rm F}(\mathcal{X})\otimes \mathcal{F}_{\rm F}( \mathcal{Y})$ by 
$\tau \Omega_{\mathcal{X}\oplus \mathcal{Y}}=\Omega_{\mathcal{X}}\otimes \Omega_{\mathcal{Y}}$  and
\begin{align}
\tau A_{\mathcal{X}\oplus \mathcal{Y}}(f_1\oplus g_1)^*\cdots A_{\mathcal{X}\oplus \mathcal{Y}}(f_n\oplus g_n)^*\Omega_{\mathcal{X}\oplus\mathcal{Y}}=B(f_1, g_1)^*\cdots B(f_n, g_n)^* \Omega_{\mathcal{X}}\otimes \Omega_{\mathcal{Y}} \label{ABIdn}
\end{align}
for $f_1, \dots, f_n\in \mathcal{X}$ and $g_1, \dots, g_n\in \mathcal{Y}$. Because 
$
\mathcal{F}_{\rm F}(\ell^2((\Lambda\sqcup \Omega) \times \{\uparrow, \downarrow\}))
$
 can be naturally identified with $\mathcal{F}_{\rm F}\Big(\big(\ell^2(\Lambda)\oplus \ell^2(\Lambda) \big)\oplus
\big( \ell^2( \Omega) \oplus \ell^2(\Omega)\big)\Big)$, we get the desired identification.
}
$n_x^c$ and $n_u^f$ stand for  the electron number operators,   and  are   respectively defined by $n_x^c=n_{x\up}^c+n_{x\down}^c$ and $n_u^f=n_{u\up}^f+n_{u\down}^f$, where $
n_{x\sigma}^c=c_{x\sigma}^*c_{x\sigma}$ and $n_{u\sigma}^f=f_{u\sigma}^*f_{u\sigma}$.
${\bs s}_x$ and ${\bs S}_u$ denote  spin operators of the conduction electrons and the localized spins,  respectively. More precisely, the spin  operators are defined by 
\begin{align}
{s}_x^{+}&=({s}_x^{-})^*=c_{x\up}^*c_{x\down},\ \ \ {s}_x^{(3)}=\frac{1}{2}(c_{x\up}^*c_{x\up}-c_{x\down}^*c_{x\down}),\\
{S}_u^{+}&=({S}_u^{-})^*=f_{u\up}^*f_{u\down},\ \ \
{S}_u^{(3)}=\frac{1}{2}(f_{u\up}^*f_{u\up}-f_{u\down}^*f_{u\down})
\end{align}
and 
\begin{align}
{\boldsymbol s}_x\cdot{\boldsymbol S}_u&=\frac{1}{2}({s}_x^{+}{S}_u^{-}+{s}_x^{-}{S}_u^{+})+{s}_x^{(3)}{S}_u^{(3)}.
\end{align}
$b_x$ and $b_x^*$ are the bosonic annihilation and creation operators at site $x\in \Lambda$ satisfying the standard commutation relations:
\begin{align}
[b_x, b_y^*]=\delta_{x, y},\ \ \ [b_x, b_y]=0.
\end{align}
By the Kato-Rellich theorem \cite[Theorem X.12]{Reed1975}, 
 ${\bf H}$ is self-adjoint on $\mathrm{dom}(N_{\rm p})
  $ and bounded from below, where $N_{\rm p}=
  \sum_{x\in\Lambda}b_x^*b_x
  $, the phonon number operator, and $\mathrm{dom}(N_{\rm p})$ indicates the domain of $N_{\rm p}$.
 \medskip

$t_{x, y}$ is the hopping matrix element, $U_{x, y}$ is the energy of the Coulomb interaction, $g_{x, y}$ is the strength of the   conductive electron-phonon interaction, and $J_{x, u}$ is the strength of the exchange interaction. 
The phonons are assumed to be dispersionless with energy $\omega_0>0$.  
Throughout the present  study, we assume  the following:
\begin{description}
\item[1. ] $g_{x,y}, t_{x,y},J_{x,u}, U_{x, y}\in\mathbb R$ for all $x,y\in\Lambda,u\in\Omega$.
\item[2. ] $g_{x,y}=g_{y,x}, t_{x,y}=t_{y,x}$ and $U_{x, y}=U_{y, x}$ for all $x,y\in\Lambda,u\in\Omega$.
\end{description}

Our principal assumptions are stated as follows:
\begin{flushleft}
{\bf Condition  (C).}
\end{flushleft}
\begin{enumerate}
\renewcommand{\theenumi}{{\bf (C.\arabic{enumi})}}
\renewcommand{\labelenumi}{\bf (C.\arabic{enumi})}
\item\label{C1}\ 
 Let $E=\{\{x,y\}\in\Lambda\times\Lambda\,|\,t_{x,y}\neq0\}$. 
The graph $(\Lambda,E)$ is   connected and  bipartite. More precisely,  
\begin{itemize}
\item for any $x, y\in \Lambda$, there is  a path $p=\{\{x_j, y_j\}\}_{j=1}^n\subset E$ such that $x_1=x$ and $y_n=y$;
\item there are disjoint   sublattices $\Lambda_1 $ and $\Lambda_2$ with  $\Lambda=\Lambda_1 \cup\Lambda_2$ such that $t_{x, y}=0$,  whenever $x, y\in \Lambda_1$ or $x, y\in \Lambda_2$.
\end{itemize}
\item\label{C2}\ For any $u\in\Omega$,  there exists an  $x\in\Lambda$ such that $J_{x,u}\neq0$. 
 If $J_{x,u}\neq0$, then $\mathrm{sgn}J_{x,u}$, the sign of $J_{x, u}$,  is independent of $x$ for each $u\in\Omega$.
 \item\label{C5}\ There are disjoint subsets $\Omega_1$ and $\Omega_2$ such that
\begin{itemize}
 \item  $\Omega=\Omega_1\cup\Omega_2$;\footnote{Note that this condition does not necessarily  mean that  $\Omega$  is bipartite.}
 
   \item  $J_{x,u}=0\ (x\in\Lambda_1,u\in\Omega_1\ or\ x\in\Lambda_2,u\in\Omega_2)$. 
   \end{itemize}
\item\label{C3}\  $|\Lambda| $ and $|\Omega|$ are even numbers. 
\item\label{C4}\ $\displaystyle\sum_{x\in\Lambda}g_{x,y}$ is independent of $y\in\Lambda$. 
\setcounter{lastenumi}{\value{enumi}}
\end{enumerate}

There is a local constraint such that every $f$ orbital
is always occupied by just one electron. Such a situation  can be expressed in term of the projection given by 
\begin{align}
P_0&=\prod_{u\in\Omega}\left[n_{u\up}^f(1-n_{u\down}^f)+(1-n_{u\up}^f)n_{u\down}^f\right]. \label{DefP_0}
\end{align}
Note that 
\begin{align}
n_{u\uparrow}^f+n^f_{u\downarrow}=1
\end{align}
  holds  on $\mathrm{ran}(P_0)$, the range of $P_0$.

The total spin operators are defined by 
\begin{align}
S_{\rm tot}^{(3)}=s_{\Lambda}^{(3)}+S_{\Omega}^{(3)},\ \ S_{\rm tot}^{\pm}=s_{\Lambda}^{\pm}+S_{\Omega}^{\pm},
\end{align}
where
\begin{align}
s_{\Lambda}^{(3)}=\sum_{x\in \Lambda}s_x,\ \ \ S^{(3)}_{\Omega}=\sum_{u\in \Omega} S_u^{(3)}, \ \ s_{\Lambda}^{\pm}=\sum_{x\in \Lambda}s_x^{\pm},\ \ 
S_{\Omega}^{\pm}=\sum_{u\in \Omega}S_u^{\pm}.
\end{align}
In addition, we set 
\begin{align}
{\bs S}_{\rm tot}^2=\frac{1}{2}\big(
S_{\rm tot}^+S_{\rm tot}^-+S_{\rm tot}^- S_{\rm tot}^+
\big)+\big(S_{\rm tot}^{(3)}\big)^2.
\end{align}
\begin{Def}\upshape
In general, if a vector $\varphi$ is an eigenvector with ${\bs S}_{\rm tot}^2\varphi=S(S+1)\varphi$,
then we say that $\varphi$ has {\it total spin $S$.}

\end{Def}
Set $N=|\Lambda|+|\Omega|$.
In the present paper, we  are interested in the ground state properties at half-filling. 
For this reason, we introduce the  subspace of $\mathcal{H}_{\rm c}\otimes \mathcal{H}_{\rm f}$ by 
\begin{align}
\mathcal{L}_N=\ker\left(S_{\rm tot}^{(3)}\right)\bigcap\ker\left(N_{\rm e}-N\right), \label{DefL_N}
\end{align}
where $N_{\rm e}=N_{\rm e}^c+N_{\rm e}^f$ is the total electron number operator with 
$N_{\rm e}^c=\sum_{x\in\Lambda}(n_{x\up}^c+n_{x\down}^c) $ and $N_{\rm e}^f=\sum_{u\in\Omega}(n_{u\up}^f+n_{u\down}^f)$.
Note that $S_{\rm tot}^{(3)}=0$ on $\mathcal{L}_N$.
\medskip\\

Taking the above requirements into account,   we introduce the following Hilbert space:
\begin{align}
\mathcal{H}=P_0\mathcal{L}_N\otimes\Hph.
\end{align}
In what follows, we will examine ground state properties of the restricted Hamiltonian $H =  {\bf H} \restriction \mathcal{H}$.
\medskip\\

The main result in this paper is the following theorem:

\begin{thm}\label{MainThm}
Assume {\bf (C)}. Let $
 U_{\mathrm{eff}, x, y}
$ be the  energy of the effective Coulomb  interaction:
 \begin{align}
 U_{\mathrm{eff}, x, y}=U_{x,y}-\omega_0^{-1}\sum_{z\in\Lambda}g_{x,z}g_{y,z}.\label{DefUeff}
 \end{align}
 Suppose  that $U_{{\rm eff}}$ is {\rm positive semi-definite}.\footnote{
More precisely, $U_{\rm eff}$ is {\it positive semi-definite}, if 
$
\sum_{x, y\in \Lambda}U_{{\rm eff}, x, y}z^*_x z_y\ge 0
$ for all ${\boldsymbol z}=\{z_x\}_{x\in \Lambda} \in \mathbb{C}^{\Lambda}$.
}  Notice that the critical  case where $U_{\rm eff} =O$, the zero matrix, satisfies this condition.
Then  we obtain the following {\rm (i)} and {\rm (ii)}:
\begin{itemize}
\item[{\rm (i)}] The ground state of $H$  is unique.
\item[{\rm (ii)}]We denote by  $\psi$ the ground state of $H$. Then $\psi$ satisfies the following: 
\begin{align}
\gamma_x \gamma_y\langle \psi,  s_x^+ s_y^-\psi\rangle>0,\ \ \ \gamma_u \gamma_v \mathrm{sgn}J_{x, u} \mathrm{sgn} J_{y, v} \langle \psi, S_u^+ S_v^-\psi\rangle >0 \label{ZigZag}
\end{align}
for every $x, y\in \Lambda$ and $u, v\in \Omega$, where $\gamma_z=-1$ for $z\in \Lambda_1$ or $\Omega_1$,  $\gamma_z=1$ for $z\in \Lambda_2$ or $\Omega_2$.
\end{itemize}
In addition, we assume one of the following conditions: 
\begin{enumerate}
\setcounter{enumi}{\value{lastenumi}}
\renewcommand{\theenumi}{{\bf (C.\arabic{enumi})}}
\renewcommand{\labelenumi}{\bf (C.\arabic{enumi})}
\item\label{C6}\ $J_{x, u} \ge 0$ for every $x\in \Lambda$ and $u\in \Omega$, the antiferromagnetic coupling.
\item\label{C7}\ $J_{x, u} \leq 0$ for every $x\in \Lambda$ and $u\in \Omega$, the ferromagnetic coupling.
\end{enumerate}
Then $\psi$ has total spin $S$ given by 
\begin{align}
S=
\begin{cases}
\frac{1}{2}\big||\Lambda_1|+|\Omega_1|-|\Lambda_2|-|\Omega_2|\big|, \ \ &\mbox{if \ref{C6} holds}, \\
\frac{1}{2}\big||\Lambda_1|+|\Omega_2|-|\Lambda_2|-|\Omega_1|\big|, \ \ &\mbox{if \ref{C7} holds}.
\end{cases}
\label{CaseDefS}
\end{align}
\end{thm}

To  explain our achievement, let us compare Theorem \ref{MainThm} with the following:
\begin{thm}\label{Miyao}
Assume {\bf (C)}.
Suppose  that $U_{{\rm eff}}$ is {\rm positive definite}.\footnote{
More precisely, $U_{\rm eff}$ is {\it positive definite}, if 
$
\sum_{x, y\in \Lambda}U_{{\rm eff}, x, y}z^*_x z_y>0
$ for all ${\boldsymbol z}=\{z_x\}_{x\in \Lambda} \in \mathbb{C}^{\Lambda}\setminus \{0\}$.
} 
Then  the assertions in Theorem \ref{MainThm} hold  true.
\end{thm}

In the previous works \cite{Miyao2016,Miyao2018}, we examined the ground state properties of the Holstein-Hubbard Hamiltonian 
under the assumption that $U_{\rm eff}$ is positive definite; once we assume that   $U_{\rm eff}$ is positive definite, then
Theorem \ref{Miyao} is an immediate consequence of the method established in  \cite{Miyao2016,Miyao2018}. In comparison with Theorem \ref{Miyao}, we only assume that $U_{\rm eff}$ is positive semi-definite
in Theorem \ref{MainThm}. 
Without the assumption of the positive definiteness of $U_{\rm eff}$, 
to prove Theorem \ref{MainThm} is a mathematically challenging problem.
One of the major achievements of the present paper is improving upon the method of \cite{Miyao2016,Miyao2018} in order to overcome this difficulty. 
\medskip

The problem of refining the assumption in Theorem \ref{Miyao} is physically important as well. In order to  briefly
illustrate this, let us consider on-site interactions: 
$
 g_{x, y}=g\delta_{x, y}, \ \ U_{x, y}=U\delta_{x, y}
$ with $U>0$. In this case,  we have $U_{{\rm eff}, x, y}=(U-g^2/\omega_0)\delta_{x, y}$. Hence if  $|g|<\sqrt{\omega_0U}$, then the assertion in  Theorem \ref{Miyao} holds. However, there is a possibility that   ground states properties of $H$ could be  dramatically changed  at $g_{\rm c}=\pm \sqrt{\omega_0 U}$. Theorem \ref{MainThm} tells us that this never happens. It is expected that 
the ground state properties  for $|g| > \sqrt{\omega_0 U}$ are different from those for $|g|\le \sqrt{\omega_0 U}$.
\medskip

A key ingredient of our analysis is order preserving operator inequalities introduced in Section \ref{OpSection}. As we will see, the  inequalities are completely different from  the standard operator inequalities which can be found in the text books on functional analysis.
In a series of works \cite{Miyao2012, Miyao2016, Miyao2018, Miyao2019}, the effectiveness of the order preserving   operator inequalities in the study of strongly correlated electron systems  has been demonstrated.  By using  the  inequalities, we can bound  from below the interaction term between the conduction electrons and the localized electrons  by  the Coulomb interaction, see Proposition \ref{proj}. This bound enables us to prove the uniqueness of ground states of $H$  under  the weaker  assumption, i.e.,   the positive semi-definiteness of $U_{\rm eff}$. In addition, the inequalities will play essential roles in deriving the formula \eqref{CaseDefS}, see Section \ref{SpinPf} for details.

\begin{rem} \label{MainRem}\upshape
\begin{itemize}
\item[1. ] By combining Theorem \ref{MainThm} with a method similar to that presented in \cite{Miyao2019,Shen1994}, we can prove that the ground state simultaneously exhibits antiferromagnetic and ferromagnetic long range orders, if there exists a constant $a>0$ such that 
$S=aN+o(N)$ as $N\to \infty$, where $S$ is given by \eqref{CaseDefS}. 
\item[2. ] We can obtain an upper bound for the charge susceptibility by arguments similar to those in \cite{Kubo1990,Miyao2016}. In addition, by using the bound, we can show the absence of  charge-density-wave-order, provided that there is  a $c>0$ such that $U_{\rm eff}\ge c$, i.e.,
$U_{\rm eff}-c$ is positive semi-definite.
\item[3. ] Tsunetsugu\rq{}s result \cite{Tsunetsugu1997} corresponds to the case that $g_{x, y} \equiv 0$ and $U_{x, y} \equiv 0$. Thus, Theorem \ref{MainThm} can be regarded as an extension of \cite{Tsunetsugu1997}.\
We also remark that \eqref{ZigZag} is an extension of \cite{Shen1996}.
\end{itemize}
\end{rem}

\begin{rem}\label{QED}\upshape
The method presented  in this paper has a variety of applications.
For instance, let us consider the Kondo lattice model with an electron-photon interaction: 
\begin{align}
\mathbf{H}_{\rm QED}=&-\sum_{x,y\in\Lambda}\sum_{\sigma=\up,\down}t_{x,y}
\exp\Bigg\{
i \int_{C_{xy}} dr\cdot A(r)
\Bigg\}
c_{x\sigma}^*c_{y\sigma}+\sum_{x\in\Lambda,u\in\Omega}J_{x,u}{\boldsymbol s}_x\cdot{\boldsymbol S}_u \no
&\quad+\sum_{x,y\in\Lambda}U_{x,y}(n_x^c-1)(n_y^c-1)+\sum_{k\in V^*}\sum_{\lambda=1,2} \omega(k) a(k, \lambda)^*a(k, \lambda).
\end{align}
Here, we assume that $\Lambda$ and $\Omega$ are embedded into the region
$V=[-L/2, L/2]^3\subset \mathbb{R}^3$. $V^*$ is defined by $V^*=(\frac{2\pi}{L}\mathbb{Z})^3$.
$a(k, \lambda)$ and $a(k, \lambda)^*$ denote the photon annihilation and creation operators, respectively. As usual,  these satisfy the following commutation relations:
\begin{align}
[a(k, \lambda), a(k\rq{}, \lambda\rq{})^*]=\delta_{k, k\rq{}}\delta_{\lambda, \lambda\rq{}},\ \ [a(k, \lambda), a(k\rq{}, \lambda\rq{})]=0.
\end{align}
$A(x)=(A_1(x), A_2(x), A_3(x))$ is the vector potential given by 
\begin{align}
A(x)=\frac{1}{\sqrt{|V|}}\sum_{k\in V^*} \frac{\chi_{\kappa}(k)}{\sqrt{2 \omega(k)}}
\varepsilon(k, \lambda)
\big(
e^{ik\cdot x} a(k, \lambda)+e^{-i k\cdot x}a(k, \lambda)^*
\big).
\end{align}
$\varepsilon(k, \lambda)$ are the polarization vectors. $C_{xy}$ is a piecewise smooth curve from $x$ to $y$.  The dispersion relation is chosen as $\omega(k)=|k|$. 
$\chi_{\kappa}$ is the indicator function of the ball of radius $\kappa>0$ centered at the origin.
Note that this kind of interaction was originally studied  by Giuliani {\it et al.} in \cite{Giuliani2012}.
Applying the method presented in this paper, we can prove that Theorem \ref{MainThm} and Remark \ref{MainRem} still hold true for $\mathbf{H}_{\rm QED}$, provided that $\{U_{x, y}\}$ is positive semi-definite.  In Section \ref{Discussion}, we further discuss possible extensions of the method presented in this paper in terms of stability classes.
\end{rem}

\begin{rem}\label{Hfp}\upshape
We can further take an interaction between the  $f$-electrons and phonons into account:
\begin{align}
\mathbf{H}_{\rm fp}=\mathbf{H}+\sum_{u, v\in \Omega}k_{u, v}n_u^f
(a_v+a_v^*)+\nu\sum_{u\in \Omega}a_u^*a_u,
\end{align}
where $a_u$ and $a_u^*$ are the annihilation and creation operators for new phonon; these satisfy the standard  commutation relations:
\begin{align}
[a_u, a_v^*]=\delta_{u, v},\ \ [a_u, a_v]=0.
\end{align}
By applying the method in the present paper, we can extend Theorem \ref{MainThm} and Remark \ref{MainRem} to $
\mathbf{H}_{\rm fp}$ with the following additional assumptions:
\begin{itemize}
\item $\{k_{u, v}\}$ is a real symmetric matrix.
\item $\sum_{v\in \Omega}k_{u, v}$ is independent of $u$.
\end{itemize}
See Section \ref{Discussion} for further discussion.
\end{rem}

\subsection{Examples}
 In this subsection, we will give some examples for better understanding of Theorem \ref{MainThm}.
\subsubsection*{Example 1}
Let us consider the case where   $\Omega=\Lambda$ with 
$\Omega_1=\Lambda_2 $ and $\Omega_2=\Lambda_1$. 
By choosing  $g_{x, y}$,  $J_{x, u}$ and $U_{x, y}$ as 
\begin{align}
J_{x, u}=J\delta_{x, u},\ \ g_{x, y}=g\delta_{x, y}, \ \ U_{x, y}=U\delta_{x, y}
\end{align}
with  $U\ge 0$, 
we can reproduce  the standard Kondo lattice model with the electron-phonon interaction:
\begin{align}
\mathbf{H}=-\sum_{x,y\in\Lambda}\sum_{\sigma=\up,\down}t_{x,y}c_{x\sigma}^*c_{y\sigma}+J\sum_{x\in\Lambda}\boldsymbol{s}_x\cdot{\boldsymbol S}_x+U \sum_{x\in\Lambda}(n_x^c-1)^2+g \sum_{x\in\Lambda} n_x^c(b_x^*+b_x)+\omega_0\sum_{x\in\Lambda}b_x^*b_x.
\end{align}
Assume that \ref{C1} is satisfied and $|\Lambda|$ is even.
In this case, the assumptions \ref{C2}--\ref{C4} are automatically fulfilled. 
If $|g| \le \sqrt{\omega_0 U}$, then  $U_{\rm eff}$ is positive semi-definite.  Notice  that the case where  $g=\pm \sqrt{\omega_0 U}$ is allowed.
It is noteworthy that, if $J>0$, then the total spin of the ground state is always equal to zero: $S=0$.
In contrast to this, if $J<0$, then we have $S=\big||\Lambda_1|-|\Lambda_2|\big|$.
\subsubsection*{Example 2}

\begin{figure}
\centering
\includegraphics[scale=0.56]{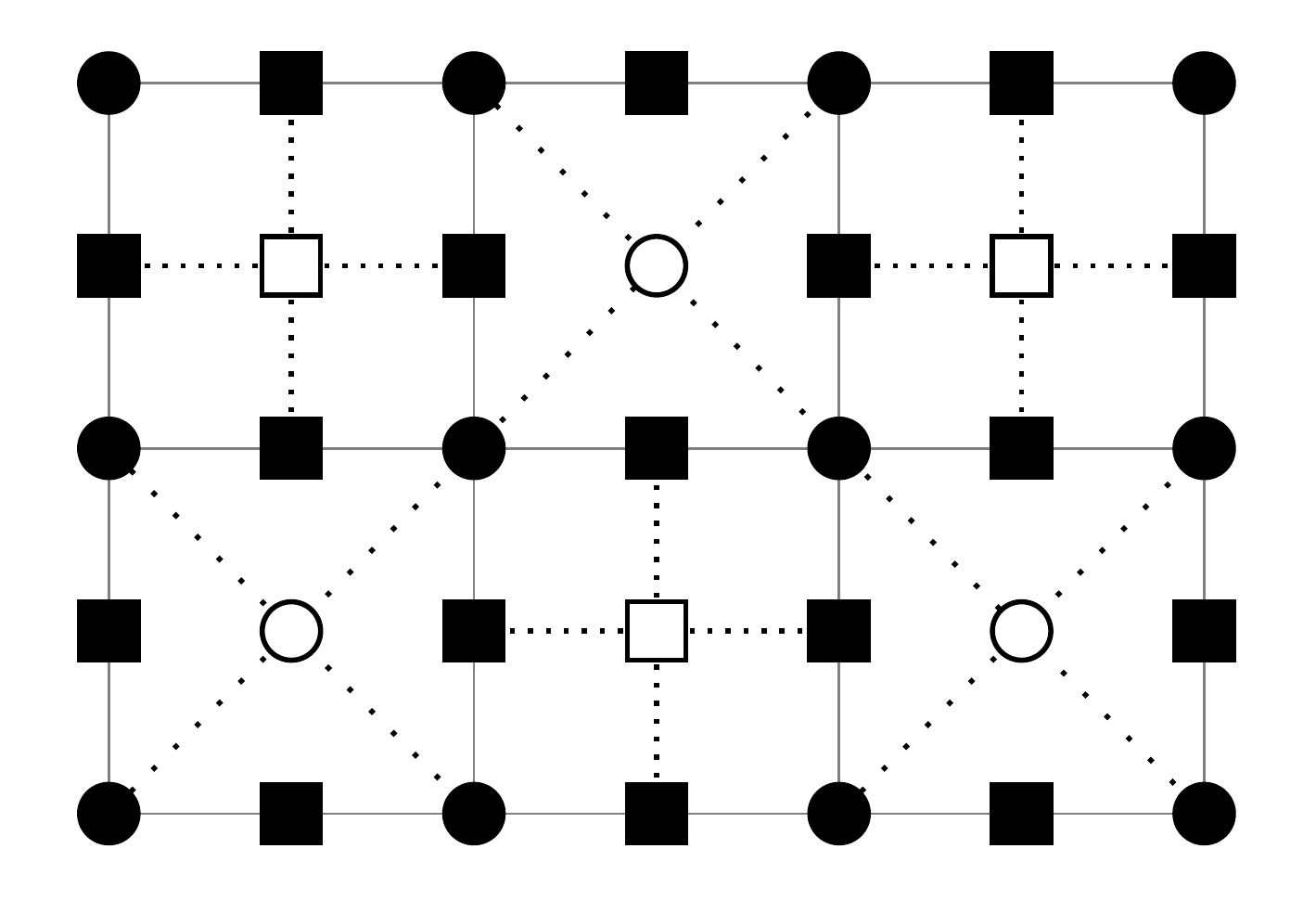}
 \caption{Filled circles and  boxes respectively  indicate the sites of $\Lambda_1$ and $\Lambda_2$. Open circles and  boxes respectively indicate the sites of $\Omega_2$ and $\Omega_1$.}
 \label{Fig1}
\end{figure}

Let us consider a two-dimensional  lattice given by Figure \ref{Fig1}.
For each $x,y\in\Lambda$ and $ u\in\Omega$, we set 
\begin{align}
t_{x,y}=
\begin{cases}
t\ \ &|x-y|=\frac{1}{2}\\
0\ \ &\text{otherwise},
\end{cases}
\quad\quad
J_{x,u}=
\begin{cases}
J\ \ &u\in\Omega_1,|x-u|=\frac{1}{2}\text{\ or\ }u\in\Omega_2,|x-u|=\frac{1}{\sqrt2}\\
0\ \ &\text{otherwise},
\end{cases}
\end{align}
where $t\neq 0 $. The   conditions \ref{C1}--\ref{C3} are satisfied. In this example, we simply assume \ref{C4}. First, let us consider the case where $J>0$. Then \ref{C6} is satisfied.
Because 
$|\Lambda_2|=2|\Lambda_1|$ and $ |\Omega_1|=|\Omega_2|=|\Lambda_1|/2$, 
the ground state has total spin 
 $S=|\Lambda_1|/2=N/8$. 
 Similarly, if $J<0$, then \ref{C7} is fulfilled and 
$S=|\Lambda_1|/2=N/8$.
\subsubsection*{Example 3}

\begin{figure}
\centering
\includegraphics[scale=0.54]{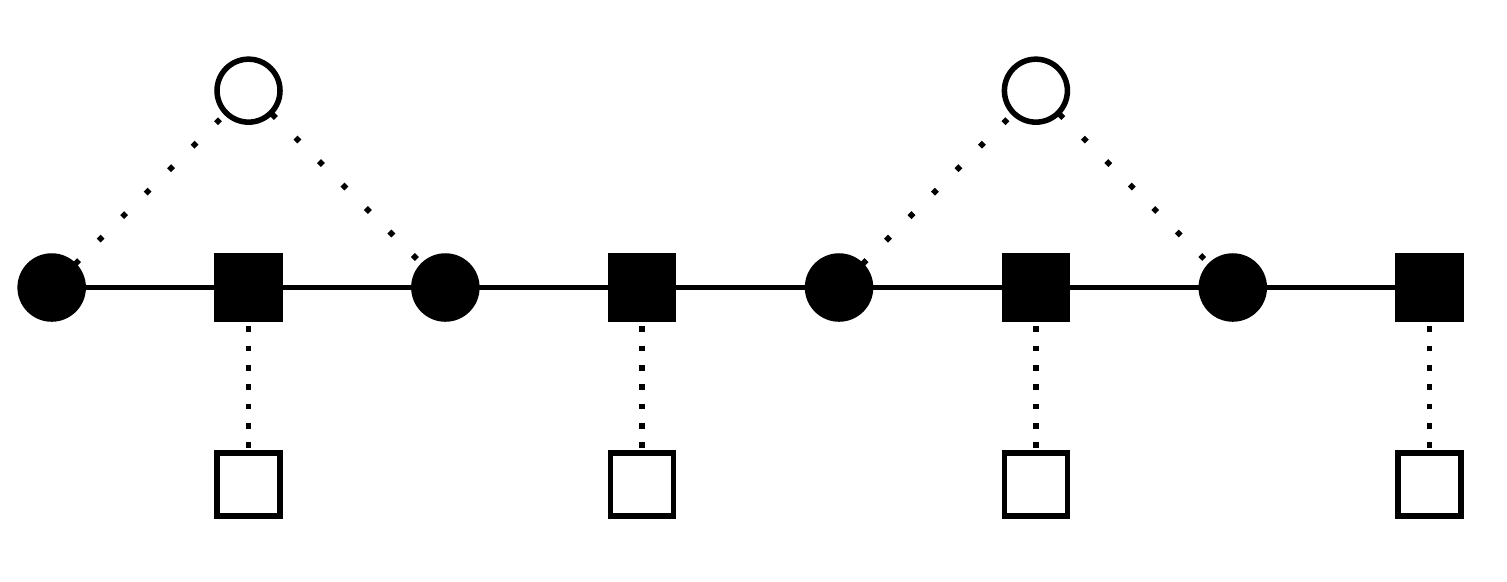}
\caption{Filled circles and  boxes respectively  indicate the sites of $\Lambda_1$ and $\Lambda_2$. Open circles and  boxes respectively indicate the sites of $\Omega_2$ and $\Omega_1$.}
 \label{Fig2}
\end{figure}
In this example, let us consider a chain given by Figure \ref{Fig2}.
We set
\begin{align}
t_{x,y}=
\begin{cases}
t\ \ &|x-y|=\frac{1}{2}\\
0\ \ &\text{otherwise},
\end{cases}
\quad\quad
J_{x,u}=
\begin{cases}
J\ \ &u\in\Omega_1,|x-u|=\frac{1}{2}\text{\ or\ }u\in\Omega_2,|x-u|=\frac{1}{\sqrt2}\\
0\ \ &\text{otherwise},
\end{cases}
\end{align}
where $t\neq 0 $.  With regard to $g_{x,y}$, we simply assume \ref{C4}.
Then we readily confirm that $|\Lambda_1|=|\Lambda_2|=|\Lambda|/2
$ and $
|\Omega_1|=|\Lambda|/2,|\Omega_2|=|\Lambda|/4$. Hence, if $J\neq 0$, then the ground state has  total spin  $S=|\Lambda|/8=N/14$, i.e., the value of $S$ is independent of the sign of $J$.

\subsection{Organization}

The organization of the present paper is as follows:
 In Section \ref{OpSection}, we present the basics of  order preserving operator inequalities.
These inequalities are essential  to express the idea of the spin reflection positivity,  mathematically.
Section \ref{UniGsPf} is devoted to the proof of the uniqueness of the ground state. In Section \ref{SpinPf},
we give the  expression for the  total spin of the ground state.
In Section \ref{Discussion}, we summarize this work and provide discussions.
 The appendices contain some auxiliary technical statements that are of independent interest.

\section{Hilbert cones and their associated operator inequalities}\label{OpSection}

\subsection{Basic definitions}

 In this section, we will  briefly review fundamental  properties 
of  Hilbert cones and their associated operator inequalities as a  preliminary.

Let $\mathcal{X}$ be a  complex Hilbert space.
We denote by $\mathscr{B}(\mathcal{X})$ the Banach space of all bounded operators on $\mathcal{X}$.

\begin{Def} \label{DefHilC}\upshape
A {\it Hilbert cone},  $\mathcal{C}$  in $\mathcal{X}$,   is  a closed convex  cone  obeying 
\begin{itemize}
\item $\i<u,v>\geq 0$ for all $u, v\in \mathcal{C}$;
\item for all $w\in \mathcal{X}$, there exist  $u,u',v,v'\in \mathcal{C}$ such that   $w=u-v+i(u'-v')$  and $\i<u,v>=\i<u',v'>=0.$
\end{itemize}
A vector $ u \in\mathcal{C}$ is said to be {\it  positive w.r.t.} $\mathcal{C}$. We write this as $u \geq 0$ w.r.t. $\mathcal{C}$. A vector $v \in\mathcal{X}$ is called {\it strictly positive w.r.t.} $\mathcal{C
}$,  whenever $\i<v, u>>0$ for all $ u \in \mathcal{C}\setminus\{0\}$. We write this as $v>0$ w.r.t. $\mathcal{C}$.
\end{Def}

The following operator inequalities  will play a major role in the present paper.

\begin{Def}\upshape 
Let $A\in\mathscr B(\mathcal{X})$. 
\begin{itemize}
\item $A$ is {\it  positivity preserving} if $A\mathcal{C}\subseteq \mathcal{C}$. We write this as $A\unrhd 0\wrt \mathcal{C}$.
\item $A$ is {\it positivity improving} if,  for all $u \in \mathcal{C
} \setminus \{0\},\ A u >0\wrt \mathcal{C}$.  We write this as $A\rhd 0\wrt \mathcal{C}$.
\end{itemize}
Remark that the notations of the operator inequalities are  borrowed  from \cite{Miura2003}.
\end{Def}

We readily confirm the following lemma:
\begin{Lemma}
Let $A,B\in\mathscr B(\mathcal{X})$.
 Suppose that  $A,B\unrhd0\wrt \mathcal{C}$. We have the following:
 \begin{itemize}
 \item[{\rm (i)}] If $a, b\ge 0$, then $aA+bB\unrhd0\wrt \mathcal{C}$;
 \item[{\rm  (ii)}] $AB\unrhd0\wrt \mathcal{C}$.
 \end{itemize}
\end{Lemma}
\begin{Proof}
For proof, see, e.g., \cite{Miura2003,Miyao2016}. \qed
\end{Proof}

Let $\mathcal{X}_{\mathbb{R}}$ be the real subspace of $\mathcal{X}$ generated by $\mathcal{C}$. From Definition \ref{DefHilC}, for all $x\in \mathcal{X}_{\mathbb{R}}$, there exist $x_+, x_-\in \mathcal{C}$ such that $x=x_+-x_-$ and $\langle x_+, x_-\rangle=0$. If $A\in \mathscr{B}(\mathcal{X})$ satisfies $A\mathcal{X}_{\mathbb{R}} \subseteq \mathcal{X}_{\mathbb{R}}$, then we say that $A$ {\it preserves the reality w.r.t. $\mathcal{C}$}.

\begin{Def} \upshape
Let $A, B\in\mathscr B(\mathcal{X})$ be reality preserving w.r.t. $\mathcal{C}$.  
If   $A-B\unrhd 0$, then we write this as  
$A\unrhd B \wrt \mathcal{C}$.

\end{Def}

Below,  we  provide  two fundamental  lemmas of the operator inequalities for later use.

\begin{Lemma}
Let $A,B,C,D\in\mathscr B(\mathcal X)$. Suppose $A\unrhd B\unrhd0\wrt\mathcal C$ and $C\unrhd D\unrhd0\wrt\mathcal C$. Then we have $AC\unrhd BD\unrhd0\wrt\mathcal C$
\end{Lemma}
\begin{Proof}
For proof, see, e.g., \cite{Miura2003,Miyao2016}. \qed
\end{Proof}
\begin{Lemma}\label{ppiexp1}
Let $A,B$ be self-adjoint operators on $\mathcal X$. Assume that $A$ is bounded from below and  that  $B\in\mathscr B(\mathcal X)$. Furthermore, suppose that $e^{-tA}\unrhd0\wrt\mathcal C$ for all $t\geq0$ and $B\unrhd0\wrt\mathcal C$. Then we have $e^{-t(A-B)}\unrhd e^{-tA}\wrt\mathcal C$ for all $t\geq0$.
\end{Lemma}

\begin{Proof}
Because  $B\unrhd0\wrt\mathcal C$, we have $e^{tB}=\sum_{n=0}^{\infty}\frac{t^n}{n!}B^n\unrhd1\wrt\mathcal C$ for all $t\geq0$. By the Trotter product formula \cite[Theorem S. 20]{Reed1981}, for all $t\geq0$, we obtain
\begin{align}
e^{-t(A-B)}=\lim_{n\to\infty}\left(e^{-\frac{t}{n}A}e^{\frac{t}{n}B}\right)^n\unrhd e^{-tA}\wrt\mathcal C.
\end{align}
\qed
\end{Proof}

\begin{Def}
\upshape
Let $A$ be a self-adjoint operator on $\mathcal{X}$,   bounded from below.
The semigroup $\{e^{-tA}\}_{t\ge 0}$ is said to be {\it ergodic} w.r.t. $\mathcal{C}$, if the following (i) and (ii) are satisfied:
\begin{itemize}
\item[(i)] $e^{-tA} \unrhd 0$ w.r.t. $\mathcal{C}$ for all $t\ge 0$;
\item[(ii)] for each $u, v\in \mathcal{C} \setminus \{0\}$, there is a $t\ge 0$
such that $\langle u, e^{-tA} v\rangle >0$. Note that $t$ could depend on $u$ and $v$.
\end{itemize}

\end{Def}

The following lemma   immediately follows  from the definitions:
\begin{Lemma}
Let $A$ be a self-adjoint operator on $\mathcal{X}$, bounded from below. If $e^{-tA}\rhd 0$ w.r.t. $\mathcal{C}$ for all $t>0$, then $\{e^{-tA}\}_{t\ge 0}$ is ergodic w.r.t. $\mathcal{C}$.
\end{Lemma}

The basic result here is:
\begin{thm}[Perron-Frobenius-Faris]\label{pff}
Let $A$ be a self-adjoint operator, bounded from below. Assume that $E(A)=\inf \mathrm{spec}(A)$ is an eigenvalue of $A$, where $\mathrm{spec}(A)$ indicates the spectrum of $A$. Let $\mathcal{V}$ be the  eigenspace corresponding to $E(A)$. If $\{e^{-tA}\}_{t\ge 0}$ is ergodic w.r.t. $\mathcal{C}$, then $\dim \mathcal{V}=1$ and $\mathcal{V}$ is spanned by a   strictly positive vector w.r.t. $\mathcal{C}$.
\end{thm}

\begin{Proof}
See \cite{Faris1972}.\qed
\end{Proof}

\subsection{Operator inequalities in $\mathscr{I}_2(\mathcal{X})$ }

Let  $\mathscr{I}_2(\mathcal{X})$ be the set of all Hilbert-Schmidt operators on $\mathcal{X}$:
 $\mathscr{I}_2(\mathcal{X})=\{\xi\in \mathscr{B}(\mathcal{X})\, |\, \tr[\xi^*\xi]<\infty\}$.
In what follows, we regard $\mathscr{I}_2(\mathcal{X})$ as a Hilbert space equipped with the inner product 
$\langle \xi, \eta\rangle_2=\tr[\xi^* \eta],\ \xi, \eta\in \mathscr{I}_2(\mathcal{X})$.
We often abbreviate the inner product by omitting the subscript $2$ if no confusion arises.

Let $\vartheta$ be an antiunitary operator on $\mathcal X$. We define the map $\Psi_\vartheta:\mathcal X\otimes\mathcal X\longrightarrow\mathscr I_2(\mathcal X)$ by
\begin{align}
\Psi_\vartheta(\phi\otimes\vartheta\psi)=|\phi\rangle\langle\psi|,\quad\phi,\psi\in\mathcal X. \label{DefTheta}
\end{align}
Since $\Psi_\vartheta$ is a unitary operator, we can identify $\mathcal X\otimes\mathcal X$ with $\mathscr{I}_2(\mathcal X)$, naturally. We write this identification as 
\begin{align}
\mathcal{X}\otimes \mathcal{X}\underset{\Psi_{\vartheta}}{=}\mathscr{I}_2(\mathcal{X}). \label{IdnSym}
\end{align}
Occasionally, we abbreviate (\ref{IdnSym}) by omitting the subscript $\Psi_{\vartheta}$ if no confusion arises.

Given  $A\in\mathscr B(\mathcal X)$, we define the left multiplication operator, $\mathcal{L}(A)$,  and the right multiplication operator, $\mathcal{R}(A)$,  as follows:
\begin{align}
\mathcal{L}(A) \xi=A\xi,\ \ \mathcal{R}(A)\xi=\xi A,\ \ \xi\in \mathscr{I}_2(\mathcal{X}).
\end{align}
Trivially, $\mathcal{L}(A)$ and $\mathcal{R}(A)$ are bounded operators on $\mathscr{I}_2(\mathcal{X})$.
In addition, we  readily confirm that 
\begin{align}
\mathcal{L}(A)\mathcal{L}(B)=\mathcal{L}(AB),\ \ \ \mathcal{R}(A)\mathcal{R}(B) =\mathcal{R}(BA).
\end{align}
Under the identification (\ref{IdnSym}), we have
\begin{align}
A\otimes 1=\mathcal L(A),\ \ \
1\otimes A = \mathcal R(\vartheta A^*\vartheta). \label{LRIden}
\end{align}
Let 
\begin{align}
\mathscr I_{2,+}(\mathcal X)=\{\xi \in\mathscr I_2(\mathcal X)\,|\,\xi\geq0\}, \label{DefI_{2, +}}
\end{align}
where the inequality in the right hand side of (\ref{DefI_{2, +}}) indicates the standard operator inequality.
It is well-known that $\mathscr{I}_{2, +}(\mathcal{X})$ is a Hilbert cone in $\mathscr{I}_2(\mathcal{X})$, see, e.g.,  \cite[Proposition 2.5]{Miyao2016}. 
 Using this fact, we can introduce a  Hilbert cone in $\mathcal{X}\otimes \mathcal{X}$ by  $\mathcal C=\Psi_\vartheta^{-1}(\mathscr I_{2,+}(\mathcal X))$. Taking the identification (\ref{IdnSym}) into account, we have the following identification:
 \begin{align}
 \mathcal{C}=\mathscr{I}_{2, +}(\mathcal{X}).
 \end{align}

\begin{Prop}\label{PPI2}
Let $A\in \mathscr{B}(\mathcal{X})$.
Then we have $\mathcal L(A)\mathcal R(A^*)\unrhd0\wrt \mathscr{I}_{2, +}(\mathcal{X})$. Hence, under  the identification (\ref{IdnSym}), we have $A\otimes \vartheta A \vartheta \unrhd 0$ w.r.t. $\mathcal{C}$.
\end{Prop}
\begin{Proof}
Take $\xi, \nu\in\mathscr{I}_{2, +}(\mathcal{X})$, arbitrarily. Then there exist  sequences  of positive numbers, $\{\xi_n\}_{n}$ and $\{\nu_n\}_n$,   and  complete orthonormal systems(CONSs)
 $\{x_n\}_n $ and $ \{y_n\}_n$ in $\mathcal{X}$ such that
$
\xi=\sum_n\xi_n|x_n\rangle\langle x_n| $ and $ \nu=\sum_n\nu_n|y_n\rangle\langle y_n|
$
hold.  Because
\begin{align}
\mathcal L(A)\mathcal R(A^*)\nu=\sum_n\nu_n|A y_n\rangle\langle Ay_n|,
\end{align}
we have
\begin{align}
\i<\xi,\mathcal L(A)\mathcal R(A^*)\nu>
&=\sum_{m,n}\xi_m\nu_n|\i<x_m,Ay_n>|^2\geq0.
\end{align}
Hence, we have $\mathcal L(A)\mathcal R(A^*)\unrhd0\wrt\mathscr{I}_{2, +}(\mathcal{X})$.\qed
\end{Proof}

\section{The uniqueness of ground states}\label{UniGsPf}

\subsection{The main result in Section \ref{UniGsPf}}

The goal of this section is to prove the first part of Theorem \ref{MainThm}, that is,
\begin{thm}\label{MainThmHalf}
Assume {\bf (C)}. 
 Suppose  that $U_{{\rm eff}}$ is {\it positive semi-definite}.
Then  we obtain the following {\rm (i)} and {\rm (ii)}:
\begin{itemize}
\item[{\rm (i)}] The ground state of $H$  is unique.
\item[{\rm (ii)}]We denote by  $\psi$ the ground state of $H$. Then $\psi$ satisfies the following: 
\begin{align}
\gamma_x \gamma_y\langle \psi,  s_x^+ s_y^-\psi\rangle>0,\ \ \  \gamma_u\gamma_v\mathrm{sgn}J_{x, u}\mathrm{sgn} J_{y, v} \langle \psi, S_u^+ S_v^-\psi\rangle >0
\end{align}
for every $x, y\in \Lambda$ and $u, v\in \Omega$.
\end{itemize}
\end{thm}
The proof of Theorem \ref{MainThmHalf} will be provided in Subsection \ref{PfUniSub}.
Our basic strategy for proving Theorem \ref{MainThmHalf} is to make use of Theorem \ref{pff}.
To realize the strategy, we employ the method of order preserving operator inequalities presented in Section \ref{OpSection}.

\subsection{Preliminaries}

\subsubsection{Useful identifications}
Let  $X=(x_1, \dots, x_n)\in \Lambda^n$, where $x_1, \dots, x_n$ are mutually different. For such an $X$,  let us define a  vector in $\bigwedge^n \ell^2(\Lambda)$ by 
\begin{align}
e_X^{ c}=\delta_{x_1}\wedge \cdots \wedge \delta_{x_n},
\end{align}
where $\delta_x$ is a  vector in $\ell^2(\Lambda)$ defined by $\delta_x(y)=\delta_{x, y}$. 
Because $\{\delta_x\}_{x\in \Lambda}$ is a  CONS in $\ell^2(\Lambda)$, 
$\{e_X^{ c}\}_{X}$ is a CONS of $\mathcal{F}_{ F}(\ell^2(\Lambda))$.  Similarly, for $U=(u_1, \dots, u_n)\in \Omega^n$,
let 
\begin{align}
e^{ f}_U=\delta_{u_1}\wedge \cdots \wedge \delta_{u_n}.
\end{align}
Then $\{
e^{ f}_U
\}_U
$ is a CONS of $\mathcal{F}_{\rm  F}(\ell^2(\Omega))$.
Trivially, $\{e^{ c}_{X_{\uparrow}} \otimes e^{ c}_{X_{\downarrow}}\}_{X_{\uparrow}, X_{\downarrow}}$ and  $\{e^{ f}_{U_{\uparrow}} \otimes e^{ f}_{U_{\downarrow}}\}_{U_{\uparrow}, U_{\downarrow}}$ are canonical CONSs in $\mathcal{H}_{\rm c}$ and $\mathcal{H}_{\rm f}$, respectively.
Hence, a canonical CONS in $\mathcal{H}_{\rm c}\otimes \mathcal{H}_{\rm f}$ is given by 
$\{
e^{ c}_{X_{\uparrow}} \otimes e^{ c}_{X_{\downarrow}} \otimes e^{ f}_{U_{\uparrow}} \otimes e^{ f}_{U_{\downarrow}}
\}_{
X_{\uparrow}, X_{\downarrow}, U_{\uparrow}, U_{\downarrow}
}
$. 

In what follows, we will freely use the following identification:
\begin{align}
\mathcal{H}_{\rm c}\otimes \mathcal{H}_{\rm f}=\mathcal{F}\otimes \mathcal{F},
\end{align}
where $\mathcal F=\mathcal F_{ \rm F}(\ell^2(\Lambda))\otimes\mathcal F_{\rm  F}(\ell^2(\Omega))=
\mathcal{F}_{ \rm F}(\ell^2(\Lambda) \oplus \ell^2(\Omega))$. (Here, the identification $\mathcal F_{ \rm F}(\ell^2(\Lambda))\otimes\mathcal F_{\rm  F}(\ell^2(\Omega))=
\mathcal{F}_{ \rm F}(\ell^2(\Lambda) \oplus \ell^2(\Omega))$ is  due to   the footnote including \eqref{ABIdn} in Subsection \ref{MainSubs}.)
 Note that this identification is implemented by the unitary operator $\tau$ given by 
\begin{align}
\tau  e^{ c}_{X_{\uparrow}} \otimes e^{ c}_{X_{\downarrow}} \otimes e^{ f}_{U_{\uparrow}} \otimes e^{ f}_{U_{\downarrow}}
=e^{ c}_{X_{\uparrow}} \otimes
 e^{ f}_{U_{\uparrow}} \otimes 
  e^{ c}_{X_{\downarrow}} \otimes
   e^{ f}_{U_{\downarrow}}.
\end{align}

Next, we define the antiunitary operator $\vartheta:\mathcal F\longrightarrow\mathcal F$ by
\begin{align}
\vartheta\left(\sum_{X, U}
c(X, U) 
e^{ c}_X\otimes e^{ f}_U
\right)=
\sum_{X, U}
\overline{c(X, U)} 
e^{ c}_X\otimes e^{ f}_U
,\ \ c(X, U)\in\mathbb C.
\end{align}
With this choice of  $\vartheta$, we can identify $\mathcal{F}\otimes \mathcal{F}$ with $\mathscr{I}_2(\mathcal{F})$ by using (\ref{IdnSym}).

To sum, we obtain the following:
\begin{align}
\mathcal{H}_{\rm c}\otimes \mathcal{H}_{\rm f} \underset{\tau}{=}\mathcal{F}\otimes \mathcal{F}\underset{\Psi_{\vartheta}}{=}\mathscr{I}_2(\mathcal{F}). \label{ImportantIden}
\end{align}
As we will see in the following sections, the above identifications play an important role.

Recall that $N=|\Lambda|+|\Omega|$.
Let $\mathcal F_N=\bigwedge^{N/2}\left(\ell^2(\Lambda)\oplus\ell^2(\Omega)\right)$.
 Then, due to   the footnote including \eqref{ABIdn} in Subsection \ref{MainSubs},  $\mathcal{L}_N$   defined by (\ref{DefL_N}) can be expressed as  
 \begin{align}
 \mathcal L_N=\mathcal F_N\otimes\mathcal F_N.\label{LNIden}
 \end{align}
 Moreover, taking (\ref{ImportantIden}) into account, we have the following identification:
 \begin{align}
 \mathcal{L}_N=\mathscr{I}_2(\mathcal{F}_N).
 \end{align}
 Let $\mathsf{c}_x$ and $\mathsf{f}_u$ be the annihilation operators on $\mathcal{F}$
  such that $
  \{\mathsf{c}_x, \mathsf{c}_y^*\}=\delta_{x, y}\ (x, y\in \Lambda)
  $, $\{\mathsf{f}_u, \mathsf{f}_v^*\}=\delta_{u, v}\ (u, v\in \Omega)$
  and $\{\mathsf{c}_x, {\sf f}_u\}=0=\{{\sf c}_x, {\sf f}_u^*\}\ (x\in \Lambda, u\in \Omega)$.
  Note that $c_{x\sigma}$ and $f_{u\sigma}$ can be rewritten as 
  \begin{align}
  c_{x\uparrow}=\mathsf{c}_x\otimes 1,\ \ f_{u\uparrow}=\mathsf{f}_u\otimes 1,\ \ 
  c_{x\downarrow}=(-1)^{\mathsf{N}} \otimes \mathsf{c}_x,\ \ f_{u\downarrow}=(-1)^{\mathsf{N}} \otimes \mathsf{f}_u, \label{FermiIdn}
  \end{align}
  where $\mathsf{N}$ is the number operator given by 
  $
  \mathsf{N}=\sum_{x\in \Lambda} \mathsf{n}_x^c+\sum_{u\in \Omega} \mathsf{n}_u^f
  $ with $\mathsf{n}_x^c=\mathsf{c}_x^*\mathsf{c}_x$
 and $\mathsf{n}_u^f=\mathsf{f}_u^*\mathsf{f}_u$.
    Using (\ref{LRIden}), we obtain the fundamental identifications:
    \begin{align}
    c_{x\uparrow}=\mathcal{L}(\mathsf{c}_x),\ \ c_{x\downarrow}=\mathcal{L}\big(
    (-1)^{\mathsf{N}} \big) \mathcal{R}(\mathsf{c}_x^*)
   ,\ \ f_{u\uparrow}=\mathcal{L}(\mathsf{f}_u),\ \ f_{u\downarrow}=\mathcal{L}\big(
    (-1)^{\mathsf{N}}
    \big)\mathcal{R}(\mathsf{f}_u^*). \label{AnCrIdn}
    \end{align}
    From these formulas, we can freely produce useful formulas. For instance,
    \begin{align}
    n_{x\uparrow}^c=\mathcal{L}(\mathsf{n}_x^c),\ \ n^c_{x\downarrow}=\mathcal{R}(\mathsf{n}_x^c),\ \ n^f_{u\uparrow}=\mathcal{L}(\mathsf{n}_u^f),\ \ n^f_{u\downarrow}
    =\mathcal{R}(\mathsf{n}_u^f). \label{NumberIden}
     \end{align}

\subsubsection{Basic Hilbert cones}
As   Theorem \ref{pff} suggests,  Hilbert cones are   important in order to prove the 
uniqueness of the ground state of $H$. 
The aim of this subsection is to introduce basic Hilbert  cones which are essential to the proof of Theorem \ref{MainThmHalf}.

We define the Hilbert cone $\mathcal  P$ in $\Hph $ by  \begin{align}
\mathcal{P}=L_+^2(\mathbb R^{|\Lambda|}), 
\end{align}
where $
L_+^2(\mathbb R^{|\Lambda|}) =\{f\in L^2(\mathbb R^{|\Lambda|})\,|\,f(\bq)\geq0\text{ a.e. $\bq$}\}.$
Note that  the number operator $N_{\rm p}$ can be identified with the Hamiltonian of the harmonic oscillators:
\begin{align}
 N_{\rm p}=\sum_{x \in \Lambda} \frac{1}{2}\big(-\Delta_{q_x}+ q_x^2-1\big),
\end{align}
where $\Delta_{q_x}$ is the Laplacian. As is well-known, it holds that 
\begin{align}
e^{-\beta  \omega_0 N_{\rm p}} \rhd 0 \label{NpPI}
\end{align}
w.r.t. $\mathcal{P}$ for all $\beta>0$. This property will be  repeatedly used in the following sections.

Using the identification (\ref{LNIden}), we introduce a  Hilbert cone, $\mathcal{L}_{N, +} $, of $\mathcal{L}_N$ by 
\begin{align}
\mathcal{L}_{N, +} =\mathscr{I}_2(\mathcal{F}_N)_+=\{\psi\in \mathcal L_N\,|\,\Psi_{\vartheta}(\psi)\geq0\}, \label{DefLN+}
\end{align}
where the inequality in (\ref{DefLN+}) means the  standard operator inequality.

Define
\begin{align}
Q_0&=\prod_{u\in\Omega}\left[n_{u\up}^fn_{u\down}^f+\big(1-n_{u\up}^f\big)\big(1-n_{u\down}^f\big)\right].  \label{DefQ0}
\end{align}
Then $Q_0$ is an orthogonal projection on $\mathcal{L}_N=\mathscr{I}_2(\mathcal{F}_N)$.

\begin{Lemma}
$Q_0\mathcal{L}_{N, +}$ is a Hilbert cone in $
 Q_0\mathcal{L}_N
$.
\end{Lemma}

\begin{Proof}
 Using (\ref{NumberIden}),  we find
\begin{align}
n_{u\up}^fn_{u\down}^f&
=\mathcal{L}(\mathsf{n}_u^f) \mathcal{R}(\mathsf{n}_u^f), \ \ 
\big(1-n_{u\up}^f\big)\big(1-n_{u\down}^f\big)=\mathcal{L}(1-\mathsf{n}_u^f) \mathcal{R}(1-\mathsf{n}_u^f).
\end{align}
 Thus,  from Proposition \ref{PPI2}, $n_{u\up}^fn_{u\down}^f \unrhd 0 $ and $\big(1-n_{u\up}^f\big)\big(1-n_{u\down}^f\big)\unrhd0\wrt\mathcal{L}_{N, +} $ hold, which imply  $Q_0\unrhd0\wrt\mathcal{L}_{N, +} $. Since $\mathcal{L}_{N, +} $ is a Hilbert cone in $\mathcal L_N$ and $Q_0\unrhd0\wrt\mathcal{L}_{N, +} $, 
 we readily confirm that $ Q_0\mathcal{L}_{N, +}$ is a Hilbert cone in $Q_0\mathcal L_N$.\qed
\end{Proof}

Next,  we define
\begin{align}
\mathcal Q&=\overline{\mathrm{coni}}\{\psi\otimes f\in Q_0\mathcal{L}_N\otimes\Hph\,|\,\psi\in  Q_0\mathcal{L}_{N, +}, f\in\mathcal  P\}, \label{DefQ}
\end{align}
where $\overline{\mathrm{coni}}\, X$ is the closure of the conical hull of $X$.
The following proposition is  crucial   in the present paper. 

\begin{Prop}\label{QCone}
$\mathcal Q$ is a Hilbert cone in $Q_0\mathcal{L}_N\otimes\Hph$.
\end{Prop}

\begin{Proof}
See Appendix \ref{PfD}.\qed
\end{Proof}

The following basic lemma is often useful.
\begin{Lemma}\label{pi}
Let $A$ be a bounded operator on $Q_0\mathcal{L}_N\otimes\Hph$.
Let $B$ be a self-adjoint operator on $Q_0\mathcal{L}_N\otimes\Hph$,  bounded from below. Assume that $e^{-tB}\unrhd0\wrt\mathcal Q$ for any $t\geq0$.
We have the following:
\begin{itemize}
  \item[{\rm (i)}]If $A$ satisfies  $\i<\phi\otimes f,A\psi\otimes g>\geq0$  for all $\phi,\psi\in  Q_0 \mathcal{L}_{N, +}$ and $f,g\in\mathcal P$, then we have $A\unrhd0\wrt\mathcal Q$.
  \item[{\rm (ii)}] If $A$ satisfies  $\i<\phi\otimes f,A\psi\otimes g>>0$  for all $\phi,\psi\in Q_0 \mathcal{L}_{N, +} \setminus\{0\}$ and $f,g\in\mathcal P\setminus\{0\}$, then we have $A\rhd0\wrt \mathcal{Q}$.
  \item[{\rm (iii)}] 
  Assume that $e^{-tB} \unrhd 0$ w.r.t. $\mathcal{Q}$ for all $t \ge 0$. In addition, 
  assume that,  for all $\phi,\psi\in Q_0 \mathcal{L}_{N, +} \setminus\{0\}$ and $f,g\in\mathcal P\setminus\{0\}$,  there exists a $t\geq0$ such that $\i<\phi\otimes f,e^{-t B}\psi\otimes g>>0$. Then $\{e^{-tB}\}_{t\geq0}$ is ergodic w.r.t. $\mathcal Q$.
  \end{itemize}
\end{Lemma}
\begin{Proof}
(i)
From the definition of $\mathcal Q$, for any $u,v\in\mathcal Q$,  there exist $\psi_n\otimes f_n $ and $\phi_n\otimes g_n\in\mathcal Q$ satisfying 
\begin{align}
u=\sum_{n\geq1}\psi_n\otimes f_n,\ \ 
v=\sum_{n\geq1}\phi_n\otimes g_n. \label{uvExp}
\end{align}
Using these expressions, we obtain
$
\i<u,Av>=\sum_{m,n\geq1}\i<\psi_m\otimes f_m,A\phi_n\otimes g_n>\geq0, 
$
which implies  that $A\unrhd0\wrt\mathcal Q$. 

(ii) Let  $u,v\in\mathcal Q\setminus\{0\}$.
Then $u$ and $v$ can be expressed as (\ref{uvExp}). 
Because $u$ and $v$ are non-zero,  there exist $k,l\in\mathbb N$ such that $\psi_k\otimes f_k\neq0$ and $\phi_l\otimes g_l\neq0$. Hence, we obtain
$
\i<u,Av>=\sum_{m,n\geq1}\i<\psi_m\otimes f_m,A\phi_n\otimes g_n>\geq\i<\psi_k\otimes f_k,A\phi_l\otimes g_l>>0, 
$
which implies  that $A\rhd0\wrt\mathcal Q$. 

(iii)
Let $u, v\in \mathcal{Q} \setminus \{0\}$. We continue to employ  the expressions \eqref{uvExp}. 
Because $u$ and $v$ are non-zero,  there exist $k,l\in\mathbb N$ such that $\psi_k\otimes f_k\neq0$ and $\phi_l\otimes g_l\neq0$.
 By the assumption, there exists a  $t\geq0$ such that $\i<\psi_k\otimes f_k,e^{-tB}\phi_l\otimes g_l>>0$. Since $e^{-tB }\unrhd0\wrt\mathcal Q$,  it holds that $\i<u,e^{-tB}v>\geq\i<\psi_k\otimes f_k,e^{-tB}\phi_l\otimes g_l>>0$.  Hence,  $\{e^{-tB}\}_{t\ge 0}$ is  ergodic w.r.t. $\mathcal{Q}$.\qed
\end{Proof}

 In what follows, we use the following identification:
 \begin{align}
 Q_0\mathcal{L}_N\otimes \Hph
 =\int^{\oplus}_{\mathbb{R}^{|\Lambda|}} Q_0\mathcal{L}_Nd\bq, \label{FiberHil}
 \end{align}
 where the right hand side of (\ref{FiberHil}) is the constant fiber direct integral \cite[Section XIII.16]{Reed1978}.

\begin{Lemma}\label{PPI3}
Let $A\in \mathscr{B}(Q_0\mathcal{L}_N\otimes \Hph)$ be a decomposable operator\footnote{As for the definition of  the decomposable operators, see, e.g., \cite[Section XIII.16]{Reed1978}.}:
\begin{align}
A=\int_{\mathbb R^{|\Lambda|}}^\oplus A(\boldsymbol q)\,d\boldsymbol q.
\end{align}
If $A(\boldsymbol q)\unrhd0\wrt Q_0\mathcal{L}_{N, +}$ for a.e. $\bq$, then we have $A\unrhd0\wrt\mathcal Q$.
\end{Lemma}

\begin{Proof}
Take $\phi,\psi\in
Q_0\mathcal{L}_{N, +}
$ and $f,g\in\mathcal P$, arbitrarily. Since $A(\boldsymbol q)\unrhd0\wrt Q_0\mathcal{L}_{N, +}$ and $f(\bq), g(\bq)\geq0$ a.e., we have
\begin{align}
\l<\phi\otimes f,A\psi\otimes g>=\int_{\mathbb R^{|\Lambda|}}f(\boldsymbol q)g(\boldsymbol q)\l<\phi,A(\boldsymbol q)\psi>\,d\boldsymbol q\geq0.
\end{align}
By Lemma \ref{pi},  we conclude that  $A\unrhd0\wrt\mathcal Q$. \qed
\end{Proof}

\begin{Lemma}\label{PPI4}
Let $A\in\mathscr B(\mathcal{L}_N)$. Assume the following:
\begin{itemize}
\item[{\rm (i)}] $A$ commutes with $Q_0$.
\item[{\rm (ii)}] $A\unrhd 0$ w.r.t. $\mathcal{L}_{N, +}$ where $\mathcal{L}_{N, +}$ is given by (\ref{DefLN+}).
\end{itemize} 
Then we have $A \restriction Q_0\mathcal{L}_{N} \unrhd0\wrt Q_0\mathcal{L}_{N, +}$, where $
A \restriction Q_0\mathcal{L}_{N}
$ is the restriction of $A$ to $
Q_0\mathcal{L}_{N}
$.
\end{Lemma}

\begin{Proof}
Since $Q_0\unrhd0\wrt\mathcal{L}_{N, +}$, we see $Q_0AQ_0\unrhd0\wrt\mathcal{L}_{N, +}$. Hence, for any $\phi,\psi\in\mathcal{L}_{N, +}$, $\i<\phi,Q_0AQ_0\psi>=\i<Q_0\phi,AQ_0\psi>\geq0$ holds. Thus, we have $A\unrhd0\wrt Q_0\mathcal{L}_{N, +}$.\qed
\end{Proof}

\begin{Lemma}\label{PPI5}
Let $A, B \in\mathscr B(\mathcal{F}_N)$. 
Assume that $A\otimes1+1\otimes\vartheta A\vartheta$ and $B\otimes\vartheta B\vartheta$ commute with $Q_0$. 
Then we have
\begin{align}
\exp\{(A\otimes1+1\otimes\vartheta A\vartheta)\restriction Q_0\mathcal{L}_N\}&\unrhd0\wrt Q_0\mathcal{L}_{N, +}, \\
B\otimes\vartheta B\vartheta\restriction Q_0\mathcal{L}_N&\unrhd0\wrt Q_0\mathcal{L}_{N, +}.
\end{align}
\end{Lemma}

\begin{Proof}
Using Proposition \ref{PPI2},  we have
\begin{align}
&e^{A\otimes1+1\otimes\vartheta A\vartheta}=e^A\otimes\vartheta e^A\vartheta\unrhd0\wrt\mathcal{ L}_{N, +}, \\
& B\otimes\vartheta B\vartheta\unrhd0\wrt\mathcal{ L}_{N, +}.
\end{align}
Thus, applying Lemma \ref{PPI4}, we obtain the desired results. 
\qed
\end{Proof}

\begin{Lemma}\label{PPIQ}
Let $A\in\mathscr B(Q_0\mathcal{L}_N)$. 
Assume that $A\unrhd0\wrt Q_0\mathcal{L}_{N, +}$.
Then we have $A\otimes1\unrhd0\wrt\mathcal{Q}$,
\end{Lemma}
\begin{Proof}
For any $\phi,\psi\in Q_0\mathcal{L}_{N, +}$ and $ f,g\in\mathcal{P}$, we observe that 
\begin{align}
\i<\phi\otimes f,A\otimes1\psi\otimes g>=\i<\phi,A\psi>\i<f,g>\geq 0. 
\end{align}
Hence, by applying Lemma \ref{pi}, we conclude that $A\otimes1\unrhd0\wrt\mathcal{Q}$.\qed
\end{Proof}

\begin{Lemma}\label{PPIQ2}
Let $A\in \mathscr{B}(Q_0\mathcal{L}_N\otimes \Hph)$.
Assume $A\unrhd0\wrt\mathcal{Q}$.
Then we have $e^A\unrhd0\wrt\mathcal{Q}$.
\end{Lemma}
\begin{Proof}
By the assumption, we obtain $A^n\unrhd0\wrt\mathcal{Q}, n=0,1,\ldots$. Thus,  we find
$
e^A=\sum_{n=0}^\infty\frac{1}{n!}A^n\unrhd0\wrt\mathcal{Q}.
$
\qed
\end{Proof}

\subsection{Basic transformations}

In order to properly apply  the theory given in Section \ref{OpSection}, we introduce
a useful transformation;
the definition of $\mathcal{U}$, i.e., (\ref{DefUTrn}) and Corollary \ref{BasicTrn} are fundamental results in this subsection.

We begin with the following lemma.
\begin{Lemma}\label{HolePart}
There exists a unitary operator $U$ on $\mathcal{L}_N$ satisfying 
\begin{align}
U^*c_{x\up}U=c_{x\up},\quad U^*f_{u\up}U=f_{u\up},\quad U^*c_{x\down}U=\gamma_xc_{x\down}^*,\quad U^*f_{u\down}U=\gamma_u\mathrm{sgn}J_{x,u}f_{u\down}^*, \label{DefUniU}
\end{align}
where
\begin{align}
\gamma_z=
\begin{cases}
-1\ \ &(z\in\Lambda_1\ or\ z\in\Omega_1)\\
1\ \ &(z\in\Lambda_2\ or\ z\in\Omega_2), \label{HPProp}
\end{cases}
\end{align}
and 
$\mathrm{sgn}J_{x,u}$ is determined by the assumption  \ref{C2}.
\end{Lemma}

\begin{Proof}
Let $U_1$ be the  unitary operator on $\mathcal{H}_{\rm c}$ such that
\begin{align}
U_1^*c_{x\up}U_1=c_{x\up},\quad U_1^*c_{x\down}U_1=\gamma_xc_{x\down}^*.
\end{align}
Note that $U_1$ is  the standard hole-particle transformation on $\mathcal{H}_{\rm c}$.

By \ref{C2}, for any $u\in\Omega$,  there exists an $x_u\in\Lambda$ satisfying $J_{x_u,u}\neq0$.  Note that $\mathrm{sgn}J_{x_u, u}$ is independent of the choice of $x_u$.
Let $U_2$ be the  unitary operator on $\mathcal{H}_{\rm f}$ such that
\begin{align}
U_2^*f_{u\up}U_2=f_{u\up},\quad U_2^*f_{u\down}U_2=\gamma_u\mathrm{sgn}J_{x_u,u}f_{u\down}^*.
\end{align}
Choosing  $U=U_1 \otimes U_2$,  we readily confirm  that $U$ satisfies the desired properties in (\ref{HPProp}).
\qed
\end{Proof}

For each $x\in \Lambda$, define self-adjoint operators, $p_x$ and $q_x$, by 
\begin{align}
p_x=\frac{i}{\sqrt2}(\overline{b_x^*-b_x}), \ \ \  q_ x=\frac{1}{\sqrt{2}} (\overline{b_x^*+b_x}), \label{Defpq}
\end{align}
where $\overline{A}$ is the closure of $A$.
As is well-known, these operators satisfy the standard commutation relation:
$[q_x, p_y]=i\delta_{x,y}$.

\begin{Lemma} \label{BigTr1}
We set 
\begin{align}
 L_c=-i\frac{\sqrt2}{\omega_0}\sum_{x,y\in\Lambda}g_{x,y}n_x^cp_y.
\end{align}
The unitary operator $e^{L_c}$ is called the {\it Lang-Firsov transformation} which was first introduced in \cite{Lang1963}.
Let $N_{\rm p}$ be the phonon number operator: $N_{\rm p}=\sum_{x\in \Lambda} b_x^*b_x$. 
Then
\begin{align}
&e^{i\frac{\pi}{2}N_{\rm p}}e^{L_c}He^{-L_c}e^{-i\frac{\pi}{2}N_{\rm p}}\no
&=-T_\up^+-T_\down^++\sum_{x\in\Lambda,u\in\Omega}J_{x,u}{\boldsymbol s}_x\cdot{\boldsymbol S}_u+\mathbb U+\omega_0N_{\rm p}-\omega_0^{-1}g^2|\Lambda|
\end{align}
holds,  where $T_{\sigma}^{\pm}, \mathbb{U}$ and $g$ are defined  respectively by 
\begin{itemize}
\item $\displaystyle 
T_\sigma^\pm=\sum_{x,y\in\Lambda}t_{x,y}c_{x\sigma}^*c_{y\sigma}\exp\left(\pm i\Phi_{x,y}\right)$ with $
\Phi_{x,y}=\frac{\sqrt2}{\omega_0}\sum_{z\in\Lambda}(g_{xz}-g_{yz})q_z;
$
\item 
$\displaystyle 
\mathbb U=\sum_{x,y\in\Lambda}U_{\mathrm{eff},xy}(n_x^c-1)(n_y^c-1)
$
with $U_{{\rm eff}, xy}$ given by (\ref{DefUeff});
\item $
g=\sum_{x\in\Lambda}g_{x,y}
$.
Note that $g$ is independent of $y$ due to  \ref{C4}.
\end{itemize}
\end{Lemma}

\begin{Proof}
Let 
$\mathbb{T}=\displaystyle \sum_{x,y\in\Lambda}\sum_{\sigma=\up,\down}t_{x,y}c_{x\sigma}c_{y\sigma}$. 
Applying  properties of basic operators  in  Appendix \ref{LFTrnSec}, we have
\begin{align}
e^{i\frac{\pi}{2}N_{\rm p}}e^{L_c}\mathbb Te^{-L_c}e^{-i\frac{\pi}{2}N_{\rm p}}
&=-T_\up^+-T_\down^+,\\
e^{i\frac{\pi}{2}N_{\rm p}}e^{L_c}\Bigg(\sum_{x\in\Lambda,u\in\Omega}J_{x,u}{\boldsymbol s}_x\cdot{\boldsymbol S}_u \Bigg)e^{-L_c}e^{-i\frac{\pi}{2}N_{\rm p}}&=\sum_{x\in\Lambda,u\in\Omega}J_{x,u}{\boldsymbol s}_x\cdot{\boldsymbol S}_u,\\
e^{i\frac{\pi}{2}N_{\rm p}}e^{L_c}\Bigg\{\sum_{x,y\in\Lambda}U_{x,y}(n_x^c-1)(n_y^c-1)\Bigg\}e^{-L_c}e^{-i\frac{\pi}{2}N_{\rm p}}&=\sum_{x,y\in\Lambda}U_{x,y}(n_x^c-1)(n_y^c-1),\\
e^{L_c}\Bigg\{\sum_{x,y\in\Lambda}g_{x,y}n_x^c(b_y^*+b_y)\Bigg\}e^{-L_c}
&=\sum_{x,y\in\Lambda}g_{x,y}n_x^c(b_y^*+b_y)-\frac{2}{\omega_0}\sum_{x,y,z\in\Lambda}g_{x,z}g_{y,z}n_x^cn_y^c,\label{Transbc}\\
e^{L_c}N_{\rm p}e^{-L_c}
&=N_{\rm p}-\frac{1}{\omega_0}\sum_{x,y\in\Lambda}g_{x,y}n_x^c(b_y^*+b_y)+\omega_0^{-2}\sum_{x,y,z\in\Lambda}g_{x,z}g_{y,z}n_x^cn_y^c. \label{Transbb}
\end{align}
Combining (\ref{Transbc}) and (\ref{Transbb}), we find
\begin{align}
&e^{i\frac{\pi}{2}N_{\rm p}}e^{L_c}\Bigg\{ \sum_{x,y\in\Lambda}g_{x,y}n_x^c(b_y^*+b_y)+\omega_0 N_{\rm p}\Bigg\}e^{-L_c}e^{-i\frac{\pi}{2}N_{\rm p}} \no
&=\omega_0 N_{\rm p}-\sum_{x,y\in\Lambda}V_{x,y}n_x^cn_y^c \no
&=\omega_0 N_{\rm p}-\sum_{x,y\in\Lambda}V_{x,y}(n_x^c-1)(n_y^c-1)-\sum_{x,y\in\Lambda}V_{x,y}(n_x^c+n_y^c)+\sum_{x,y\in\Lambda}V_{x,y} \no
&=\omega_0 N_{\rm p}-\sum_{x,y\in\Lambda}V_{x,y}(n_x^c-1)(n_y^c-1)-\omega_0^{-1}\sum_{x,y,z\in\Lambda}g_{x,z}g_{y,z}(n_x^c+n_y^c)+\omega_0^{-1}\sum_{x,y,z\in\Lambda}g_{x,z}g_{y,z} \no
&=\omega_0 N_{\rm p}-\sum_{x,y\in\Lambda}V_{x,y}(n_x^c-1)(n_y^c-1)-\omega_0^{-1}g\sum_{x,z\in\Lambda}g_{x,z}n_x^c-\omega_0^{-1}g\sum_{y,z\in\Lambda}g_{y,z}n_y^c+\omega_0^{-1}g^2|\Lambda| \no
&=\omega_0 N_{\rm p}-\sum_{x,y\in\Lambda}V_{x,y}(n_x^c-1)(n_y^c-1)-2\omega_0^{-1}g^2\sum_{x\in\Lambda}n_x^c+\omega_0^{-1}g^2|\Lambda| \no
&=\omega_0 N_{\rm p}-\sum_{x,y\in\Lambda}V_{x,y}(n_x^c-1)(n_y^c-1)-\omega_0^{-1}g^2|\Lambda|,
\end{align}
where $V_{x,y}=\omega_0^{-1} \sum_{z\in \Lambda}g_{x, z} g_{y, z}$.
 Therefore, we finally obtain
\begin{align}
&e^{i\frac{\pi}{2}N_{\rm p}}e^{L_c}He^{-L_c}e^{-i\frac{\pi}{2}N_{\rm p}}\no
&=-T_\up^+-T_\down^++\sum_{x\in\Lambda,u\in\Omega}J_{x,u}{\boldsymbol s}_x\cdot{\boldsymbol S}_u+\mathbb U+\omega_0N_{\rm p}-\omega_0^{-1}g^2|\Lambda|.
\end{align}
\qed
\end{Proof}

\begin{Lemma}\label{BigTr2}
Set
\begin{align}
H'=-T_\up^+-T_\down^-+\sum_{x\in\Lambda,u\in\Omega}J_{x,u}{\boldsymbol s}_x\cdot{\boldsymbol S}_u+\mathbb{U}.
\end{align}
Then we have
\begin{align}
U^*H'U=R-\frac{1}{2}\sum_{x\in\Lambda,u\in\Omega}|J_{x,u}|\left(c_{x\up}^*f_{u\up}c_{x\down}^*f_{u\down}+f_{u\up}^*c_{x\up}f_{u\down}^*c_{x\down}\right)-\tilde{\mathbb{U}}, \label{BigTr2-1}
\end{align}
where
\begin{align}
R
&=-T_\up^+-T_\down^-+\frac{1}{4}\sum_{x\in\Lambda,u\in\Omega}J_{x,u}(n_x^c-1)(n_u^f-1)+\sum_{x,y\in\Lambda}U_{\mathrm{eff},x,y}(n_{x\up}^cn_{y\up}^c+n_{x\down}^cn_{y\down}^c),\\
\tilde{\mathbb{U}}
&=2\sum_{x,y\in\Lambda}U_{\mathrm{eff},x,y}n_{x\up}^cn_{y\down}^c.
\end{align}
\end{Lemma}

\begin{Proof}
By  using \ref{C1} and \ref{C5}, we have
\begin{align}
U^*(T_\up^++T_\down^+)U
&=T_\up^++\sum_{x,y\in\Lambda}t_{x,y}\gamma_x\gamma_yc_{x\down}c_{y\down}^*\exp(i\Phi_{x,y})\no
&=T_\up^+-\sum_{x,y\in\Lambda}t_{x,y}c_{x\down}c_{y\down}^*\exp(i\Phi_{x,y}) \no
&=T_\up^++\sum_{x,y\in\Lambda}t_{x,y}c_{y\down}^*c_{x\down}\exp(-i\Phi_{y,x}) \no
&=T_\up^++T_\down^- \label{UKine},\\
U^*\mathbb{U}U
&=\sum_{x,y\in\Lambda}U_{\mathrm{eff},x,y}(n_{x\up}^c-n_{x\down}^c)(n_{y\up}^c-n_{y\down}^c)\no
&=\sum_{x,y\in\Lambda}U_{\mathrm{eff},x,y}(n_{x\up}^cn_{y\up}^c+n_{x\down}^cn_{y\down}^c)-2\sum_{x,y\in\Lambda}U_{\mathrm{eff},x,y}n_{x\up}^cn_{y\down}^c,
\end{align}
and
\begin{align}
&U^*\sum_{x\in\Lambda,u\in\Omega}J_{x,u}{\boldsymbol s}_x\cdot{\boldsymbol S}_uU\no
&=\sum_{x\in\Lambda,u\in\Omega}J_{x,u}U^*\Big(\frac{1}{2}s_x^+S_u^-+\frac{1}{2}s_x^-S_u^++s_x^{(3)}S_u^{(3)}\Big)U\no
&=\sum_{x\in\Lambda,u\in\Omega}J_{x,u}U^*\Big\{\frac{1}{2}c_{x\up}^*c_{x\down}f_{u\down}^*f_{u\up}+\frac{1}{2}c_{x\down}^*c_{x\up}f_{u\up}^*f_{u\down}+\frac{1}{4}(n_{x\up}^c-n_{x\down}^c)(n_{u\up}^f-n_{u\down}^f)\Big\}U\no
&=-\frac{1}{2}\sum_{x\in\Lambda,u\in\Omega}|J_{x,u}|\left(c_{x\up}^*f_{u\up}c_{x\down}^*f_{u\down}+f_{u\up}^*c_{x\up}f_{u\down}^*c_{x\down}\right)+\frac{1}{4}\sum_{x\in\Lambda,u\in\Omega}J_{x,u}(n_x^c-1)(n_u^f-1). \label{USpinS}
\end{align}
Combining (\ref{UKine}) and (\ref{USpinS}), we conclude (\ref{BigTr2-1}). 
\qed
\end{Proof}

Define 
\begin{align}
\mathcal U=e^{-L_c}e^{-i\frac{\pi}{2}N_{\rm p}}U. \label{DefUTrn}
\end{align}
Note that 
\begin{align}
\mathcal{U}^*P_0\mathcal{U}=Q_0. \label{PQU}
\end{align}
Hence, $\mathcal U^*H\mathcal U$ acts on $Q_0 \mathcal{L}_N \otimes \Hph$.
Applying Lemmas \ref{BigTr1} and \ref{BigTr2}, we obtain the following:

\begin{Coro}\label{BasicTrn}
Let
\begin{align}
\mathbb{J}=\frac{1}{2}\sum_{x\in\Lambda,u\in\Omega}|J_{x,u}|\left(c_{x\up}^*f_{u\up}c_{x\down}^*f_{u\down}+f_{u\up}^*c_{x\up}f_{u\down}^*c_{x\down}\right). \label{DefJ}
\end{align}
We have
\begin{align}
\mathcal U^*H\mathcal U
=R-\mathbb{J}-\tilde{\mathbb{U}}+\omega_0N_{\rm p}-\omega_0^{-1}g^2|\Lambda|.
\end{align}
\end{Coro}

\subsection{
Positivity preserving property of  $
e^{-\beta \mathcal{U}^* H \mathcal{U} } 
$ 
}

The goal in this subsection is to prove the following proposition:
\begin{Prop}\label{PPSemi}
Suppose that $U_{\mathrm{eff}}$ is positive semi-definite.
For all $\beta \ge 0$, one has 
$
e^{-\beta \mathcal{U}^* H \mathcal{U} } \unrhd 0
$ w.r.t. $\mathcal{Q}$.
\end{Prop}
A role of Proposition \ref{PPSemi} is as follows: We wish to employ Theorem \ref{pff} (the Perron-Frobenius-Faris theorem) to prove the the uniqueness of the ground state of $H$. 
Proposition \ref{PPSemi} is a basic input in order to apply Theorem \ref{pff}.
The proof of Proposition \ref{PPSemi} will be given in the end of this subsection.

Before we proceed to the proof of Proposition \ref{PPSemi}, we remark that the following:
By using arguments similar to those of the proof of Proposition \ref{PPSemi},  we obtain 
\begin{Lemma}\label{PPSemi2}
Suppose that $U_{\mathrm{eff}}$ is positive semi-definite.
For all $\beta \ge 0$, one has 
$
e^{-\beta (R-\frac{1}{2} \mathbb{J}+\omega_0 N_{\rm p}) } \unrhd 0
$ w.r.t. $\mathcal{Q}$.
\end{Lemma}
Note that Lemma \ref{PPSemi2} will be  repeatedly used in Subsection \ref{PfUniSub}.

Now,  we return to the proof of Proposition \ref{PPSemi}. 

\begin{Lemma}\label{PPI6}
For each $x,y\in\Lambda$ and $\boldsymbol q=(q_z)_{z\in\Lambda}\in\mathbb R^{|\Lambda|}$, 
 define
\begin{align}
R(\boldsymbol q)
&=-\sum_{\substack{x,y\in\Lambda}}t_{x,y}c_{x\up}^*c_{y\up}\exp\left( i\Phi_{x,y}(\boldsymbol q)\right)
-\sum_{\substack{x,y\in\Lambda}}t_{x,y}c_{x\down}^*c_{y\down}\exp\left( -i\Phi_{x,y}(\boldsymbol q)\right) \no
&\quad+\frac{1}{4}\sum_{x\in\Lambda,u\in\Omega}J_{x,u}(n_x^c-1)(n_u^f-1)
+\sum_{x,y\in\Lambda}U_{\mathrm{eff},x,y}(n_{x\up}^cn_{y\up}^c+n_{x\down}^cn_{y\down}^c),      
\end{align}
where $
\Phi_{x,y}(\boldsymbol q)=\frac{\sqrt2}{\omega_0}\sum_{z\in\Lambda}(g_{xz}-g_{yz})q_z.
$
 Then we have $e^{-\beta R(\boldsymbol q)}\unrhd0\wrt Q_0\mathcal{L}_{N, +}$ for any $\boldsymbol q\in\Lambda\in\mathbb R^{|\Lambda|}$ and $\beta\geq0$.
\end{Lemma}

\begin{Proof}
By the definition of $Q_0$, $n_{u\up}^f=n_{u\down}^f$ holds on $Q_0\mathcal L_N$.  Hence, by \eqref{NumberIden}, 
\begin{align}
&\sum_{x\in\Lambda,u\in\Omega}J_{x,u}(n_x^c-1)(n_u^f-1)
+4\sum_{x,y\in\Lambda}U_{\mathrm{eff},x,y}(n_{x\up}^cn_{y\up}^c+n_{x\down}^cn_{y\down}^c)\no
&=\sum_{x\in\Lambda,u\in\Omega}J_{x,u}\left(n_x^cn_u^f-n_x^c-n_u^f+1\right)
+4\sum_{x,y\in\Lambda}U_{\mathrm{eff},x,y}(n_{x\up}^cn_{y\up}^c+n_{x\down}^cn_{y\down}^c)\no
&=\sum_{x\in\Lambda,u\in\Omega}J_{x,u}\left(2n_{x\up}^cn_{u\up}^f+2n_{x\down}^cn_{u\down}^f-n_{x\up}^c-n_{u\up}^f-n_{x\down}^c-n_{u\down}^f+1\right)
+4\sum_{x,y\in\Lambda}U_{\mathrm{eff},x,y}(n_{x\up}^cn_{y\up}^c+n_{x\down}^cn_{y\down}^c)\no
&=\sum_{x\in\Lambda,u\in\Omega}J_{x,u}\left(2n_{x\up}^cn_{u\up}^f-n_{x\up}^c-n_{u\up}^f+\frac{1}{2}\right)
+4\sum_{x,y\in\Lambda}U_{\mathrm{eff},x,y}n_{x\up}^cn_{y\up}^c \no
&\quad\quad+\sum_{x\in\Lambda,u\in\Omega}J_{x,u}\left(2n_{x\down}^cn_{u\down}^f-n_{x\down}^c-n_{u\down}^f+\frac{1}{2}\right)
+4\sum_{x,y\in\Lambda}U_{\mathrm{eff},x,y}n_{x\down}^cn_{y\down}^c \no
&=\mathcal{L}\left(\mathsf{J_n}\right)+\mathcal{R}\left(\vartheta\mathsf{J_n}\vartheta \right)\label{J_nExp}
\end{align}
on $Q_0\mathcal L_N$, where
$
\mathsf{J_n}=\sum_{x\in\Lambda,u\in\Omega}J_{x,u}\left(2\mathsf n_{x}^c\mathsf n_{u}^f-\mathsf n_{x}^c-\mathsf n_{u}^f+\frac{1}{2}\right)+4\sum_{x,y\in\Lambda}U_{\mathrm{eff},x,y}\mathsf{n}_{x}^c\mathsf{n}_{y}^c
$.
We set
\begin{align}
\mathsf{J_c}(\bq)=-\sum_{x,y\in\Lambda} t_{x,y} \mathsf c_{x}^*\mathsf c_{y}\exp(i\Phi_{x,y}(\boldsymbol q)). \label{DefJ_c(q)}
\end{align}
 Then using (\ref{J_nExp}), we find that 
\begin{align}
R(\boldsymbol q)
&=\mathcal{L}\left(
\mathsf{J_c}(\bq)
\right)
+\mathcal{R}\left(\vartheta
\mathsf{J_c}(\bq)
\vartheta\right)
+\frac{1}{4}\mathcal{L}\left(
\mathsf{J_n}
\right)
+\frac{1}{4}\mathcal{R}\left(\vartheta
\mathsf{J_n}
\vartheta\right)
\end{align}
holds on $Q_0\mathcal L_N$. 
Thus, we can write $R(\bq)$ as 
$
R(\bq)=\mathsf{R}(\bq)\otimes 1+1\otimes \vartheta \mathsf{R}(\bq) \vartheta
$ with 
\begin{align}
\mathsf{R}(\bq )=\mathsf{J_c}(\bq)+\frac{1}{4}\mathsf{J_n}.\label{DefSfR(q)}
\end{align}
 Using this expression and  Lemma \ref{PPI5}, we can  conclude that  $e^{-\beta R(\boldsymbol q)}\unrhd0\wrt Q_0\mathcal{L}_{N, +}$ for each  $\boldsymbol q\in\Lambda\in\mathbb R^{|\Lambda|}$ and $\beta\geq0$.\qed
\end{Proof}
 
 \subsubsection*{Proof of Proposition \ref{PPSemi}}

By Lemmas \ref{PPI3} and \ref{PPI6}, we have
\begin{align}
e^{-\beta R}=\int_{\mathbb R^{|\Lambda|}}^\oplus e^{-\beta R(\boldsymbol q)}\,d\boldsymbol q\unrhd0\wrt\mathcal Q. \label{RPP}
\end{align}

Next, we will show that 
\begin{align}
\tilde{\mathbb{U}}\unrhd0\wrt\mathcal{Q}. \label{UPP}
\end{align}
Note that  $\tilde{\mathbb{U}}$ commutes with $Q_0$. Hence, taking Lemmas \ref{PPI4} and \ref{PPIQ} into account, it suffices to prove that  $\tilde{\mathbb{U}}\unrhd0\wrt\mathcal{L}_{N,+}$.
Using the identifications \eqref{FermiIdn}, we can express $
\tilde{\mathbb{U}}
$ as $
\tilde{\mathbb{U}}=2 \sum_{x, y\in \Lambda} U_{\mathrm{eff},x,y}\mathsf{n}_x^c\otimes \vartheta \mathsf{n}_y^c \vartheta
$. Hence, by Proposition \ref{PPI2}, we conclude that 
$\tilde{\mathbb{U}}\unrhd0\wrt\mathcal{Q}$.

Recall the definition of $\mathbb{J}$, i.e., \eqref{DefJ}.
Using arguments similar to those in the proof of \eqref{UPP}, we can show
\begin{align}
\mathbb{J}\unrhd0\wrt\mathcal{Q}.
\end{align}
Hence, by applying Lemma \ref{PPIQ2},  we readily confirm that 
\begin{align}
\exp\left[\frac{\beta}{n}(\mathbb{J}+\tilde{\mathbb{U}})\right]&\unrhd0\wrt\mathcal Q\label{HopPP}
\end{align}
for all $\beta \ge 0$ and $n\in \mathbb{N}$.
By the Trotter product formula \cite[Theorem S.20]{Reed1981}, we have
\begin{align}
\exp\left[-\beta \mathcal{U}^* H \mathcal{U}\right]
&=\exp\left[-\beta R+\beta  \mathbb{J}+\beta \tilde{\mathbb{U}}-\beta\omega_0N_{\rm p}+\beta\omega_0^{-1}g^2|\Lambda|\right]\no
&=e^{\beta\omega_0^{-1}g^2|\Lambda|}{\rm s}\text{-}\lim_{n\to\infty}\left\{\exp\left[-\frac{\beta}{n}R\right]\exp\left[\frac{\beta}{n} (\mathbb{J}+\tilde{\mathbb{U}})\right]\exp\left[-\frac{\beta}{n}\omega_0N_{\rm p}\right]\right\}^n. \label{TroKatoA}
\end{align}
Using (\ref{NpPI}), (\ref{RPP}) and (\ref{HopPP}), 
we see that the right hand side of (\ref{TroKatoA}) is positivity preserving w.r.t. $\mathcal{Q}$ for all $\beta\geq 0$.\qed

\subsection{Useful operator inequalities}
For later use, we will  prove some operator  inequalities here.

Let
\begin{align}
F&=\{(x,u)\in\Lambda\times\Omega\,|\,J_{x,u}\neq0\},\\
F_x&=\{u\in\Omega\,|\,J_{x,u}\neq0\}.
\end{align}

\begin{Lemma}
We have the following equalities:
\begin{itemize}
\item[{\rm (i)}]
\begin{align}
&N=2\sum_{u\in\Omega}n_{u\up}^fn_{u\down}^f+\sum_{x\in\Lambda}\sum_{u\in F_x}|F_x|^{-1}(n_{x\up}^c+n_{x\down}^c)n_{u\up}^fn_{u\down}^f+\sum_{x\in\Lambda}\sum_{u\in F_x}|F_x|^{-1}(n_{x\up}^c+n_{x\down}^c)(1-n_{u\up}^f)(1-n_{u\down}^f). \label{eq:4.3}
\end{align}
\item[{\rm (ii)}]
\begin{align}
&n_{x\up}^c(1-n_{u\up}^f)n_{x\down}^c(1-n_{u\down}^f)+(1-n_{x\up}^c)n_{u\up}^f(1-n_{x\down}^c)n_{u\down}^f+(n_{x\up}^c+n_{x\down}^c)n_{u\up}^fn_{u\down}^f=n_{x\up}^cn_{x\down}^c+n_{u\up}^fn_{u\down}^f. \label{eq:4.1}
\end{align}
\item[{\rm (iii)}] 
\begin{align}
&n_{x\up}^c(1-n_{u\up}^f)n_{x\down}^c(1-n_{u\down}^f)+(1-n_{x\up}^c)n_{u\up}^f(1-n_{x\down}^c)n_{u\down}^f+(n_{x\up}^c+n_{x\down}^c)(1-n_{u\up}^f)(1-n_{u\down}^f)+1\no
&=(1+n_{x\up}^c)(1+n_{x\down}^c)(1-n_{u\up}^f)(1-n_{u\down}^f)+n_{u\up}^fn_{u\down}^f+(1-n_{x\up}^c)n_{u\up}^f(1-n_{x\down}^c)n_{u\down}^f.\label{eq:4.2}
\end{align}
\end{itemize}

\end{Lemma}

\begin{Proof}
By the definition of $Q_0$, i.e., (\ref{DefQ0}),  we have
\begin{align}
 n_{u\up}^f&=n_{u\down}^f, \label{nu=nd}\\
1&=n_{u\up}^fn_{u\down}^f+(1-n_{u\up}^f)(1-n_{u\down}^f)\label{eq:4.4}
\end{align}
on $Q_0\mathcal{L}_N$.

(i) Recalling that $N_{\rm e}=N$ on $Q_0 \mathcal{L}_N$, we obtain
\begin{align}
N&=N_{\rm e}\no
&=\sum_{x\in\Lambda}(n_{x\up}^c+n_{x\down}^c)+\sum_{u\in\Omega}(n_{u\up}^f+n_{u\down}^f)\no
&=\sum_{x\in\Lambda}\sum_{u\in F_x}|F_x|^{-1}(n_{x\up}^c+n_{x\down}^c)\big\{n_{u\up}^fn_{u\down}^f+(1-n_{u\up}^f)(1-n_{u\down}^f)\big\}+2\sum_{u\in\Omega}n_{u\up}^fn_{u\down}^f\no
&=
\mbox{the right hand side  of (\ref{eq:4.3})}.
\end{align}
In the third equality, we have used (\ref{nu=nd}) and (\ref{eq:4.4}).

(ii) We observe
\begin{align}
&n_{x\up}^c(1-n_{u\up}^f)n_{x\down}^c(1-n_{u\down}^f)+(1-n_{x\up}^c)n_{u\up}^f(1-n_{x\down}^c)n_{u\down}^f+(n_{x\up}^c+n_{x\down}^c)n_{u\up}^fn_{u\down}^f\no
&=n_{x\up}^c(1-n_{u\up}^f)n_{x\down}^c(1-n_{u\down}^f)+(1+n_{x\up}^cn_{x\down}^c)n_{u\up}^fn_{u\down}^f\no
&\overset{\eqref{eq:4.4}}{=}n_{x\up}^cn_{x\down}^c+n_{u\up}^fn_{u\down}^f.
\end{align}

(iii) We have
\begin{align}
&n_{x\up}^c(1-n_{u\up}^f)n_{x\down}^c(1-n_{u\down}^f)+(1-n_{x\up}^c)n_{u\up}^f(1-n_{x\down}^c)n_{u\down}^f+(n_{x\up}^c+n_{x\down}^c)(1-n_{u\up}^f)(1-n_{u\down}^f)+1\no
&\overset{\eqref{eq:4.4}}{=}(1+n_{x\up}^c+n_{x\down}^c+n_{x\up}^cn_{x\down}^c)(1-n_{u\up}^f)(1-n_{u\down}^f)+n_{u\uparrow}^fn_{u\downarrow}^f+(1-n_{x\up}^c)n_{u\up}^f(1-n_{x\down}^c)n_{u\down}^f\no
&=(1+n_{x\up}^c)(1+n_{x\down}^c)(1-n_{u\up}^f)(1-n_{u\down}^f)+n_{u\up}^fn_{u\down}^f+(1-n_{x\up}^c)n_{u\up}^f(1-n_{x\down}^c)n_{u\down}^f.
\end{align}
\qed
\end{Proof}

The following proposition is essential for the proof of  Theorem \ref{MainThmHalf}:

\begin{Prop}\label{proj}
One obtains
\begin{align}
\frac{8}{J^2}\big\{J(|\Lambda|+|\Omega|)+\mathbb J\big\}^2\unrhd\sum_{x\in\Lambda}n_{x\up}^cn_{x\down}^c+\sum_{u\in\Omega}n_{u\up}^fn_{u\down}^f\wrt\mathcal Q,
\end{align}
where  $J=\min_{(x,u)\in F}|J_{x,u}|$.
\end{Prop}

\begin{Proof}
Let $V_{x,u}=c_{x\up}^*f_{u\up}c_{x\down}^*f_{u\down}$.  Then we have
\begin{align}
V_{x,u}V_{x,u}^*+V_{x,u}^*V_{x,u}=n_{x\up}^c(1-n_{u\up}^f)n_{x\down}^c(1-n_{u\down}^f) +(1-n_{x\up}^c)n_{u\up}^f(1-n_{x\down}^c)n_{u\down}^f. \label{EqVcf}
\end{align}
Because of Lemmas \ref{PPI5}  and  \ref{PPIQ},  it holds that $V_{x,u}\unrhd 0 $ and $V_{x,u}^*\unrhd0\wrt\mathcal Q$. Hence,  we find
\begin{align}
\mathbb J^2&\unrhd\frac{1}{4}\sum_{x\in\Lambda,u\in\Omega}|J_{x,u}|^2\left(V_{x,u}+V_{x,u}^*\right)^2\no
&\unrhd\frac{J^2}{4}\sum_{(x,u)\in F}\left(V_{x,u}V_{x,u}^*+V_{x,u}^*V_{x,u}\right)\no
&\overset{\eqref{EqVcf}}{=}\frac{J^2}{4}\sum_{(x,u)\in F}\Big\{ n_{x\up}^c(1-n_{u\up}^f)n_{x\down}^c(1-n_{u\down}^f)+(1-n_{x\up}^c)n_{u\up}^f(1-n_{x\down}^c)n_{u\down}^f\Big\}\wrt\mathcal Q.
\end{align}
 Using  $\sum_{(x,u)\in F}=\sum_{x\in\Lambda}\sum_{u\in F_x}$ and recalling that $N=|\Lambda|+|\Omega|$, we obtain
\begin{align}
&2\sum_{(x,u)\in F}\Big\{ n_{x\up}^c(1-n_{u\up}^f)n_{x\down}^c(1-n_{u\down}^f)+(1-n_{x\up}^c)n_{u\up}^f(1-n_{x\down}^c)n_{u\down}^f\Big\}+2(|\Lambda|+|\Omega|)\no
&\overset{\eqref{eq:4.3}}{=}2\sum_{(x,u)\in F}\Big\{ n_{x\up}^c(1-n_{u\up}^f)n_{x\down}^c(1-n_{u\down}^f)+(1-n_{x\up}^c)n_{u\up}^f(1-n_{x\down}^c)n_{u\down}^f\Big\}+|\Lambda|+|\Omega|+2\sum_{u\in\Omega}n_{u\up}^fn_{u\down}^f+\no
&\quad\quad+\sum_{x\in\Lambda}\sum_{u\in F_x}|F_x|^{-1}(n_{x\up}^c+n_{x\down}^c)n_{u\up}^fn_{u\down}^f+\sum_{x\in\Lambda}\sum_{u\in F_x}|F_x|^{-1}(n_{x\up}^c+n_{x\down}^c)(1-n_{u\up}^f)(1-n_{u\down}^f)\no
&\overset{\eqref{eq:4.1}}{\unrhd}\sum_{x\in\Lambda}\sum_{u\in F_x}|F_x|^{-1}\left(n_{x\up}^cn_{x\down}^c+n_{u\up}^fn_{u\down}^f\right)+|\Lambda|+2\sum_{u\in\Omega}n_{u\up}^fn_{u\down}^f+\sum_{x\in\Lambda}\sum_{u\in F_x}|F_x|^{-1}(n_{x\up}^c+n_{x\down}^c)(1-n_{u\up}^f)(1-n_{u\down}^f)+\no
&\quad\quad+\sum_{(x,u)\in F}\Big\{ n_{x\up}^c(1-n_{u\up}^f)n_{x\down}^c(1-n_{u\down}^f)+(1-n_{x\up}^c)n_{u\up}^f(1-n_{x\down}^c)n_{u\down}^f\Big\}\no
&\unrhd\sum_{x\in\Lambda}n_{x\up}^cn_{x\down}^c+\sum_{u\in\Omega}n_{u\up}^fn_{u\down}^f+\no
&\quad+\sum_{x\in\Lambda}\sum_{u\in F_x}|F_x|^{-1}\Big\{ 
(n_{x\up}^c+n_{x\down}^c)(1-n_{u\up}^f)(1-n_{u\down}^f)+n_{x\up}^c(1-n_{u\up}^f)n_{x\down}^c(1-n_{u\down}^f)+(1-n_{x\up}^c)n_{u\up}^f(1-n_{x\down}^c)n_{u\down}^f+1\Big\}\no
&\overset{\eqref{eq:4.2}}{=}\sum_{x\in\Lambda}n_{x\up}^cn_{x\down}^c+\sum_{u\in\Omega}n_{u\up}^fn_{u\down}^f+\no
&\quad+\sum_{x\in\Lambda}\sum_{u\in F_x}|F_x|^{-1}\Big\{(1+n_{x\up}^c)(1+n_{x\down}^c)(1-n_{u\up}^f)(1-n_{u\down}^f)+n_{u\up}^fn_{u\down}^f+(1-n_{x\up}^c)n_{u\up}^f(1-n_{x\down}^c)n_{u\down}^f\Big\}\no
&\unrhd\sum_{x\in\Lambda}n_{x\up}^cn_{x\down}^c+\sum_{u\in\Omega}n_{u\up}^fn_{u\down}^f\wrt\mathcal Q.
\end{align}
Hence,  we get
\begin{align}
\frac{8}{J^2}\big\{ J(|\Lambda|+|\Omega|)+\mathbb J\big\}^2\unrhd\frac{8}{J^2}\mathbb J^2+2(|\Lambda|+|\Omega|)\unrhd\sum_{x\in\Lambda}n_{x\up}^cn_{x\down}^c+\sum_{u\in\Omega}n_{u\up}^fn_{u\down}^f\wrt\mathcal Q,
\end{align}
where we have used the fact  $\mathbb{J}\unrhd0\wrt\mathcal Q$ in the first inequality.\qed
\end{Proof}

\subsection{Proof of Theorem \ref{MainThmHalf}}\label{PfUniSub}
For later use, we introduce a useful complete orthonormal system (CONS) in $\mathcal{H}_{\rm c}\otimes \mathcal{H}_{\rm f}$ as follows:
Let  
$|0\rangle_{ c} $  be the Fock vacuum in $\mathcal{H}_{\rm c}$:
$|0\rangle_{ c}
=(1,0,\ldots)
$. Similarly, let 
$|0\rangle_{ f}$ be the Fock vacuum in $\mathcal{H}_{\rm f}$.
Set $|0\rangle=|0\rangle_c\otimes |0 \rangle_f$.
Note that $c_{x \sigma}|0\rangle_c=0$ and $f_{u \sigma}|0\rangle_{ f}=0$.
Define  $\mathcal S_{ c}=\{0,1\}^\Lambda$ and $\mathcal S_{ f}=\{0,1\}^\Omega$.
For $\bsigma_{ c}=\{\sigma_{c, x}\}_{x\in \Lambda}\in\mathcal{ S}_{ c}$, we define
\begin{align}
{\bs c}^*_{\up}(\bsigma_c)=\prod^{\prime}_{x\in \Lambda} (c_{x \up}^*)^{\sigma_{{\rm c},x}},\ \ \ {\bs c}^*_{\down}(\bsigma_c)=\prod^{\prime}_{x\in \Lambda} (c_{x \down}^*)^{\sigma_{{\rm c},x}},
\end{align}
where $\prod_{x\in \Lambda}^{\prime}$ indicates the ordered product according to an arbitrarily fixed order in $\Lambda$.
Similarly, 
for $\bsigma_{ f}=\{\sigma_{f, u}\}_{u\in \Omega}\in\mathcal{S}_{ f}$, we define
${\bs f}_{\up}^*({\bsigma_f})$ and ${\bs f}_{\down}^*(\bsigma_f)$. Needless to say, we fix an arbitrarily fixed order in $\Omega$  to define ${\bs f}_{\up}^*({\bsigma_f})$ and ${\bs f}_{\down}^*(\bsigma_f)$.
 Given  $\bsigma_{ c}, \bsigma_c^{\prime}\in \mathcal{ S}_{ c}$ and  $\bsigma_{ f}, \bsigma_f^{\prime}\in \mathcal{ S}_{ f}$, let
  \begin{align}
  \big|\bsigma_{c},\bsigma_{c}^{\prime},\bsigma_{f},
\bsigma_{f}^{\prime }\big\rangle={\bs c}_{\up}^*(\bsigma_c) {\bs c}_{\down}^*(\bsigma_{c}^{\prime}) {\bs f}_{\up} (\bsigma_f) {\bs f}_{\down}^*(\bsigma_f^{\prime})|0\rangle\in \mathcal{H}_{\rm c}\otimes \mathcal{H}_{\rm f}.
  \end{align}
 For $\bsigma_c\in \mathcal{S}_c$ and $\bsigma_f\in \mathcal{S}_f$, we 
 set
 $|\bsigma_c|=\sum_{x\in \Lambda} \sigma_{c, x}$ and $|\bsigma_f|=\sum_{u\in \Omega}\sigma_{f, u}.$
Note that 
\begin{align}
\big\{
\big|\bsigma_{c},\bsigma_{c}^{\prime},\bsigma_{f},
\bsigma_{f}^{\prime }\big\rangle\, |\, \bsigma_{c},\bsigma_{c}^{\prime }\in \mathcal{S}_{ c},\ \bsigma_{f},
\bsigma_{f}^{\prime}\in \mathcal{S}_{f},\ |\bsigma_{c}|+|\bsigma_f|=N/2, \ |\bsigma'_{c}|+|\bsigma'_f|=N/2
\big\}
\end{align}
is a CONS of $\mathcal{L}_N$.
Taking this into consideration,  we define 
\begin{align}
\SN
=\{(\bsigma_c, \bsigma_f)\in \mathcal{S}_c\times \mathcal{S}_f\, |\, |\bsigma_c|+|\bsigma_f|=N/2\}.
\end{align}

\begin{Lemma}\label{one step}
Let 
 $(\bsigma_c, \bsigma_f), (\bsigma'_c, \bsigma'_f)\in \SN$.
 Let $g, h\in\mathcal{P}\setminus\{0\}$.
Set
\begin{align}
S(t)&=\l<\bsigma_c,\bsigma_c,\bsigma_f,\bsigma_f, g\left|e^{-t\left(R-\frac{1}{2}\mathbb{J}+\omega_0N_{\rm p}\right)}\right|\bsigma_c',\bsigma_c',\bsigma_f',\bsigma_f',h>,\ \ 0<t<1, 
\end{align}
where $
\big|\bsigma_c,\bsigma_c,\bsigma_f,\bsigma_f,g\big\rangle
= \big|\bsigma_c,\bsigma_c,\bsigma_f,\bsigma_f\big\rangle \otimes g
$.

 Assume  either 
 \begin{itemize}
 \item[{\rm (i)}]  there exist $x,y\in\Lambda$ such that $t_{x,y}\neq0$ and
$
\big|\bsigma_c,\bsigma_c,\bsigma_f,\bsigma_f\big\rangle=c_{x\up}^*c_{y\up}c_{x\down}^*c_{y\down}\big|\bsigma_c',\bsigma_c',\bsigma_f',\bsigma_f'\big\rangle,
$
\end{itemize}
 or 
 \begin{itemize}
\item[{\rm (ii)}]
there exist  $x\in\Lambda, u\in\Omega$ such that $J_{x,u}\neq0$ and
$
\big|\bsigma_c,\bsigma_c,\bsigma_f,\bsigma_f \big\rangle=(c_{x\up}^*f_{u\up}c_{x\down}^*f_{u\down}+f_{u\up}^*c_{x\up}f_{u\down}^*c_{x\down}) \big|\bsigma_c',\bsigma_c',\bsigma_f',\bsigma_f'\big\rangle.
$
\end{itemize}
Then there exists a $\gamma(g,h)>0$ depending on $g$ and $h$ such that if $0<t<\gamma(g,h)$,  then $S(t)>0$ holds.
\end{Lemma}

\begin{Proof}
See Appendix \ref{B}.\qed
\end{Proof}

As we will see below,
Lemma \ref{one step} plays an important role in the proof of Theorem \ref{MainThmHalf}. To properly use Lemma \ref{one step}, the following lemma is needed.

\begin{Lemma}\label{Connectivity}
For each $(\bsigma_c,\bsigma_f),(\bsigma_c',\bsigma_f')\in \SN$, 
 there exist $(\bsigma_{c, 1},\bsigma_{f, 1}),\ldots,(\bsigma_{c, n},\bsigma_{f, n})\in\SN, x_1\ldots,x_{n+1},y_1,\ldots,y_{n+1}\in\Lambda
 $ and $u_1,\ldots,u_{n+1}\in\Omega$ such that 
   any one of the following conditions holds for each  $j=0,1, \ldots,n$:
\begin{itemize}
\item[{\rm (i)}]  $ t_{x_{j+1},y_{j+1}}\neq0$ and 
\begin{align}
\big|\bsigma_{c, j},\bsigma_{c, j},\bsigma_{f, j},\bsigma_{f, j}\big\rangle=c_{x_{j+1}\up}^*c_{y_{j+1}\up}c_{x_{j+1}\down}^*c_{y_{j+1}\down}
\big|\bsigma_{c, j+1},\bsigma_{c, j+1},\bsigma_{f, j+1},\bsigma_{f, j+1} \big\rangle; 
\end{align}
\item[{\rm (ii)} ] $
J_{x_{j+1},u_{j+1}}\neq0
$ and 
\begin{align}
\big|\bsigma_{c,j},\bsigma_{c,j},\bsigma_{f,j},\bsigma_{f,j}\big\rangle=c_{x_{j+1}\up}^*f_{u_{j+1}\up}c_{x_{j+1}\down}^*f_{u_{j+1}\down} \big|\bsigma_{c,j+1},\bsigma_{c,j+1},\bsigma_{f,j+1},\bsigma_{f,j+1}  \big\rangle; 
\end{align}
\item[{\rm (iii)} ]  $J_{x_{j+1},u_{j+1}}\neq 0$ and 
\begin{align}
\big|\bsigma_{c,j},\bsigma_{c,j},\bsigma_{f,j},\bsigma_{f,j} \big\rangle=f_{u_{j+1}\up}^*c_{x_{j+1}\up}f_{u_{j+1}\down}^*c_{x_{j+1}\down} \big|\bsigma_{c,j+1},\bsigma_{c,j+1},\bsigma_{f,j+1},\bsigma_{f,j+1}  \big\rangle. 
\end{align}

  \end{itemize}
 In the above, we have used the following notations:  $\bsigma_{c,0}=\bsigma_c, \bsigma_{f,0}=\bsigma_f, \bsigma_{c,n+1}=\bsigma_c', \bsigma_{f,n+1}'=\bsigma_f'$.
\end{Lemma}

\begin{Proof}
For readers\rq{} convenience, we provide a sketch of the proof.
We divide the proof into two steps.

{\bf Step 1.}
Choose  $\bsigma_c, \bsigma_c^{\prime}\in \mathcal{S}_c $
with 
$\sum_{x\in\Lambda}\sigma_{c,x}=\sum_{x\in\Lambda}\sigma_{c,x}'=|\Lambda|/2$.
Because  the graph $(\Lambda,E)$ is connected by the assumption \ref{C1},  we can prove the following:
 There exist $\bsigma_{c, 1},\ldots,\bsigma_{c, n}\in\mathcal{S}_c,x_1\ldots,x_{n+1},y_1,\ldots,y_{n+1}\in\Lambda$ such that following (a) and (b) hold for each  $j=0,\ldots,n$:
 \begin{itemize}
\item[(a)] $t_{x_{j+1},y_{j+1}}\neq 0$; 
\item[(b)]
$
\big|\bsigma_{c, j},\bsigma_{c, j}\big\rangle=c_{x_{j+1}\up}^*c_{y_{j+1}\up}c_{x_{j+1}\down}^*c_{y_{j+1}\down}
\big|\bsigma_{c, j+1},\bsigma_{c, j+1} \big\rangle.
$
\end{itemize}
As for  the proof, see, e.g., \cite{Freericks1995,Miyao2012,Tasaki2020}.

{\bf Step 2.} Let 
$\Xi=\Lambda\cup\Omega$ and  let $E'=\{\{x,y\}\subset\Xi\,|\,t_{x,y}\neq0\}\cup\{\{x,u\}\subset\Xi\,|\,J_{x,u}\neq0\}$.
By using  the assumptions 
\ref{C1} and  \ref{C3}, the extended graph $(\Xi, E')$ is connected.
Thus, the assertion in Lemma \ref{Connectivity} follows from the property stated in {\bf Step 1}.
\qed
\end{Proof}

The following lemma is necessary for the proof of Theorem \ref{MainThmHalf}.

\begin{Lemma}\label{product1}
Let  $n\in \mathbb{N} $ and $\beta>0$. For each $j=1, \dots, n+1$,  let $\{G_j(s)\}_{s\ge 0}$ be a family of bounded  self-adjoint operators on $L^2(\mathbb{R}^{|\Lambda|})$.
Assume the following:
\begin{itemize}
\item[{\rm (i)}] $G_j(s)\unrhd0\wrt\mathcal P$ for all $s \ge 0$ and $j=1, \dots, n+1$.
\item[{\rm (ii)}] For any given  $g,h\in\mathcal P\setminus\{0\}$ and $j=1,\ldots,n$, there exists a $\gamma_j(g,h)>0$ such that if $0<s<\gamma_j(g,h)$, then $\i<g,G_j(s)h>>0$ holds.
\item[{\rm (iii)}] For any given $g,h\in\mathcal P\setminus\{0\}$, there exists a  $\gamma_{n+1}>0$, independent of $ g$ and $h$,  such that if $0<s<\gamma_{n+1}$, then $\i<g,G_{n+1}(s)h>>0$ holds.
\end{itemize}
Then, for any given  $g,h\in\mathcal P\setminus\{0\} $ and $\beta>0$, there exist positive numbers $s_1,\ldots,s_n$  with $\sum_{j=1}^ns_j<\beta$ such that 
\begin{align}
\i<g,G_1(s_1)G_2(s_2)\cdots G_n(s_n)G_{n+1}(s)^kh>>0.
\end{align}
 holds for any $k\in\mathbb{N}$ and $0<s<\gamma_{n+1}$.
\end{Lemma}

\begin{Proof}
If $0<s_1<\min\{\gamma_1(g,h),\beta/n\}$,  then $\i<g,G_1(s_1)h>>0 $ holds
 due to  the condition (ii). Hence, using (i), we conclude that  $G_1(s_1)g\in\mathcal P\setminus\{0\}$. For $j=2,\ldots,n$, choose $s_j$ such that $0<s_j<\min\{\gamma_j(G_{j-1}(s_{j-1})\cdots G_1(s_1)g,h), \beta/n\}$.
 Then $\i<g,G_1(s_1)\cdots G_{j}(s_{j})h>>0$ holds, which implies that $G_{j}(s_{j})\cdots G_1(s_1)g\in\mathcal P\setminus\{0\}$.
 By induction on $j$,
 there are positive numbers $s_1,\ldots,s_n$ with $\sum_{j=1}^ns_j<\beta$ such that 
  $G_{n}(s_{n})\cdots G_1(s_1)g\in\mathcal P\setminus\{0\}$ holds.
  Because of the condition (iii), it holds that  $G_{n+1}(s)\rhd0\wrt\mathcal P$,  if $0<s<\gamma_{n+1}$.
  Hence, we have $G_{n+1}(s)^k\rhd0\wrt\mathcal P$.
  Therefore, for any $k\in\mathbb{N}$,
   $\i<g,G_1(s_1)G_2(s_2)\cdots G_n(s_n)G_{n+1}(s)^kh>>0$ holds, provided that $0<s<\gamma_{n+1}$. \qed
\end{Proof}

To apply Lemma \ref{product1} in the proof of Theorem \ref{MainThmHalf}, the following lemma is useful.

\begin{Lemma}\label{nostep}
Let $\bsigma\in\SN$ and $g,h\in\mathcal P\setminus\{0\}$. Set $\alpha=2\sum_{x,y\in\Lambda}|t_{x,y}|+\sum_{x\in\Lambda,u\in\Omega}|J_{x,u}|+2\sum_{x,y\in\Lambda}|U_{\mathrm{eff},x,y}|+\frac{1}{2}\|\mathbb{J}\|$.
If $0<t<e^{-\alpha}$, then we have 
\begin{align}
\i<\bsigma,\bsigma,g|e^{-t(R-\frac{1}{2}\mathbb{J}+\omega_0N_{\rm p})}|\bsigma,\bsigma,h>>0.
\end{align}
\end{Lemma}

\begin{Proof}
By using the Duhamel formula, we have
\begin{align}
&\i<\bsigma,\bsigma,g|e^{-t(R-\frac{1}{2}\mathbb{J}+\omega_0N_{\rm p})}|\bsigma,\bsigma,h>\no
&=\l<g,e^{-t\omega_0N_{\rm p}}h>\no
&\hspace{1cm}+\sum_{n\geq1}(-t)^n\int_{0\le s_1 \le \cdots \le s_m \le 1}\l<\bsigma,\bsigma,g\Big|e^{-s_1t\omega_0N_{\rm p}}\left(R-\frac{1}{2}\mathbb{J}\right)\cdots\left(R-\frac{1}{2}\mathbb{J}\right)e^{-(1-s_n)t\omega_0N_{\rm p}}\Big|\bsigma,\bsigma,h>\,ds_n\cdots ds_1\no
&\geq\l<g,e^{-t\omega_0N_{\rm p}}h>-\sum_{n\geq1}\frac{t^n}{n!}\left(2\sum_{x,y\in\Lambda}|t_{x,y}|+\sum_{x\in\Lambda,u\in\Omega}|J_{x,u}|+2\sum_{x,y\in\Lambda}|U_{\mathrm{eff},x,y}| +\frac{1}{2}\|\mathbb{J}\|\right)^n\l<g,e^{-t\omega_0N_{\rm p}}h>\no
&\geq\l<g,e^{-t\omega_0N_{\rm p}}h>-t\sum_{n\geq1}\frac{\alpha^n}{n!}\l<g,e^{-t\omega_0N_{\rm p}}h>\no
&\geq \left(1-te^{\alpha}\right)\l<g,e^{-t\omega_0N_{\rm p}}h>, \label{LowerDuh}
\end{align}
where in the first inequality, we have used (\ref{PPProperty}).
Because $e^{-t \omega_0 N_{\rm p}} \rhd 0$ w.r.t. $\mathcal{P}$, we have 
$
\i<g,e^{-t\omega_0N_{\rm p}}h>>0
$. Hence, the right hand side of (\ref{LowerDuh}) is strictly positive.
\qed
\end{Proof}

\begin{thm}\label{mt}
Suppose that $U_{\mathrm{eff}}$ is positive semi-definite. Define $\widehat{ H}=\mathcal U^*H\mathcal U+\omega_0^{-1}g^2|\Lambda|-\frac{1}{2}JN$.  Then we obtain  $e^{-\beta\widehat{H}}\rhd0\wrt\mathcal Q$ for all $\beta>0$.
\end{thm}

\begin{Proof}
By applying Corollary \ref{BasicTrn}, we have the following expression: 
\begin{align}
\widehat{H}=R-\mathbb{J}-\tilde{\mathbb{U}}+\omega_0N_{\rm p}-\frac{1}{2}JN.
\end{align}

Choose 
$\psi, \phi\in Q_0 \mathcal{L}_{N, +} \setminus \{0\}$ and $g, h\in \mathcal{P}\setminus \{0\}$, arbitrarily.
Because $\tr[\Psi_{\vartheta}(\psi)]>0$ and $\tr[\Psi_{\vartheta}(\phi)]>0$, we see that there exist $(\bsigma_c,\bsigma_f),(\bsigma_c',\bsigma_f')\in\SN$ satisfying $\i<\psi|\bsigma_c,\bsigma_c,\bsigma_f,\bsigma_f>\neq0 $ and $\i<\phi|\bsigma_c',\bsigma_c',\bsigma_f',\bsigma_f'>\neq0$. 
With this in mind, 
we set  $\psi_\bsigma=\i<\psi|\bsigma_c,\bsigma_c,\bsigma_f,\bsigma_f> $ and $\phi_{\bs \sigma'}=\i<\phi|\bsigma_c',\bsigma_c',\bsigma_f',\bsigma_f'>$. Since $\psi,\phi\in  Q_0 \mathcal{L}_{N, +} $,  it holds that $\psi_\bsigma>0$ and  $\phi_{\bsigma'}> 0$.  By the  Duhamel formula, we have
\begin{align}
&\l<\psi\otimes g,e^{-\beta\widehat{H}}\phi\otimes h>\no
&=\sum_{m\geq0}2^{-m}\int_{0\le s_1 \le \cdots \le s_m \le \beta}\l<\psi\otimes g,e^{-s_1(R-\frac{1}{2}\mathbb{J}+\omega_0N_{\rm p})}(X+2\tilde{\mathbb{U}})\cdots (X+2\tilde{\mathbb{U}})e^{-(\beta-s_m)(R-\frac{1}{2}\mathbb{J}+\omega_0N_{\rm p})}\phi\otimes h>\,ds_m\cdots ds_1,
\end{align}
where $X=JN+\mathbb{J}$.
In the proof of Proposition \ref{PPSemi}, we have already proved that 
 $\tilde{\mathbb{U}}\unrhd0$ and $X\unrhd 0 \wrt\mathcal Q$. In addition, by using  arguments similar to those of the proof of Proposition \ref{PPSemi}, we can show that  $e^{-s(R-\frac{1}{2}\mathbb{J}+\omega_0N_{\rm p})}\unrhd0\wrt\mathcal Q$ for each $s\geq0$. Therefore,  we obtain that  
\begin{align}
\l<\psi\otimes g,e^{-s_1(R-\frac{1}{2}\mathbb{J}+\omega_0N_{\rm p})}Y_1 \cdots Y_{n-1} e^{-(\beta-s_n)(R-\frac{1}{2}\mathbb{J}+\omega_0N_{\rm p})}\phi\otimes h>\geq0
\end{align}
holds, provided that  $0\le s_1\le \cdots \le s_n \le \beta$,  where $Y_i=X\ \mathrm{or}\ 2\tilde{\mathbb U}$.
  Hence, we obtain the following lower bound: 
\begin{align}
&\l<\psi\otimes g,e^{-\beta\widehat{H}}\phi\otimes h>\no
&\geq2^{-m}\int_{0\le s_1 \le \cdots \le s_m \le \beta}\l<\psi\otimes g,e^{-s_1(R-\frac{1}{2}\mathbb{J}+\omega_0N_{\rm p})}X \cdots Xe^{-(\beta-s_m)(R-\frac{1}{2}\mathbb{J}+\omega_0N_{\rm p})}\phi\otimes h>\,ds_m\cdots ds_1. \label{BigIntegral}
\end{align}
Because the integrand  of the right hand side of \eqref{BigIntegral} is continuous in $s_1, \dots, s_m$ with $0\le s_1\le \cdots \le \beta$,  it suffices to prove that there exist $m\in\mathbb N $ and $ s_1,\ldots,s_m\in \mathbb{R}$ with $0 \leq  s_1 \le \cdots \le s_m\le \beta$ satisfying
\begin{align}
\l<\psi\otimes g,e^{-s_1(R-\frac{1}{2}\mathbb{J}+\omega_0N_{\rm p})}X \cdots Xe^{-(\beta-s_m)(R-\frac{1}{2}\mathbb{J}+\omega_0N_{\rm p})}\phi\otimes h>>0. \label{DuhaSP}
\end{align}

To prove  (\ref{DuhaSP}), we first derive a useful operator inequality:
By applying Proposition \ref{proj}, we see that,  for each $(\bsigma,\bsigma')\in\SN$, 
\begin{align}
\left(\frac{8}{J^2}\right)^{\frac{N}{2}}X^{N}&\unrhd\left(\sum_{x\in\Lambda}n_{x\up}^cn_{x\down}^c+\sum_{u\in\Omega}n_{u\up}^fn_{u\down}^f\right)^{\frac{N}{2}}\no
&\unrhd\left(\sum_{x\in\Lambda}n_{x\up}^cn_{x\down}^c\right)^{\frac{|\Lambda|}{2}}\left(\sum_{u\in\Omega}n_{u\up}n_{u\down}\right)^{\frac{|\Omega|}{2}}\no
&\unrhd\prod_{x\in\Lambda}(n_{x\up}^cn_{x\down}^c)^{\sigma_x}\prod_{u\in\Omega}(n_{u\up}^fn_{u\down}^f)^{\sigma_u'}\no
&=\left|\bsigma,\bsigma,\bsigma',\bsigma'\right\rangle\left\langle\bsigma,\bsigma,\bsigma',\bsigma'\right|\wrt\mathcal Q. \label{LowerOp}
\end{align}
The inequality (\ref{LowerOp}) is essential for the proof as we will see below.

Fix $k\in \mathbb{N} $, arbitrarily. 
Set $m=N(n+2+k)$ and define a  function  $F$  by 
\begin{align}
F(s_1,\ldots,s_m)=\left(\frac{8}{J^2}\right)^{\frac{m}{2}}\l<\psi\otimes g,e^{-s_1(R-\frac{1}{2}\mathbb{J}+\omega_0N_{\rm p})}X \cdots Xe^{-(\beta-s_m)(R-\frac{1}{2}\mathbb{J}+\omega_0N_{\rm p})}\phi\otimes h>.
\end{align}
Let $\{
(\bsigma_{c, 1}, \bsigma_{f, 1}), \dots, (\bsigma_{c, 1}, \bsigma_{f, n})
\}\subseteq \SN$ be a sequence given in Lemma \ref{Connectivity}. Recall that this sequence \lq\lq{}connects\rq\rq{} $(\bsigma_c, \bsigma_f)$ and $(\bsigma_c\rq{}, \bsigma_f\rq{})$ as stated in Lemma \ref{Connectivity}. For notational simplicity, we set
$|\bsigma_0\rangle=|\bsigma_c,\bsigma_c,\bsigma_f,\bsigma_f\rangle,|\bsigma_j\rangle=|\bsigma_{c,j},\bsigma_{c,j},\bsigma_{f,j},\bsigma_{f,j}\rangle,\ j=1, \dots, n$,  and $|\bsigma_{n+1}\rangle=|\bsigma_{c}\rq{}, \bsigma_{c}\rq{},\bsigma_{f}\rq{},\bsigma_f\rq{}\rangle$. 
Choose strictly positive numbers $t_1, \dots, t_{n+1}$ such that  $0<\varepsilon<\beta$,  where $\varepsilon=\sum_{j=1}^{n+1}t_j$. We have
\begin{align}
&F\left(\underbrace{0,\ldots,0}_N,\underbrace{t_1,\ldots,t_1}_N, \underbrace{t_1+t_2,\ldots,t_1+t_2}_N, \ldots,\varepsilon,\ldots,\varepsilon,\varepsilon+\frac{\beta-\varepsilon}{k},\ldots,\varepsilon+\frac{\beta-\varepsilon}{k},\ldots,\beta,\ldots,\beta\right)\no
&=\l<\psi\otimes g,\left(\frac{8}{J^2}\right)^{\frac{N}{2}}X^{N}e^{-t_1(R-\frac{1}{2}\mathbb{J}+\omega_0N_{\rm p})}\cdots e^{-t_{n+1}(R-\frac{1}{2}\mathbb{J}+\omega_0N_{\rm p})}\left(\frac{8}{J^2}\right)^{\frac{N}{2}}X^{N}\phi\otimes h>\no
&
\overset{\eqref{LowerOp}}{\geq}
\bigg<\psi\otimes g,\prod_{j=0}^n\left(|\bsigma_j\rangle\langle\bsigma_j|e^{-t_{j+1}(R-\frac{1}{2}\mathbb{J}+\omega_0N_{\rm p})}\right)\left(|\bsigma_{n+1}\rangle\langle\bsigma_{n+1}|e^{-\frac{\beta-\varepsilon}{k}(R-\frac{1}{2}\mathbb{J}+\omega_0N_{\rm p})}\right)^k|\bsigma_{n+1}\rangle\langle\bsigma_{n+1}|\phi\otimes h\bigg>\no
&=\psi_\bsigma\phi_{\bsigma'}\l<\bsigma_0,g\left|\prod_{j=0}^n\left(|\bsigma_j\rangle\langle\bsigma_j|e^{-t_{j+1}(R-\frac{1}{2}\mathbb{J}+\omega_0N_{\rm p})}\right)\left(|\bsigma_{n+1}\rangle\langle\bsigma_{n+1}|e^{-\frac{\beta-\varepsilon}{k}(R-\frac{1}{2}\mathbb{J}+\omega_0N_{\rm p})}\right)^k|\bsigma_{n+1}\rangle\langle\bsigma_{n+1}|\right|\bsigma_{n+1},h>, \label{FLowerB}
\end{align}
where 
in the first inequality,   we used the inequality (\ref{LowerOp}); 
in addition, we have used the fact that each $
| \bsigma_j\rangle
$ is positive w.r.t. $Q_0\mathcal{L}_{N, +}.$

Let $K_t(\bq, \bq\rq{})$ be the kernel operator of $e^{-t(R-\frac{1}{2}\mathbb{J}+\omega_0N_{\rm p})}$ given in Proposition \ref{PropKernel}.
In terms of  $K_t(\bq, \bq\rq{})$, 
 we have the following expressions:
\begin{align}
\langle\bsigma_{j-1},g|e^{-t(R-\frac{1}{2}\mathbb{J}+\omega_0N_{\rm p})}|\bsigma_{j},h\rangle
&=\int g(\boldsymbol q)h(\boldsymbol q')\langle\bsigma_{j-1}|K_t(\boldsymbol q,
\boldsymbol q')|\bsigma_{j}\rangle d\boldsymbol qd\boldsymbol q',\ \ j=1, \dots, n+1, \label{Kernel1} \\
\langle\bsigma_{n+1},g|e^{-t(R-\frac{1}{2}\mathbb{J}+\omega_0N_{\rm p})}|\bsigma_{n+1},h\rangle
&=\int g(\boldsymbol q)h(\boldsymbol q')\langle\bsigma_{n+1}|K_t(\boldsymbol q,\boldsymbol q')|\bsigma_{n+1}\rangle d\boldsymbol qd\boldsymbol q'.\label{Kernel2}
\end{align}
With this mind, 
 we define $K_j(t)\in\mathscr B(L^2(\mathbb R^{|\Lambda|}))$ by 
\begin{align}
\l<g,K_j(t)h>
&=\mbox{the right hand side of \eqref{Kernel1}}
, \ \ j=1, \dots, n+1,\\
\l<g,K_{n+2}(t)h>
&=\mbox{the right hand side of \eqref{Kernel2}}.
\end{align}
Note that  $K_t(\bq, \bq\rq{}) \unrhd 0$ w.r.t. $Q_0\mathcal{L}_{N, +}$ holds  due to  Proposition \ref{PropKernel}. Hence,   we have $\langle\bsigma_{j-1}|K_t(\boldsymbol q,
\boldsymbol q')|\bsigma_{j}\rangle \ge 0$ and $\langle\bsigma_{n+1}|K_t(\boldsymbol q,\boldsymbol q')|\bsigma_{n+1}\rangle\ge 0$  for a.e. $\bq, \bq\rq{}$, which imply that $K_j(t) \unrhd 0$ w.r.t. $\mathcal{P}$ for all $t\ge 0$ and $j=1, \dots, n+2$.
Rewriting  the right hand side of (\ref{FLowerB}) by using $K_j(t)$, we get 
\begin{align}
&F\left(0,\ldots,0,t_1,\ldots,t_1,\ldots,\varepsilon,\ldots,\varepsilon,\varepsilon+\frac{\beta-\varepsilon}{k},\ldots,\varepsilon+\frac{\beta-\varepsilon}{k},\ldots,\beta,\ldots,\beta\right)\no
&\geq\psi_\bsigma\phi_{\bsigma'}\l<g,K_1(t_1)K_2(t_2)\cdots K_{n+1}(t_{n+1})K_{n+2}\left(\frac{\beta-\varepsilon}{k}\right)^kh>.\label{FKInq}
\end{align}
By Lemmas \ref{one step} and \ref{Connectivity}, we see that for any $g,h\in\mathcal P\setminus\{0\}$,  $\i<g,K_j(t)h>>0$ holds,  provided that  $0<t<\gamma(g, h)$. Because  $\varepsilon<\beta$, there exists a  $k\in\mathbb N$ such that $\frac{\beta-\varepsilon}{k}<e^{-\alpha}$. In the remainder of the proof,
we assume that $k$ satisfies this inequality.
We are aiming to apply Lemma \ref{product1} with the  correspondence
$G_j(t)=K_j(t)$.  For this purpose, we have to check the assumptions (i)-(iii) of  Lemma \ref{product1}. We readily check (i) and (ii); by using Lemma \ref{nostep}, 
we can confirm that  the assumption (iii) is satisfied. 
Hence, from Lemma \ref{product1}, there exist $t_1,\ldots,t_{n+1}>0$ with $\sum_{j=1}^{n+1} t_j<\beta$ such that $\i<g,K_1(t_1)\cdots K_{n+1}(t_{n+1})K_{n+2}\left(\frac{\beta-\varepsilon}{k}\right)^kh>>0$ holds. Hence, by (\ref{FKInq}) , we have
\begin{align}
F\left(0,\ldots,0,t_1,\ldots,t_1,\ldots,\varepsilon,\ldots,\varepsilon,\varepsilon+\frac{\beta-\varepsilon}{k},\ldots,\varepsilon+\frac{\beta-\varepsilon}{k},\ldots,\beta,\ldots,\beta\right)>0.
\end{align}
Therefore, for any $\psi\otimes g,\phi\otimes h\in\mathcal Q\setminus\{0\}$, $\i<\psi\otimes g,e^{-\beta\widehat{H}}\phi\otimes h>>0$ holds. By using Lemma \ref{pi} {\rm (ii)}, we finally  conclude that  $e^{-\beta\widehat{H}}\rhd0\wrt\mathcal Q$ for all $\beta>0$.\qed
\end{Proof}

\subsubsection*{Proof of Theorem \ref{MainThmHalf}}

 Applying Theorems \ref{pff} and \ref{mt}, we immediately obtain  (i).
In addition, the ground state, $\psi$,  can be chosen  such that 
$\psi>0\wrt\mathcal{U}\mathcal{Q}$.
Put $\phi=\mathcal{U}^*\psi$. Trivially, $\phi>0\wrt\mathcal{Q}$  holds.
By the definition of $\mathcal{U}$, i.e., \eqref{DefUTrn}, we find that 
\begin{align}
\gamma_x\gamma_y\mathcal{U}^*s_x^+s_y^-\mathcal{U}
&=\gamma_x\gamma_yU^*s_x^+s_y^-U
=c_{x\up}^*c_{y\up}c_{x\down}^*c_{y\down}\unrhd0\wrt\mathcal{Q},\\
\gamma_u\gamma_v\mathrm{sgn}J_{x,u}\mathrm{sgn}J_{y,v}\mathcal{U}^*S_u^+S_v^-\mathcal{U}
&=\gamma_u\gamma_v\mathrm{sgn}J_{x,u}\mathrm{sgn}J_{y,v}U^*S_u^+S_v^-U
=f_{u\up}^*f_{v\up}f_{u\down}^*f_{v\down}\unrhd0\wrt\mathcal{Q}.
\end{align}
Because $c_{x\up}^*c_{y\up}c_{x\down}^*c_{y\down}\phi\neq 0 $ and 
$f_{u\up}^*f_{v\up}f_{u\down}^*f_{v\down}\phi\neq 0$, we have
\begin{align}
\gamma_x\gamma_y\i<\psi,s_x^+s_y^-\psi>
&=\gamma_x\gamma_y\i<\phi,\mathcal{U}^*s_x^+s_y^-\mathcal{U}\phi>\no
&=\i<\phi,c_{x\up}^*c_{y\up}c_{x\down}^*c_{y\down}\phi>>0,\\
\gamma_u\gamma_v\mathrm{sgn}J_{x,u}\mathrm{sgn}J_{y,v}\i<\psi,S_u^+S_v^-\psi>
&=\gamma_u\gamma_v\mathrm{sgn}J_{x,u}\mathrm{sgn}J_{y,v}\i<\phi,\mathcal{U}^*S_u^+S_v^-\mathcal{U}\phi>\no
&=\i<\phi,f_{u\up}^*f_{v\up}f_{u\down}^*f_{v\down}\phi>>0.
\end{align}
This  completes the proof of Theorem \ref{MainThmHalf}.
\qed

\section{The total spin of the ground state}\label{SpinPf}

\subsection{The main result in  Section \ref{SpinPf}}

We already proved  the uniqueness of the ground state of $H$ in Theorem \ref{MainThmHalf}.
Our  goal in this section is to prove the following theorem.

\begin{thm}\label{MainSpin}
Assume {\bf (C)}. Assume that  $U_{\mathrm{eff}}$ is positive semi-definite. 
Then we have the following {\rm (i)} and {\rm (ii)}:
\begin{itemize}
\item[\rm (i)]
If \ref{C6} holds, then the ground state of $H$ has total spin $S=\frac{1}{2}\big||\Lambda_1|+|\Omega_1|-|\Lambda_2|-|\Omega_2|\big|$.
\item[\rm (ii)]
If \ref{C7} holds, then the ground state of $H$ has total spin $S=\frac{1}{2}\big||\Lambda_1|+|\Omega_2|-|\Lambda_2|-|\Omega_1|\big|$.
\end{itemize}
\end{thm}

\subsection{Strategy}
Here, we  briefly explain our strategy of the proof of  (i) of Theorem \ref{MainSpin}.
As for (ii) of Theorem \ref{MainSpin}, we will provide a proof  in Subsection \ref{Pf(ii)Go}.

Recall the definition of $P_0$, i.e., (\ref{DefP_0}).
The following proposition plays   a key role in the remainder of this section.
\begin{Prop}\label{Overlap}
Let $\mathcal{X}$ be  any one of $P_0 \mathcal{L}_N,\ \mathcal{L}_N $  and $P_0 \mathcal{L}_N\otimes \Hph$.
Let $\mathcal{C}$ be a Hilbert cone in $\mathcal{X}$.
Consider positive self-adjoint operators $\mathsf{H}_0$ and $\mathsf{H}$ acting on $\mathcal{X}$. 
Assume the following:
\begin{itemize}
\item[{\rm (i)}] $\mathsf{H}_0$ and $\mathsf{H}$ commute with the total spin operators $S_{\rm tot}^{(3)}, S_{\rm tot}^{(+)}$ and $S_{\rm tot}^{(-)}$.
\item[{\rm (ii)}] $\{ e^{-\beta \mathsf{H}_0}\}_{\beta \ge 0}$ and $\{e^{-\beta \mathsf{H}}\}_{\beta \ge 0}$ are ergodic w.r.t. $\mathcal{C}$. Hence, the ground state of each of $\mathsf{H}_0$ and $\mathsf{H}$ is unique and strictly positive w.r.t. $\mathcal{C}$ due to  Theorem  \ref{pff}.

\end{itemize}
We denote by $S_0$ (resp. $S$) the total spin of the ground state of $\mathsf{H}_0$ (resp. $\mathsf{H}$).
Then we have $S_0=S$.

\end{Prop}
\begin{Proof}
Let $\psi_0$ (resp. $\psi$) be the unique ground state of $\mathsf{H}_0$ (resp. $\mathsf{H}$). By the assumption (ii), $\psi_0$
and $\psi$ are strictly positive w.r.t. $\mathcal{C}$. Because ${\bs S}_{\rm tot}^2$ is self-adjoint, we have
\begin{align}
S_0(S_0+1)\langle \psi_0, \psi\rangle=\langle {\bs S}_{\rm tot}^2 \psi_0, \psi\rangle=
\langle \psi_0,  {\bs S}_{\rm tot}^2 \psi\rangle=S(S+1) \langle \psi_0, \psi\rangle.
\end{align}
Because $\langle \psi_0, \psi\rangle>0$, we conclude that $S_0=S$. \qed
\end{Proof}

Note that the method of   nonzero overlap between ground states has been extensively 
used in many-electron systems, see, e.g.,  \cite{Shen1996,Tian2004,Tsunetsugu1997,TSU1997}.  In \cite{Miyao2019}, this method is further  extended  and applied to electron-phonon interacting systems. 
Proposition \ref{Overlap} is a mathematically  abstracted form of the method,  which is essentially proved  in \cite{Miyao2019}.
 
We divide the proof  of Theorem \ref{MainSpin} into two steps:

\subsubsection*{Step 1:}

Define  a self-adjoint operator on $\mathcal{L}_N$ by  
\begin{align}
K_1&=\frac{1}{2}\sum_{x, y\in \Lambda}|t_{x,y}|^2({\boldsymbol s}_x^+\cdot{\boldsymbol s}_y^-+{\boldsymbol s}_x^-\cdot{\boldsymbol s}_y^+)
+\sum_{x\in \Lambda, u\in \Omega}|J_{x,u}|^2({\boldsymbol s}_x^+\cdot{\boldsymbol S}_u^-+{\boldsymbol s}_x^-\cdot{\boldsymbol S}_u^+)\no
&\quad+\sum_{x\in\Lambda}\left(n_{x\up}^c-\frac{1}{2}\right)\left(n_{x\down}^c-\frac{1}{2}\right)+\sum_{u\in\Omega}\left(n_{u\up}^f-\frac{1}{2}\right)\left(n_{u\down}^f-\frac{1}{2}\right).
\end{align}
First, we wish to examine the ground state properties of the restricted  Hamiltonian:
\begin{align}
{\bs K}=K_1\restriction P_0\mathcal L_N.
\end{align}
Note 
that 
\begin{align}
U^*{\bs K} U=U^* K_1U \restriction Q_0\mathcal L_N, \label{Q_0P_0}
\end{align}
where $U$ is given by Lemma \ref{HolePart}.

In Subsection \ref{PfStep1}, we will prove the following proposition as  a basic input.
\begin{Prop}\label{tsl4}
Assume {\bf (C)} and \ref{C6}. We have
\begin{align}
e^{-\beta U^*{\bs K}U}\rhd0\wrt Q_0\mathcal{L}_{N, +} \label{WKWPI}
\end{align}
for every $\beta>0$.
Hence,  the ground state of ${\bs K}$ is unique.
Furthermore, the ground state of ${\bs K}$ has total spin $S=\frac{1}{2}\big||\Lambda_1|+|\Omega_1|-|\Lambda_2|-|\Omega_2|\big|$. 
\end{Prop}

\begin{rem}\upshape
The readers would guess that since  the form of ${\bs K}$ is  
similar to that  of the Heisenberg Hamiltonian, $H_{\rm Heis}$,  magnetic properties of the  ground state of ${\bs K}$ are readily confirmed  by the Marshall-Lieb-Mattis theorem \cite{Lieb1962, Marshall1955}. On the contrary, because the Hilbert space on which ${\bs K}$ acts is different from the one on which $H_{\rm Heis}$ acts, we cannot directly  apply the Marshall-Lieb-Mattis theorem to ${\bs K}$.  In  Subsection \ref{PfStep1}, we will explain how to  overcome this difficulty.
\end{rem}

\subsubsection*{Step 2:}
In Subsection \ref{PfStep3}, we will prove (i) of  Theorem \ref{MainSpin} by using Theorem \ref{mt} and Proposition \ref{tsl4}. As we will see, a variant of Proposition \ref{Overlap}  is essential for the proof.

\subsection{Step 1: Proof of Proposition \ref{tsl4}} \label{PfStep1}

Recall the definition of $\mathcal{L}_{N, +}:  \mathcal{L}_{N, +}=\{\psi\in\mathcal L_N\,|\,\Psi_{\vartheta}(\psi)\geq0\}$.
As a first step, we prepare an abstract lemma:
\begin{Lemma}\label{tsl1}
For $A_1,\ldots,A_n\in\mathscr B(\mathcal F_N) $ and $c_n\in\mathbb C$, we have
\begin{align}
\exp\left[\sum_{k=1}^n\left(1+|c_k|^2A_k\otimes\vartheta A_k\vartheta\right)\right]
\unrhd\exp\left[\sum_{k=1}^n\left(c_kA_k\otimes1+c_k^*1\otimes\vartheta A_k\vartheta\right)\right]\wrt  \mathcal{L}_{N, +}.
\end{align}
\end{Lemma}

\begin{Proof}
For each $m\in\mathbb N$, one obtains, by applying  Proposition \ref{PPI2},
\begin{align}
\left(\frac{1}{\sqrt m}-\frac{c_k}{\sqrt m}A_k\right)\otimes\vartheta\left(\frac{1}{\sqrt m}-\frac{c_k^*}{\sqrt m}A_k\right)\vartheta\unrhd0\wrt\mathcal{L}_{N, +},
\end{align}
which implies
\begin{align}
\exp\left[\left(\frac{1}{\sqrt m}-\frac{c_k}{\sqrt m}A_k\right)\otimes\vartheta\left(\frac{1}{\sqrt m}-\frac{c_k^*}{\sqrt m}A_k\right)\vartheta\right]\unrhd1\wrt\mathcal{L}_{N, +}.
\end{align}
In addition, by using Proposition \ref{PPI2} again, we have
\begin{align}
\exp\left[\frac{c_k}{m}A_k\otimes1+\frac{c_k^*}{m}1\otimes\vartheta A_k\vartheta\right]
=
\exp\left[\frac{c_k}{m}A_k\right]\otimes \vartheta
\exp\left[\frac{c_k}{m} A_k\right]\vartheta
\unrhd0\wrt\mathcal{L}_{N, +}.
\end{align}
Hence, 
\begin{align}
\exp\left[\frac{1}{m}+\frac{|c_k|^2}{m}A_k\otimes\vartheta A_k\vartheta\right]
&=
 \exp\left[\left(\frac{1}{\sqrt m}-\frac{c_k}{\sqrt m}A_k\right)\otimes\vartheta\left(\frac{1}{\sqrt m}-\frac{c_k^*}{\sqrt m}A_k\right)
 \vartheta\right]
 \exp\left[\frac{c_k}{m}A_k\otimes1+\frac{c_k^*}{m}1\otimes\vartheta A_k\vartheta\right]\no
 &\unrhd \exp\left[\frac{c_k}{m}A_k\otimes1+\frac{c_k^*}{m}1\otimes\vartheta A_k\vartheta\right]\wrt\mathcal{L}_{N, +}.
\end{align}
Therefore, by applying the Trotter product formula, one finds 
\begin{align}
\exp\left[\sum_{k=1}^n\left(c_kA_k\otimes1+c_k^*1\otimes\vartheta A_k\vartheta\right)\right]
&=\lim_{m\to\infty}\left(\prod_{k=1}^ne^{\frac{1}{m}\left(c_kA_k\otimes1+c_k^*1\otimes\vartheta A_k\vartheta\right)}\right)^m\no
&\unlhd\lim_{m\to\infty}\left(\prod_{k=1}^n\exp\left[\frac{1}{m}+\frac{|c_k|^2}{m}A_k\otimes\vartheta A_k\vartheta\right]\right)^m\no
&=\exp\left[\sum_{k=1}^n\left(1+|c_k|^2A_k\otimes\vartheta A_k\vartheta\right)\right]\wrt\mathcal{L}_{N, +}.
\end{align}
\qed
\end{Proof}

As an application of  Lemma \ref{tsl1}, we obtain: 
\begin{Lemma}\label{tsl2}
Assume {\bf (C)} and \ref{C6}.
Define 
\begin{align}
H_\mathrm{H}&=-\sum_{x,y\in\Lambda}t_{x,y}(c_{x\up}^*c_{y\up}+c_{x\down}^*c_{y\down})
-\sum_{x\in\Lambda,u\in\Omega}J_{x,u}\left(c_{x\up}^*f_{u\up}+c_{x\down}^*f_{u\down}+f_{u\up}^*c_{x\up}+f_{u\down}^*c_{x\down}\right)\no
&\quad+\sum_{x\in\Lambda}\left(n_{x\up}^c-\frac{1}{2}\right)\left(n_{x\down}^c-\frac{1}{2}\right)+\sum_{u\in\Omega}\left(n_{u\up}^f-\frac{1}{2}\right)\left(n_{u\down}^f-\frac{1}{2}\right).
\end{align}
  Then we have
\begin{align}
e^{-\beta U^*K_1U}e^{|\Lambda|^2+|\Lambda||\Omega|}\unrhd e^{-\beta U^*H_\mathrm{H}U}\rhd0\wrt\mathcal{L}_{N, +} \label{KHInq}
\end{align}
for all $\beta>0$.
Hence,  the ground state of $ K_1 $ is unique.
Furthermore the ground state of $ K_1$ has total spin $S=\frac{1}{2}\big||\Lambda_1|+|\Omega_1|-|\Lambda_2|-|\Omega_2|\big|$. 
\end{Lemma}

\begin{Proof}
First, we observe
\begin{align}
U^*H_\mathrm{H}U
&=-\sum_{x,y\in\Lambda}t_{x,y}(c_{x\up}^*c_{y\up}+c_{x\down}^*c_{y\down})
    -\sum_{x\in\Lambda,u\in\Omega}J_{x,u}\left(c_{x\up}^*f_{u\up}+f_{u\down}^*c_{x\down}+f_{u\up}^*c_{x\up}+c_{x\down}^*f_{u\down}\right)\no
&\quad-\sum_{x\in\Lambda}\left(n_{x\up}^c-\frac{1}{2}\right)\left(n_{x\down}^c-\frac{1}{2}\right)
          -\sum_{u\in\Omega}\left(n_{u\up}^f-\frac{1}{2}\right)\left(n_{u\down}^f-\frac{1}{2}\right) \label{WH_HW}
          \end{align}
          and 
          \begin{align}
U^*K_1U
&=-\sum_{x,y\in\Lambda}|t_{x,y}|^2c_{x\up}^*c_{y\up}c_{x\down}^*c_{y\down}
                -\sum_{x\in \Lambda, u\in \Omega}|J_{x,u}|^2(c_{x\up}^*f_{u\up}c_{x\down}^*f_{u\down}+f_{u\up}^*c_{x\up}f_{u\down}^*c_{x\down})\no
                &\quad-\sum_{x\in\Lambda}\left(n_{x\up}^c-\frac{1}{2}\right)\left(n_{x\down}^c-\frac{1}{2}\right)
                -\sum_{u\in\Omega}\left(n_{u\up}^f-\frac{1}{2}\right)\left(n_{u\down}^f-\frac{1}{2}\right). \label{WK_1W}
\end{align}
Without loss of generality, we may assume $\beta=1$.
Using (\ref{FermiIdn}) and (\ref{WH_HW}), we can  apply Lemma \ref{tsl1} to $U^*H_\mathrm{H}U$ and obtain 
\begin{align}
&\exp\left[-U^*H_\mathrm{H}U\right]\no
&=\exp\left[\sum_{x,y\in\Lambda}t_{x,y}(c_{x\up}^*c_{y\up}+c_{x\down}^*c_{y\down})
    +\sum_{x\in\Lambda,u\in\Omega}J_{x,u}\left(c_{x\up}^*f_{u\up}+f_{u\down}^*c_{x\down}+f_{u\up}^*c_{x\up}+c_{x\down}^*f_{u\down}\right)
    \right.\no
    &\hspace{14pc}+\left.\sum_{x\in\Lambda}\left(n_{x\up}^c-\frac{1}{2}\right)\left(n_{x\down}^c-\frac{1}{2}\right)
          +\sum_{u\in\Omega}\left(n_{u\up}^f-\frac{1}{2}\right)\left(n_{u\down}^f-\frac{1}{2}\right)\right]\no
&\unlhd\exp\left[\sum_{x,y\in\Lambda}\left(1+|t_{x,y}|^2c_{x\up}^*c_{y\up}c_{x\down}^*c_{y\down}\right)
          +\sum_{x\in\Lambda,u\in\Omega}\big\{1+|J_{x,u}|^2\left(c_{x\up}^*f_{u\up}c_{x\down}^*f_{u\down}+f_{u\up}^*c_{x\up}f_{u\down}
          ^*c_{x\down}\right)\big\}\right.\no
          &\hspace{14pc}+\left.\sum_{x\in\Lambda}\left(n_{x\up}^c-\frac{1}{2}\right)\left(n_{x\down}^c-\frac{1}{2}\right)
          +\sum_{u\in\Omega}\left(n_{u\up}^f-\frac{1}{2}\right)\left(n_{u\down}^f-\frac{1}{2}\right)\right]\no
&=\exp\left[-U^*K_1U\right]e^{|\Lambda|^2+|\Lambda||\Omega|} \ \wrt \ \mathcal{L}_{N, +}, \label{HuSmall}
\end{align}
where in the second equality, we have used  (\ref{WK_1W}).
Because  $H_{\rm H}$ is a Hubbard Hamiltonian on the connected bipartite lattice $\Lambda\cup \Omega$, 
we can apply a generalized version  of  Lieb's theorem presented in \cite{Miyao2012,Miyao2019}  to $H_{\rm H}$. 
Thus, we find  that   $e^{-\beta U^*H_\mathrm{H}U}\rhd 0\wrt \mathcal{L}_{N, +}$ for all $\beta>0$. 
Combining this fact  with (\ref{HuSmall}),
we obtain the inequality (\ref{KHInq}).

In order to  specify  the value of the total spin of the ground state, we recall Lieb\rq{}s theorem for readers\rq{} convenience: 
Lieb\rq{}s theorem claims that with a bipartite lattice and a half-filled band, the ground state
of the repulsive Hubbard model has total spin 
\begin{align}
S= \frac{1}{2}\big||A|-|B|\big|, \label{LiebS}
\end{align} where $|A|$ (resp. $|B|$) is the
number of 
sites in the $A$-sublattice (resp. $B$-sublattice),
 see  \cite{Lieb1989} for details. Because  $H_{\mathrm{H}}$ is a   Hubbard Hamiltonian on the bipartite lattice with $A=\Lambda_1\sqcup \Omega_1$ and $B=\Lambda_2\sqcup \Omega_2$,
 the ground state of $H_{\mathrm{H}}$ has total spin 
  $S=\frac{1}{2}\big||\Lambda_1|+|\Omega_1|-|\Lambda_2|-|\Omega_2|\big|$.
Hence, due to 
 Proposition \ref{Overlap},   the  ground state of $K_1$ has total spin $S=\frac{1}{2}\big||\Lambda_1|+|\Omega_1|-|\Lambda_2|-|\Omega_2|\big|$ as well.
\qed
\end{Proof}

To complete the proof of Proposition \ref{tsl4}, the following lemma is useful:

\begin{Lemma}\label{Reduction}
Let $H_0$ be a self-adjoint operator acting in $\mathcal{L}_N$. Assume that 
\begin{itemize}
\item[{\rm (i)}] $e^{-\beta H_0} \rhd 0$ w.r.t. $\mathcal{L}_{N, +}$ for all $\beta>0$;
\item[{\rm (ii)}] $H_0$ commutes with $Q_0$.
\end{itemize}
Then we obtain $\exp(-\beta H_0\restriction Q_0 \mathcal{L}_N) \rhd 0$ w.r.t. $Q_0\mathcal{L}_{N, +}$
 for all $\beta>0$.
\end{Lemma}
\begin{Proof}
Take $Q_0\varphi_1, Q_0\varphi_2\in Q_0\mathcal{L}_{N, +}\setminus \{0\}$, arbitrarily.
Because $Q_0\unrhd 0$ w.r.t. $\mathcal{L}_{N, +}$, we have $Q_0\varphi_1 \ge 0$ and $Q_0\varphi_2 \ge 0$
w.r.t. $\mathcal{L}_{N, +}$ as vectors in $\mathcal{L}_N$. Using this, we have
\begin{align}
\big\langle Q_0\varphi_1, e^{-\beta H_0\restriction Q_0 \mathcal{L}_N} Q_0 \varphi_2\big\rangle_{Q_0 \mathcal{L}_N}
=\big\langle Q_0\varphi_1, e^{-\beta H_0} Q_0 \varphi_2\big\rangle_{\mathcal{L}_N}>0,
\end{align}
where in the first equality, we have used the assumption (ii), and in the first inequality, we have used the assumption (i). This completes the proof. \qed

\end{Proof}

\subsubsection*{Proof of Proposition \ref{tsl4}}
Taking (\ref{Q_0P_0}) into consideration, we can apply 
 Lemma \ref{Reduction} with $H_0=U^*K_1U$ and  obtain (\ref{WKWPI}). Hence, the ground state, $\varphi_{\rm g}$,  of ${\bs K}$ is unique and strictly positive w.r.t. $
UQ_0\mathcal{L}_{N, +}
$. Let $\psi$ be the ground state of $
K_1
$. By Lemma \ref{tsl2}, $\psi$ has total spin 
$S=\frac{1}{2}\big||\Lambda_1|+|\Omega_1|-|\Lambda_2|-|\Omega_2|\big|$.  Because
$K_1$ commutes with $P_0$, $P_0\psi$
 is the ground state of  ${\bs K}$.
 Hence, due to the uniqueness, $\varphi_{\rm g}$ and $P_0 \psi$ are  identical.
   In addition, since ${\bs S}_{\rm tot}^2$ commutes with $P_0$,
   the total spin of 
  $P_0\psi$ 
  coincides with that of $\psi$.
  \qed

\subsection{Step 2: Proof of  (i) of Theorem \ref{MainSpin}}\label{PfStep3}
Set
\begin{align}
L_2=\boldsymbol{K}+\omega_0 N_{\rm p}.
\end{align}
Trivially, $L_2$ is self-adjoint on $\mathrm{dom}(N_{\rm p})$ and bounded from below.
Recall the definition of $\mathcal{Q}$, i.e., (\ref{DefQ}).
\begin{Lemma}\label{ts5}
Assume {\bf (C)} and \ref{C6}. 
 Then we have
\begin{align}
\exp\left[-\beta U^*L_2U\right]\rhd0\wrt\mathcal Q
\end{align}
for any $\beta>0$.
Hence, the ground state of $L_2$ is unique.
In addition, the ground state of $L_2$ has total spin $S=\frac{1}{2}\big| |\Lambda_1|+|\Omega_1|-|\Lambda_2|-|\Omega_2|\big|$.

\end{Lemma}

\begin{Proof}
Since $\boldsymbol{K}$ commutes with $N_{\rm p}$ and $\exp(-\beta N_{\rm p})\rhd 0 \wrt \mathcal{P}$ for any $\beta>0$, we have
\begin{align}
\exp\left[-\beta U^*L_2U\right]=\exp\left[-\beta U^*\boldsymbol{K}U\right]e^{-\beta\omega_0 N_{\rm p}}\rhd0\wrt\mathcal Q,
\end{align}
where we have used (\ref{NpPI})  and (\ref{WKWPI}).
Let $\psi$ be the ground state of $\boldsymbol{K}$ and let $\eta_0$ be the bosonic
 Fock vacuum in $\Hph$. Trivially, the vector $\psi\otimes \eta_0$ is the ground state of $L_2$. 
Since the vector $\psi $ has total spin $S=\frac{1}{2}\big| |\Lambda_1|+|\Omega_1|-|\Lambda_2|-|\Omega_2|\big|$ due to Proposition \ref{tsl4}, $\psi\otimes \eta_0$ has the same total spin.
\qed
\end{Proof}

The following lemma is a variant  of Proposition \ref{Overlap}. 
\begin{Lemma}\label{Overlap2}
We set $\mathcal{ X}=Q_0 \mathcal{L}_N\otimes \Hph$.
Let 
$A $ and $B$ be  positive  self-adjoint operators on 
$
\mathcal{X}
$.
 Let $V_1$ and $V_2$ be unitary operators on $\mathcal{X}$.
 We assume the following: 
\begin{itemize}
\item[{\rm (i)}] $A$ and $B$ commute with the total spin operators $S_{\rm tot}^{(3)}, S_{\rm tot}^{(+)}$ and $S_{\rm tot}^{(-)}$.
\item[{\rm (ii)}] Let 
$V=V_1V_2$.
$\{ e^{-\beta V^*AV} \}_{\beta \ge 0}$ and $\{e^{-\beta V_2^*BV_2}\}_{\beta \ge 0}$ are ergodic w.r.t. $\mathcal{Q}$. Hence, the ground state of each of $V^*AV$ and $V_2^*BV_2$ is unique and strictly positive w.r.t. $\mathcal{Q}$ due to Theorem \ref{pff}. 
\item[{\rm (iii)}] $V_1$ commutes with $\boldsymbol{S}_{\rm{}tot}^2$.
\end{itemize}
We denote by $S_A$ (resp. $S_B$) the total spin of the ground state of $A$ (resp. $B$).
Then we have $S_A=S_B$.
\end{Lemma}

\begin{Proof}
We denote by $\psi_A$ (resp. $\psi_B$) the ground state of $V^*AV$ (resp. $V_2^*BV_2$).
 By the assumption (ii), $\psi_A$ and $\psi_B$ are strictly positive w.r.t. $\mathcal{Q}$.
 Because $V\psi_A$ (resp. $V_2\psi_B$) is the ground state of $A$ (resp. $B$), we have
\begin{align}
\boldsymbol{S}_{\rm{}tot}^2V\psi_A&=S_A(S_A+1)V\psi_A,\\
\boldsymbol{S}_{\rm{}tot}^2V_2\psi_B&=S_B(S_B+1)V_2\psi_B.
\end{align}
Applying the assumption (iii),  we readily confirm that  $\boldsymbol{S}_{\rm tot}^2V_2\psi_A=S_A(S_A+1)V_2\psi_A$.
Using the strict positivity of $\psi_A$ and $\psi_B$, we have
$\i<V_2\psi_A,V_2\psi_B>=\i<\psi_A,\psi_B>>0$.
Therefore, by applying  the method of nonzero overlap between the ground states, we have
\begin{align}
S_A(S_A+1)\langle V_2 \psi_A, V_2\psi_B\rangle=\langle {\bs S}_{\rm tot}^2 V_2 \psi_A, V_2\psi_B\rangle=
\langle V_2\psi_A, {\bs S}_{\rm tot}^2 V_2\psi_B\rangle=S_B(S_B+1) \langle V_2\psi_A,  V_2\psi_B\rangle,
\end{align}
which implies that  $S_A=S_B$. 
\qed
\end{Proof}

\subsubsection*{Completion of the proof of (i) of  Theorem  \ref{MainSpin}}

Taking Theorem \ref{mt} and Lemma \ref{ts5} into consideration, we can apply 
 Lemma \ref{Overlap2} with 
$V_1=e^{-L_c}e^{-i\frac{\pi}{2} N_{\mathrm{p}}}$,  $V_2=U$,  $V=V_1 V_2=\mathcal{U}$,
$A=H$ and $ B= L_2$.  \qed

\subsection{Proof of (ii) of Theorem  \ref{MainSpin}}\label{Pf(ii)Go}
The idea of  proof of (ii) of Theorem \ref{MainSpin} is parallel to that of the proof of (i).
Therefore, we will provide a sketch only.

Corresponding to $K_1$ and $H_{\mathrm{H}}$ in the previous subsections, we consider the following Hamiltonians:
\begin{align}
K'
&=\frac{1}{2}\sum_{x, y\in \Lambda}|t_{x,y}|^2({\boldsymbol s}_x^+\cdot{\boldsymbol s}_y^-+{\boldsymbol s}_x^-\cdot{\boldsymbol s}_y^+)
+\sum_{x\in \Lambda_1, u\in \Omega_1}({\boldsymbol s}_x^+\cdot{\boldsymbol S}_u^-+{\boldsymbol s}_x^-\cdot{\boldsymbol S}_u^+)\no
&\quad+\sum_{x\in \Lambda_2, u\in \Omega_2}({\boldsymbol s}_x^+\cdot{\boldsymbol S}_u^-+{\boldsymbol s}_x^-\cdot{\boldsymbol S}_u^+)
+\sum_{x\in\Lambda}\left(n_{x\up}^c-\frac{1}{2}\right)\left(n_{x\down}^c-\frac{1}{2}\right)+\sum_{u\in\Omega}\left(n_{u\up}^f-\frac{1}{2}\right)\left(n_{u\down}^f-\frac{1}{2}\right).
\end{align}
and 
\begin{align}
H_{\mathrm{H}}'
&=\sum_{x,y\in\Lambda}\sum_{\sigma=\up,\down}t_{x,y}c_{x\sigma}^*c_{y\sigma}
+\sum_{x\in\Lambda_1, u\in\Omega_1}\sum_{\sigma=\up,\down}(c_{x\sigma}^*f_{u\sigma}+f_{u\sigma}^*c_{x\sigma})\no
&\quad+\sum_{x\in\Lambda_2, u\in\Omega_2}\sum_{\sigma=\up,\down}(c_{x\sigma}^*f_{u\sigma}+f_{u\sigma}^*c_{x\sigma})
+\sum_{x\in\Lambda}\left(n_{x\up}^c-\frac{1}{2}\right)\left(n_{x\down}^c-\frac{1}{2}\right)+\sum_{u\in\Omega}\left(n_{u\up}^f-\frac{1}{2}\right)\left(n_{u\down}^f-\frac{1}{2}\right).
\end{align}

The following lemma corresponds to Lemma \ref{tsl2}:
\begin{Lemma}
Assume {\bf (C)} and \ref{C7}.
We have
\begin{align}
e^{- U^*K'U}e^{|\Lambda|^2+2|\Lambda_1||\Omega_1|+2|\Lambda_2||\Omega_2|} 
\unrhd e^{-U^*H_{\mathrm{H}}'U} 
\rhd 0 \wrt\mathcal{L}_{N,+} \label{UHUUKUINQ}
\end{align}
Hence, the ground state of $K'$ is unique. Furthermore,  the ground state of $K'$ has total spin $S=\frac{1}{2}\big||\Lambda_1|+|\Omega_2|-|\Lambda_2|-|\Omega_1|\big|$.
\end{Lemma}

\begin{Proof} The basic idea of proof is similar to that of the proof of Lemma \ref{tsl2}.

Because 
$J_{x,u}\leq0$, it holds that $\mathrm{sgn}J_{x,u}=-1$. Hence, the unitary operator $U$ in Lemma \ref{HolePart} satisfies
\begin{align}
U^*c_{x\up}U=c_{x\up},\quad U^*f_{u\up}U=f_{u\up},\quad U^*c_{x\down}U=\gamma_xc_{x\down}^*,\quad U^*f_{u\down}U=-\gamma_uf_{u\down}^*.
\end{align}
Hence, we obtain 
\begin{align}
U^*K' U
&=-\sum_{x, y\in \Lambda}|t_{x,y}|^2c_{x\up}^*c_{y\up}c_{x\down}^*c_{y\down}
-\sum_{x\in \Lambda_1, u\in \Omega_1}(c_{x\up}^*f_{u\up}c_{x\down}^*f_{u\down} + f_{u\up}^*c_{x\up}f_{u\down}^*c_{x\down})\no
&\quad -\sum_{x\in \Lambda_2, u\in \Omega_2}(c_{x\up}^*f_{u\up}c_{x\down}^*f_{u\down} + f_{u\up}^*c_{x\up}f_{u\down}^*c_{x\down})
-\sum_{x\in\Lambda}\left(n_{x\up}^c-\frac{1}{2}\right)\left(n_{x\down}^c-\frac{1}{2}\right)-\sum_{u\in\Omega}\left(n_{u\up}^f-\frac{1}{2}\right)\left(n_{u\down}^f-\frac{1}{2}\right)
\end{align}
and 
\begin{align}
U^*H_{\mathrm{H}}'U
&=\sum_{x,y\in\Lambda}\sum_{\sigma=\up,\down}t_{x,y}c_{x\sigma}^*c_{y\sigma}
+\sum_{x\in\Lambda_1, u\in\Omega_1}\sum_{\sigma=\up,\down}(c_{x\sigma}^*f_{u\sigma}+f_{u\sigma}^*c_{x\sigma})\no
&\quad+\sum_{x\in\Lambda_2, u\in\Omega_2}\sum_{\sigma=\up,\down}(c_{x\sigma}^*f_{u\sigma}+f_{u\sigma}^*c_{x\sigma})
-\sum_{x\in\Lambda}\left(n_{x\up}^c-\frac{1}{2}\right)\left(n_{x\down}^c-\frac{1}{2}\right)-\sum_{u\in\Omega}\left(n_{u\up}^f-\frac{1}{2}\right)\left(n_{u\down}^f-\frac{1}{2}\right).\label{H'UU}
\end{align}
Using (\ref{FermiIdn}) and \eqref{H'UU}, we can  apply Lemma \ref{tsl1} to $U^*H'_\mathrm{H}U$ and obtain
\begin{align}
\exp[-U^*H_{\mathrm{H}}'U]
&\unlhd\exp\Big[- U^*K'U+|\Lambda|^2+2|\Lambda_1||\Omega_1|+2|\Lambda_2||\Omega_2|
\Big] \ \ \wrt \ \mathcal{L}_{N, +}. \label{HKINQ}
\end{align}
Because 
$H_{\mathrm{H}}'$ is a Hubbard Hamiltonian on the bipartite lattice with $A=\Lambda_1\sqcup \Omega_2$ and $B=\Lambda_2\sqcup \Omega_1$, the property  $\exp[-U^*H_{\mathrm{H}}'U]\rhd0$ is already proved in \cite{Miyao2012,Miyao2019}. 
Combining this with \eqref{HKINQ}, we obtain  \eqref{UHUUKUINQ}.
 Furthermore, because of  \eqref{LiebS}, the  ground state of $H'_{\rm H}$ has total spin $S=\frac{1}{2}\big||\Lambda_1|+|\Omega_2|-|\Lambda_2|-|\Omega_1|\big|$.
 Hence,  by applying Proposition \ref{Overlap}, we conclude that the ground state of $K'$  has total spin $S=\frac{1}{2}\big||\Lambda_1|+|\Omega_2|-|\Lambda_2|-|\Omega_1|\big|$, too.
\qed
\end{Proof}

The following proposition corresponds to Proposition \ref{tsl4}:

\begin{Prop}\label{boldKImp} 
Assume {\bf (C)} and \ref{C7}.
Set $\boldsymbol{K}'=K'\restriction P_0\mathcal{L}_N$.  One obtains that 
\begin{align}
e^{- U^*{\bs K}'U}\rhd0\wrt Q_0\mathcal{L}_{N,+}.
\end{align}
Hence, the ground state of ${\bs K}'$ is unique. In addition, the ground state of ${\bs K}'$ has total spin $S=\frac{1}{2}\big||\Lambda_1|+|\Omega_2|-|\Lambda_2|-|\Omega_1|\big|$.
\end{Prop}
\begin{Proof}
By using the arguments similar to those of the proof of Proposition \ref{tsl4}, we can prove Proposition \ref{boldKImp}.
\qed
\end{Proof}

Using a method of proof similar to that applied to  Lemma \ref{ts5}, we obtain the following:
\begin{Lemma}\label{L'_2Imp}
Assume {\bf (C)} and \ref{C7}.
Set $
L_2'
=\boldsymbol{K}'+\omega_0N_{\mathrm{p}}.
$
Then we have
\begin{align}
e^{- U^*L_2'U}\rhd0\wrt \mathcal{Q}.
\end{align}
Hence, the ground state of $L_2'$ is unique. In addition, the ground state of $L_2'$ has total spin $S=\frac{1}{2}\big||\Lambda_1|+|\Omega_2|-|\Lambda_2|-|\Omega_1|\big|$.
\end{Lemma}

\subsubsection*{Completion of the proof of (ii) of  Theorem  \ref{MainSpin}}

Taking Theorem \ref{mt} and Lemma \ref{L'_2Imp} into consideration, we can apply 
 Lemma \ref{Overlap2} with 
$V_1=e^{-L_c}e^{-i\frac{\pi}{2} N_{\mathrm{p}}}$,  $V_2=U$,  $V=V_1 V_2=\mathcal{U}$,
$A=H$ and $ B= L_2'$.   \qed

\section{Discussion}\label{Discussion}
In the present paper, we proved that the ground state of the KLM with the electron-phonon interaction, ${\bf H}$, is  unique 
  and it has total spin $S$ given by \eqref{CaseDefS}.
Note that the value of $S$ is equal to that  of the total spin of the ground state of the antiferromagnetic 
Heisenberg model, $H_{\rm Heis}$, on the coupled  lattice $\Lambda\sqcup \Omega$.\footnote{
To be precise, bipartite structure of the lattice should be   specified:
The KLM with  antiferromagnetic coupling corresponds to  $H_{\rm Heis}$ on $\Lambda\sqcup \Omega$ with 
the bipartite structure $\Lambda\sqcup \Omega=A\sqcup B$, where $A=\Lambda_1\sqcup \Omega_1$ and $B=\Lambda_2\sqcup \Omega_2$; in contrast to this, the KLM with  ferromagnetic coupling corresponds to   $H_{\rm Heis}$ on $\Lambda\sqcup \Omega=A\sqcup B$ with $A=\Lambda_1\sqcup \Omega_2$ and $B=\Lambda_2\sqcup \Omega_1$. This is the reason why the value of $S$ depends on the type of  coupling, see \eqref{CaseDefS}.
}
This is not just a coincidence; the reason behind this agreement is examined in detail in \cite{Miyao2019};
in  the context of the theory established in \cite{Miyao2019}, $H_{\rm Heis}$, ${\bf H}$,  ${\bf H}_{\rm QED}$ and $\mathbf{H}_{\rm fp}$  belong to 
the {\it Marshall-Lieb-Mattis stability class}, $\mathscr{U}_{\rm MLM}$, on $\Lambda\sqcup \Omega$. Here, recall that $\mathbf{H}_{\rm QED}$ and $\mathbf{H}_{\rm fp}$ are defined in Remarks \ref{QED} and \ref{Hfp}, respectively.
Every Hamiltonian in  $\mathscr{U}_{\rm MLM}$ was  proved to have  the common   total spin $S$ in the ground state; in addition, it was shown that $\mathscr{U}_{\rm MLM}$ contains  at least a  countably infinite number of Hamiltonians.
Within $\mathscr{U}_{\rm MLM}$, we can consider the KLM with additional  interactions which are more complicated than the electron-phonon and   electron-photon interactions examined in this paper;  a simple example is the combination of the two interactions: 
\begin{align}
\mathbf{H}_{\rm ep, ep}=&-\sum_{x,y\in\Lambda}\sum_{\sigma=\up,\down}t_{x,y}
\exp\Bigg\{
i \int_{C_{xy}} dr\cdot A(r)
\Bigg\}
c_{xs}^*c_{ys}+\sum_{x\in\Lambda,u\in\Omega}J_{x,u}{\boldsymbol s}_x\cdot{\boldsymbol S}_u+ \no
&+\sum_{x,y\in\Lambda}U_{x,y}(n_x^c-1)(n_y^c-1)
+\sum_{x,y\in\Lambda}g_{x,y}n_x^c(b_y^*+b_y)+\no
&+\sum_{k\in V^*}\sum_{\lambda=1,2} \omega(k) a(k, \lambda)^*a(k, \lambda)+\omega_0\sum_{x\in\Lambda}b_x^*b_x.
\end{align}

\subsection*{Acknowledgements}
T.M. was  supported by 
JSPS KAKENHI Grant Numbers 18K03315, 20KK0304.
 We are  grateful to   the anonymous referee for the  constructive comments and suggestions,
which helped considerably to improve the presentation of the manuscript.

\appendix
\section{Basic properties of the Lang-Firsov transformation}\label{LFTrnSec}

In this appendix, we review some basic properties of the Lang-Firsov transformation.

For each  $\theta\in\mathbb R$,  we have
\begin{align}
e^{i\theta N_{\rm p}}b_xe^{-i\theta N_{\rm p}}=e^{-i\theta}b_x.
\end{align}
Hence, 
\begin{align}
e^{i\frac{\pi}{2}N_{\rm p}}q_xe^{-i\frac{\pi}{2}N_{\rm p}}=p_x,\quad e^{i\frac{\pi}{2}N_{\rm p}}p_xe^{-i\frac{\pi}{2}N_{\rm p}}=-q_x,
\end{align}
where $p_x$ and $q_x$ are defined by (\ref{Defpq}).

Next, we set
\begin{align}
L_c=-i\frac{\sqrt2}{\omega_0}\sum_{x,y\in\Lambda}g_{x,y}n_x^cp_y.
\end{align}
Then we readily confirm that 
\begin{align}
e^{L_c}c_{x\sigma}e^{-L_c}&= \exp\left(i\frac{\sqrt2}{\omega_0}\sum_{y\in\Lambda}g_{x,y}p_y\right)c_{x\sigma},\\
e^{L_c}f_{u\sigma}e^{-L_c} &= f_{u\sigma},\\
  e^{L_c}b_xe^{-L_c} & =b_x-\frac{1}{\omega_0}\sum_{y\in\Lambda}g_{y,x}n_y^c.
\end{align}

\section{Feynman-Kac formulas for kernel operators}
\subsection{Strong product integrations}

As a preliminary, we briefly review  strong product integrations (for details, see \cite{Dollard1984}).

Let $M_n(\mathbb{C})$ be the space of $n\times n$ matrices with
complex entries. Let $A(\cdot): [0, a]\to M_n(\mathbb{C})$ be
continuous. Let $P=\{s_0, s_1, \dots, s_n\}$ be  a partition of $[0, a]$
and $\mu(P)=\max_j\{s_j-s_{j-1}\}$.  The {\it strong product integration of
$A$}
is defined by 
\begin{align}
\prod_0^{a} e^{A(s)ds}:=\lim_{\mu(P)\to
 0}e^{A(s_1)(s_1-s_0)} e^{A(s_2)(s_2-s_1)} \cdots   e^{A(s_n)(s_n-s_{n-1})}
 .\label{DefStrInt}
\end{align} 
Note that the limit is independent of any partition $P$.
The following estimate will be useful:
\begin{align}
\Bigg\|
\prod_0^{a} e^{A(s)ds}-1 -\int_0^a ds A(s)
\Bigg\|\le 
e^{\int_0^a ds \|A(s)\|} -1 -\int_0^a ds \|A(s)\|.\label{ProdInq}
\end{align} 
As for the proof of \eqref{ProdInq},  see \cite{Dollard1984}.

\subsection{Kernel operators}
Under identification \eqref{FiberHil}, each $\psi\in Q_0\mathcal{L}_N\otimes \Hph
$ can be expressed as 
$\psi=\int_{\mathbb{R}^{|\Lambda|}}^{\oplus } \psi(\bphi)\, d\bphi$, where
$\psi(\bphi) \in  Q_0\mathcal{L}_N$ for a.e. $\bphi$.

\begin{Def}
{\rm 
Let $A$ be a bounded linear operator on $
Q_0\mathcal{L}_N\otimes \Hph
$. If there exists a
 $\mathscr{B}(
 Q_0\mathcal{L}_N
 )$-valued map $(\bphi,
 \bphi')\mapsto K(\bphi,
 \bphi')$
 such that 
\begin{align}
(A\psi)(\bphi)=\int_{\mathbb{R}^{|\Lambda|}} K(\bphi, \bphi') \psi(\bphi')d\bphi' \
 \   \mbox{$\forall\psi \in 
  Q_0\mathcal{L}_N\otimes \Hph
 $ },
\end{align} 
then we say that {\it $A$ has a kernel operator $K$. } We denote by $A(\bphi,
 \bphi')$ the kernel
 operator of $A$  if it exists. Trivially,  it holds that 
\begin{align}
\langle \vphi, A\psi\rangle=\int_{\mathbb{R}^{|\Lambda|}\times \mathbb{R}^{|\Lambda|}} d\bphi d\bphi'
\big \langle\vphi(\bphi), A(\bphi, \bphi') \psi(\bphi')
 \big\rangle_{
 Q_0\mathcal{L}_N
 }. 
\end{align} 
}
\end{Def}   

The following lemma  is often useful.
\begin{Lemma}\label{kernelPP}
Let $A$ be a bounded linear operator on $
Q_0\mathcal{L}_N\otimes \Hph
$. Suppose that $A$ has a kernel operator. If $A\unrhd 0$ w.r.t. $\mathcal{Q}$,  then $A(\bphi, \bphi') \unrhd 0$ w.r.t. $Q_0\mathcal{L}_{N, +}$ for a.e. $\bphi, \bphi'$.
\end{Lemma}
\begin{Proof}
Let $\phi, \psi\in Q_0\mathcal{L}_{N, +}$ and let $f, g\in \mathcal{P}$. Since $\phi \otimes f\ge 0$ and $\psi\otimes g \ge 0$ \wrt $\mathcal{Q}$, we have
\begin{align}
0\le \langle \phi\otimes f, A\psi\otimes g\rangle=\int_{\mathbb{R}^{|\Lambda|} \times \mathbb{R}^{|\Lambda|}} f(\bphi) g(\bphi)
\langle \phi, A(\bphi, \bphi')\psi\rangle_{Q_0\mathcal{L}_N}.
\end{align}
Because $f$ and $g$ are arbitrary, we find that $\langle \phi, A(\bphi, \bphi') \psi\rangle_{Q_0\mathcal{L}_N} \ge 0$. Since $\phi $ and $\psi$ are arbitrary,
we conclude the desired assertion in the lemma. \qed
\end{Proof}

\subsection{Feynman-Kac formulas  for kernel operators}
In this subsection,  we will express   kernel operators  of
$\exp\{-\beta(R+\omega_0N_{\mathrm{p}})\}$ and $\exp\{-\beta(R- \frac{1}{2}\mathbb{J}+\omega_0N_{\mathrm{p}})\}$ in terms of  functional
integral representation.
To this end, we recall some basic facts  concerning the Wiener  process (see  \cite{Simon2005} for details).
Let $(A, \mathcal{M}, P)$ be the  probability space for the 
$|\Lambda|$-dimensional Brownian bridge 
 $\{{\boldsymbol\alpha}(s)\, |\, 0\le s
\le 1\}=\{\{\alpha_x(s)\}_{x\in \Lambda}\, |\, 0\le s \le 1\}$, i.e., the Gaussian process with covariance
\begin{align}
\int_A \alpha_x(s)\alpha_y(t)\, dP=\delta_{xy}s(1-t) \label{Bridge}
\end{align}  
for $0\le s \le t \le 1$ and $x, y\in \Lambda$. Define, for each $\bphi, \bphi'\in
\mathbb{R}^{|\Lambda|}$,
\begin{align}
\bomega(s)=(
1-\beta^{-1}s)
\bphi+\beta^{-1}s \bphi'+\sqrt{\beta}{\boldsymbol \alpha}(\beta^{-1}s).\label{DefOmega}
\end{align} 
The conditional Wiener measure $d\mu_{\bphi, \bphi'; \beta}$ is
given by 
\begin{align}
d \mu_{\bphi, \bphi'; \beta}=P_{\beta}(\bphi, \bphi')dP,
\end{align} 
where $P_{\beta}(\bphi,
\bphi')=(2\pi\beta)^{-|\Lambda|/2}\exp\big(-\frac{1}{2\beta}|\bphi-\bphi'|^2\big)$.

For each $\bfphi\in A$, $\bomega(\bfphi)$ indicates a function $s
\mapsto \bomega(s)(\bfphi)$, the sample path $\bomega(\cdot)(\bfphi)$
associated with $\bfphi$.
 Let 
\begin{align}
G_{\beta}(\bomega(\bfphi))=
\prod_0^{\beta}
e^{-\omega_0^{-1}\mathsf{R}(\bomega(s)(\bfphi))\, d s},\label{GOp}
\end{align} 
where the right hand side  of (\ref{GOp}) is a  strong product integration (see
 \eqref{DefStrInt})  and $\mathsf{R}(\bq)$ is defined by (\ref{DefSfR(q)}). Because    $\bomega(s)(\bfphi)$ is continuous
in $s$
for all $\bfphi\in A$,  
the right hand side
 of (\ref{GOp}) exists.

\begin{Prop}\label{KMkernel}
$e^{-\omega_0^{-1}\beta (R+\omega_0 N_{\rm p})}$ has a kernel operator given by 
\begin{align}
e^{-\omega_0^{-1}\beta (R+\omega_0N_{\rm p}) }(\bphi, \bphi')
=&\int d \mu_{\bphi, \bphi'; \beta}\, \mathcal{L}\Big[
G_{\beta}(\bomega)
\Big]
 \mathcal{R}\Big[
G_{\beta}(\bomega)^*\,
\Big]\, 
e^{-\int_0^{\beta}d s\,  V(\bomega(s))}, \label{FKformula1}
\end{align} 
   where 
$
V(\bphi)=\frac{1}{2}\sum_{x\in \Lambda}(q_x^2-1)
$.
\end{Prop}
\begin{Proof}
See \cite[Proposition 3.7]{Miyao2016}. 
Note that, in \cite{Miyao2016}, $\phi_x$ and $\pi_x$ are defined by 
$\phi_x=\frac{1}{\sqrt{2\omega_0}}\overline{(b_x^*+b_x)}$ and $\pi_x=i\sqrt{\frac{\omega_0}{2}}\overline{(b_x^*-b_x)}$, respectively, which are slightly different from \eqref{Defpq}. This is the reason why we need the factor $\omega_0^{-1}$ in the right hand side of \eqref{GOp} and  in the left hand side of \eqref{FKformula1}.
\qed
\end{Proof}

Next, we will express a kernel operator of $
e^{-\omega_0^{-1}\beta(R-\frac{1}{2}\mathbb{J}+\omega_0 N_{\rm p})}
$
 in terms of the Wiener process. Note that because $\bomega(s)(\bfphi)$ is continuous in $s$ for all $\bfphi\in A$, the following strong product integration exists:
  \begin{align}
  \prod_0^{\beta}e^{-\omega_0^{-1}(R(\bomega(s)(\bfphi))-\frac{1}{2}\mathbb{J})ds}.
  \end{align}

\begin{Prop}\label{PropKernel}
$e^{-\omega_0^{-1}\beta(R-\frac{1}{2}\mathbb{J}+\omega_0 N_{\rm p})}$
 has a kernel operator given by 
\begin{align}
e^{-\omega_0^{-1}\beta (R-\frac{1}{2}\mathbb{J}+\omega_0N_{\rm p}) }(\bphi, \bphi')
=&\int d \mu_{\bphi, \bphi'; \bd}\, \bigg[
\prod_0^{\bd}e^{-\omega_0^{-1}(R(\bomega(s))-\frac{1}{2}\mathbb{J})ds}
\bigg]
\, 
e^{-\int_0^{\bd}d s\,  V(\bomega(s))}. \label{FKformula}
\end{align} 
In addition, $e^{-\beta (R-\frac{1}{2}\mathbb{J}+\omega_0N_{\rm p}) }(\bphi, \bphi') \unrhd 0$ w.r.t. $Q_0 \mathcal{L}_{N, +}$ for a.e. $\bphi, \bphi'$.
\end{Prop}
\begin{Proof}
First,  recall the following fact \cite[Theorem 4.8]{Simon2005}:
\begin{align}
&\Big\langle f_0,  e^{-\beta N_{\mathrm{p}}/n}f_1  e^{-\beta
  N_{\mathrm{p}}/n}f_2 \cdots f_n\Big\rangle_{L^2(\mathbb{R}^{|\Lambda|})}\no
=&\int_{\mathbb{R}^{|\Lambda|}\times \mathbb{R}^{|\Lambda|}} d\bphi d\bphi'\int d\mu_{\bphi,
 \bphi'; \bd}
f_0(\bphi)^* f_1\big(\bomega(\tfrac{\bd}{n})\big)
 f_2\big(\bomega(\tfrac{2\bd}{n})\big) \cdots f_{n-1}\big(\bomega(\tfrac{(n-1)\bd}{n})\big) f_n(\bphi')
\,  e^{-\int_0^{\bd}ds V(\bomega(s))} \label{FKFormula}
\end{align} 
for  $f_0, f_n\in L^2(\mathbb{R}^{|\Lambda|})$ and $f_1, \dots, f_{n-1}\in
L^{\infty}(\mathbb{R}^{|\Lambda|})$.
 By using  (\ref{FKFormula}) and the Trotter--Kato product formula, we have 
\begin{align}
\big\langle \vphi,  e^{-\omega^{-1}_0\beta(
R-\frac{1}{2}\mathbb{J}+\omega_0 N_{\rm p})
}\psi\big\rangle
=&\lim_{n\to \infty} \Big\langle\vphi, \Big(
 e^{- \beta   N_{\mathrm{p}}/n}  e^{-\omega_0^{-1}\beta 
 (
 R-\frac{1}{2}\mathbb{J}
 ) /n}
\Big)^n \psi\Big\rangle\no
=& \lim_{n\to \infty} \int_{\mathbb{R}^{|\Lambda|}\times \mathbb{R}^{|\Lambda|}}d\bphi d\bphi'
\int d\mu_{\bphi, \bphi'; \bd} e^{-\int_0^{\bd}ds V(\bomega(s))}\no
&\times \Big\langle
\vphi(\bphi),
  e^{-\tfrac{\beta}{n} \omega_0^{-1}\big(R(\bomega(\tfrac{\bd}{n}))-\frac{1}{2}\mathbb{J}\big)}
  e^{-\tfrac{\beta}{n}\omega_0^{-1}\big(R(\bomega(\tfrac{2\bd}{n}))-\frac{1}{2}\mathbb{J}\big)}
\cdots
  e^{-\tfrac{\beta}{n} \omega_0^{-1}
  \big(R(\bomega(\tfrac{2\bd}{n}))-\frac{1}{2}\mathbb{J}\big)
  }
\psi(\bphi')
\Big\rangle_{Q_0\mathcal{L}_N}.
\end{align} 
By applying the dominated convergence theorem and using \eqref{DefStrInt}, we
obtain (\ref{FKformula}).  By  the fact $
e^{-\beta(R-\frac{1}{2}\mathbb{J}+\omega_0 N_{\rm p})} \unrhd 0
$ w.r.t. $\mathcal{Q}$ and  Lemma \ref{kernelPP}, we conclude that the kernel operator $e^{-\beta (R-\frac{1}{2}\mathbb{J}+\omega_0N_{\rm p}) }(\bphi, \bphi')$
preserves the positivity w.r.t. $Q_0\mathcal{L}_{N, +}$  for a.e. $\bphi, \bphi'$. \qed
\end{Proof}

\section{Proof of  Lemma  \ref{one step}}\label{B}

To prove the Lemma \ref{one step}, we need some preliminaries.

Let $(\bsigma_c,\bsigma_f),(\bsigma_c',\bsigma_f')\in\SN$. Assume that  there exist $x,y\in\Lambda$ such that $t_{x,y}\neq0$ and
$
\big|\bsigma_c,\bsigma_c,\bsigma_f,\bsigma_f\big\rangle=c_{x\up}^*c_{y\up}c_{x\down}^*c_{y\down}\big|\bsigma_c',\bsigma_c',\bsigma_f',\bsigma_f'\big\rangle.
$
Let   $g,h\in\mathcal Q\setminus\{0\}$. 
Using the Feynman-Kac formula(Proposition \ref{KMkernel}), we have
\begin{align}
&\l<\bsigma_c,\bsigma_c,\bsigma_f,\bsigma_f,g\left|e^{-\omega_0^{-1}t(R+\omega_0 N_{\mathrm{p}})}\right|\bsigma_c',\bsigma_c',\bsigma_f',\bsigma_f',h>\no
=&\int d\bq d\bq\rq{} g(\bq) h(\bq\rq{})\int_A d\mu_{\bq, \bq\rq{}; t}  e^{-\int_0^{t} ds V(\bome(s))}
\bigg|
\Big\langle\bsigma_c, \bsigma_f\Big|G_{t}(\bome)\Big|\bsigma_c\rq{}, \bsigma_f\rq{}\Big\rangle
\bigg|^2,\label{FKF}
\end{align}
where 
$G_t(\bome)$ is defined by \eqref{GOp}.

Our aim is to estimate the right hand side of (\ref{FKF}) from below.
For this purpose, we recall some  facts from \cite{Miyao2016}: 
For a given  $z\in \Lambda$, we set
\begin{align}
a_z(\{x, y\})=\sqrt{2} \omega_0^{-1} (g_{xz}-g_{yz }). 
\end{align}
Let 
\begin{align}
\mathcal{Y}=\Big\{
(\bq, \bq\rq{})\in \mathbb{R}^{|\Lambda|}\times \mathbb{R}^{|\Lambda|}\, \Big|\, 
\mbox{$\sum_{z\in \Lambda} a_z(\{x, y\})(q_z-q_z\rq{}) \in 2\pi\mathbb{Z}$}
\Big\}.
\end{align}
 Next, let 
\begin{align}
W_t=\big \{{\bs \varphi} \in A\, \big|\, |\bome(s)({\bs \varphi})-(1-t^{-1}s)\bq-t^{-1}s \bq\rq{}| \le t^{1/4}\ \mbox{ for all $s\in [0, t]$} \big\}.
\end{align}
Note that $\mu_{\bq, \bq\rq{}; t}(W_t)>0$  holds for each $\bq, \bq\rq{}\in \mathcal{Y}^c$, the complement of $\mathcal{Y}$.
In \cite[Appendix C]{Miyao2016}, we have proved the following:

\begin{Lemma}
For each $\bq, \bq\rq{} \in \mathcal{Y}^c$ and ${\bs \varphi} \in W_t$,  there exist strictly positive numbers $a, b, c$ such that  
\begin{align}
\bigg|\Big\langle\bsigma_c, \bsigma_f\Big|G_t(\bome(\bfphi))\Big|\bsigma_c\rq{}, \bsigma_f\rq{}\Big\rangle\bigg| \ge a |t_{x, y}|\xi_{x y} t-b t^{5/4}-ct^2, \label{LWBD}
\end{align}
where 
\begin{align}
\xi_{x y}=2 \bigg|
\frac{\sin \theta_{xy}}{\theta_{xy}}
\bigg|,\ \ \ \theta_{xy}=\frac{1}{2}\sum_{z\in \Lambda} a_z(\{x, y\})(q_z\rq{}-q_z).
\end{align}
Note that $\xi_{xy}>0$  holds for all $\bq, \bq\rq{} \in \mathcal{Y}^c$.
\end{Lemma}

\begin{Proof}
We first recall an important bound  from \cite[Lemma C. 1]{Miyao2016}:
Let $(\bphi, \bphi')\in \mathcal{Y}^c$.
  There exist $t_0>0$ and $C>0$ such that, for all
 $t \in (0, t_0)$ and $\bfphi\in W_{ t}$,  it holds that 
\begin{align}
\Bigg|
t^{-1} \int_0^{t} ds \exp\Big\{
i \Phi_{x, y}\big(\bomega(s)(\bfphi)\big)
\Big\}
\Bigg|
\ge \xi_{xy} -C t^{1/4}.
\end{align} 
 Hence, we obtain 
 \begin{align}
 \Bigg|\Big\langle\bsigma_c, \bsigma_f \Big|
\int_0^{t}d s\,  
\mathsf{J}_{\rm c}\big(\bomega(s)(\bfphi)\big)
\Big|\bsigma_c\rq{}, \bsigma_f\rq{}\Big\rangle
\Bigg|\ge |t_{x, y}|\xi_{xy} t  -|t_{x, y}|C t^{5/4}, \label{Order1}
 \end{align}
 where $\mathsf{J}_{\rm c}(\bphi)$ is defined  by \eqref{DefJ_c(q)}.

 By applying 
 \eqref{ProdInq},  we find that 
\begin{align}
&\Bigg|\Big\langle\bsigma_c, \bsigma_f \Big|
\Bigg[ G_{t}\big(\bomega(s)(\bfphi)\big)
-\int_0^{t}d s\,  
\mathsf{J}_{\rm c}\big(\bomega(s)(\bfphi)\big)
\Bigg]\Big|\bsigma_c\rq{}, \bsigma_f\rq{}\Big\rangle
\Bigg|\no
=&\Bigg|
\Big\langle\bsigma_c, \bsigma_f \Big|
\Bigg[ G_{t}\big(\bomega(s)(\bfphi)\big)-1
-\int_0^{t}d s\,  
\mathsf{R}\big(\bomega(s)(\bfphi)\big)
\Bigg]
\Big|\bsigma_c\rq{}, \bsigma_f\rq{}\Big\rangle
\Bigg|\no
\le&
  \bigg(
\int_0^{t} d s\, \big\|\mathsf{R}
\big(\bomega(s)(\bfphi)\big)
\big\|
\bigg)^2\no
\le& {\rm Const. } \ t^2. \label{Order2}
\end{align} 
In the first equality, we have used the facts $
 \langle\bsigma_c, \bsigma_f |\bsigma_c\rq{}, \bsigma_f\rq{}\rangle=0
 $  and  $
 \langle\bsigma_c, \bsigma_f |\mathsf{R}(\bq)|\bsigma_c\rq{}, \bsigma_f\rq{}\rangle=
 \langle\bsigma_c, \bsigma_f|\mathsf{J}_{\rm c}(\bq) |\bsigma_c\rq{}, \bsigma_f\rq{}\rangle
 $.

 Combining \eqref{Order1} and \eqref{Order2}, we obtain the desired assertion in the lemma.
\qed
\end{Proof}

\begin{Lemma}\label{Diffi}
Let $(\bsigma_c,\bsigma_f),(\bsigma_c',\bsigma_f')\in\SN$. Assume that  there exist $x,y\in\Lambda$ such that $t_{x,y}\neq0$ and
$
\big|\bsigma_c,\bsigma_c,\bsigma_f,\bsigma_f\big\rangle=c_{x\up}^*c_{y\up}c_{x\down}^*c_{y\down}\big|\bsigma_c',\bsigma_c',\bsigma_f',\bsigma_f'\big\rangle.
$
For any  $g,h\in\mathcal Q\setminus\{0\}$, there exists a $\gamma(g, h)>0$ 
depending on $g$ and $h$ such that if 
$0<t<\gamma(g,h)$, then
\begin{align}
\l<\bsigma_c,\bsigma_c,\bsigma_f,\bsigma_f,g\left|e^{-t(R+\omega_0 N_{\mathrm{p}})}\right|\bsigma_c',\bsigma_c',\bsigma_f',\bsigma_f',h>>0 \label{PIC}
\end{align}
holds.
\end{Lemma}

\begin{Proof}
For a given $\varepsilon>0$, we set 
\begin{align}
\mathcal{Y}_{\varepsilon}=\big\{
(\bq, \bq\rq{})\, |\, \mathrm{dist}\big((\bq, \bq\rq{}); \mathcal{Y}\big)<\varepsilon
\big\},
\end{align}
where $
\mathrm{dist}\big((\bq, \bq\rq{}); \mathcal{Y}\big)=\inf\{|(\bq, \bq')-{\bs y}|\, |\, {\bs y} \in \mathcal{Y}\}
$, the distance between $(\bq, \bq')$ and $\mathcal{Y}$.
Since $g$ and $h$ are nonzero, there exist compact sets, $K_g$ and $ K_h$, with nonzero Lebesgue measures such that 
$K_g\subseteq \mathrm{supp} g$ and $K_h\subseteq \mathrm{supp} h$.
Therefore, $\mathcal{Y}^c_{\varepsilon}\cap (K_g\times K_h)$ is a compact set with  nonzero Lebesgue measure, provided that $\varepsilon$ is small enough. With this  setting, let $\tilde{\xi}_{x y}
=\min_{
(\bq, \bq\rq{})\in \mathcal{Y}^c_{\varepsilon}\cap (K_g\times K_h)
} \xi_{xy}
$. Note that $\tilde{\xi}_{xy}$ is strictly positive.
 Hence, there exists a $\gamma>0$ such that if $0<t<\gamma$, then 
 $a |t_{x, y}|\tilde{\xi}_{x y} t -bt^{5/4}-ct^2>0$. Combining this with \eqref{LWBD}, we get
 \begin{align}
 \mbox{the right hand side of (\ref{FKF})}
 \ge  C (a|t_{x, y}|\tilde{\xi}_{x y} t -bt^{5/4}-ct^2)>0,
 \end{align}
 provided that $0<t<\gamma$,
 where
 \begin{align}
 C=\int_{
 \mathcal{Y}^c_{\varepsilon}\cap (K_g\times K_h)
 } d\bq d\bq\rq{} g(\bq) h(\bq\rq{})\int_{W_t} d\mu_{\bq, \bq\rq{}; t}  e^{-\int_0^t ds V(\bome(s))}>0.
 \end{align}
By construction, $\gamma$ depends on $g $ and $h$. \qed
 
\end{Proof}

\subsubsection*{Proof of Lemma \ref{one step}}

For notational simplicity, 
we set $|\bsigma\rangle=|\bsigma_c,\bsigma_c,\bsigma_f,\bsigma_f\rangle
$ and $
|\bsigma'\rangle=|\bsigma_c',\bsigma_c',\bsigma_f',\bsigma_f'\rangle$.

Assume first that (i) holds. Because  $e^{-t(R+\omega_0N_{\rm p})}\unrhd0\wrt\mathcal Q$ for all $t\geq0$  and  $\mathbb{J} \unrhd 0$ w.r.t. $\mathcal{Q}$, we have, by Lemma \ref{ppiexp1},
\begin{align}
e^{-t(R-\frac{1}{2}\mathbb{J}+\omega_0N_{\rm p})} \unrhd e^{-t(R+\omega_0 N_{\rm p})} \unrhd 0
\end{align}
w.r.t. $\mathcal{Q}$ for all $t\ge 0$. Combining this with Lemma \ref{Diffi}, we obtain
\begin{align}
S(t) \ge \l<\bsigma, g\left|e^{-t(R+\omega_0 N_{\mathrm{p}})}\right|\bsigma', h>>0,
\end{align}
provided that $0<t<\gamma(g, h)$.

 Next, assume that (ii) holds.  
 Applying  the Duhamel formula, we have
 \begin{align}
 S(t)&=\l<\bsigma, g\left|e^{-t\left(R-\frac{1}{2}\mathbb{J}+\omega_0N_{\rm p}\right)}\right|\bsigma',h>\no
&=\sum_{n\geq0}\int_{0\le s_1 \le \cdots \le s_n\le 1}\left<\bsigma, g\left|e^{-s_1t\omega_0N_{\rm p}}\left\{-t(R-\frac{1}{2}\mathbb{J})\right\}\cdots\left\{-t(R-\frac{1}{2}\mathbb{J})\right\}e^{-(1-s_n)t\omega_0N_{\rm p}}\right|\bsigma',h\right>\,ds_n\cdots ds_1. \label{S(t)Duha}
 \end{align}
 Since $\i<\bsigma,g\left|e^{-st\omega_0N_{\rm p}}Re^{-(1-s)t\omega_0N_{\rm p}}\right|\bsigma',h>=0$, we have
\begin{align}
&\int_0^1\l<\bsigma,g\left|e^{-st\omega_0N_{\rm p}}\left\{-t(R-\frac{1}{2}\mathbb{J})\right\}e^{-(1-s)t\omega_0 N_{\rm p}}\right|\bsigma',h>\,ds\no
&=\frac{t}{2}\int_0^1\l<\bsigma,g\left|e^{-st\omega_0N_{\rm p}}\mathbb{J}e^{-(1-s)t\omega_0N_{\rm p}}\right|\bsigma',h>\,ds\no
&=\frac{t|J_{x,u}|}{4}\int_0^1\l<\bsigma, g\left|e^{-st\omega_0N_{\rm p}}(c_{x\up}^*f_{u\up}c_{x\down}^*f_{u\down}+f_{u\up}^*c_{x\up}f_{u\down}^*c_{x\down})e^{-(1-s)t\omega_0N_{\rm p}}\right|\bsigma',h>\,ds\no
&=\frac{t|J_{x,u}|}{4}\l<g,e^{-t\omega_0N_{\rm p}}h>.
\end{align}
Because $e^{-t\omega_0 N_{\rm p}} \unrhd 0$ w.r.t. $\mathcal{P}$ for all $t\ge 0$, it holds that 
\begin{align}
\big|
\big(
 e^{-t\omega_0 N_{\rm p}} f
\big)(\bq)
\big| \le ( e^{-t\omega_0 N_{\rm p}} |f|)(\bq)\label{PPProperty}
\end{align}
for all $f\in L^2(\mathbb{R}^{|\Lambda|})$. Applying this bound,  we have
\begin{align}
&\left|\sum_{n\geq2}\int\left<\bsigma,g\left|e^{-s_1t\omega_0N_{\rm p}}\left\{-t(R-\frac{1}{2}\mathbb{J})\right\}\cdots\left\{-t(R-\frac{1}{2}\mathbb{J})\right\}e^{-(1-s_n)t\omega_0N_{\rm p}}\right|\bsigma',h\right>\,ds_n\cdots ds_1\right|\no
&\leq t^2\sum_{n\geq2}\frac{1}{n!}\left(2\sum_{x,y\in\Lambda}|t_{x,y}|+\sum_{x\in\Lambda,u\in\Omega}|J_{x,u}|+2\sum_{x,y\in\Lambda}|U_{\mathrm{eff},x,y}|+\frac{1}{2}\|\mathbb{J}\|\right)^n\l<g,e^{-t\omega_0N_{\rm p}}h>\no
&\leq t^2e^\alpha\l<g,e^{-t\omega_0N_{\rm p}}h>,
\end{align}
where
$
\alpha=2\sum_{x,y\in\Lambda}|t_{x,y}|+\sum_{x\in\Lambda,u\in\Omega}|J_{x,u}|+2\sum_{x,y\in\Lambda}|U_{\mathrm{eff},x,y}|+\frac{1}{2}\|\mathbb{J}\|.
$
Inserting the above bounds into \eqref{S(t)Duha},  we find that 
\begin{align}
S(t)
\geq\frac{t|J_{x,u}|}{4}\l<g,e^{-t\omega_0N_{\rm p}}h>-t^2e^\alpha\l<g,e^{-t\omega_0N_{\rm p}}h>
=\l<g,e^{-t\omega_0N_{\rm p}}h>t\left(\frac{|J_{x,u}|}{4}-e^\alpha t\right)>0,
\end{align}
provided that $t<|J_{x,u}|e^{-\alpha}/4$.
This completes the proof of Lemma \ref{one step}. 
\qed

\section{Proof of Proposition \ref{QCone}}\label{PfD}
In Appendix \ref{PfD}, we will prove Proposition \ref{QCone}.
For this purpose, let  $(M,\Sigma,\mu)$ be a $\sigma$-finite measure space. 
We assume that $L^2(M)$ is separable.
Let $\mathcal X$ be a separable Hilbert space, and let $\mathcal{C} \subset \mathcal X$ be a Hilbert cone.

Define 
\begin{align}
\mathcal A&=\left\{\left.\int_M^\oplus F(x)\,d\mu(x)\in\mathcal X\otimes L^2(M)\,\right|\,F(x)\in \mathcal{C}\ \mbox{$\mu$-a.e.}\right\}.
\end{align}
As is well-known, $\mathcal{A}$ is a Hilbert cone in $\mathcal X\otimes L^2(M)$, see, e.g.,  \cite{Bs1976} and \cite[Proof of Proposition 4.2]{Miyao2016}.

\begin{Prop} \label{usefulDCONE} One obtains
\begin{align}
\mathcal A=
\overline{\mathrm{coni}}\{\phi\otimes f\in\mathcal X\otimes L^2(M)\,|\,\phi\in \mathcal{C},f\in L_+^2(M)\}, \label{P=Q}
\end{align}
 where $L_+^2(M)$ is a canonical Hilbert  cone in $L^2(M):\ L^2_+(M)=\{f\in L^2(M),\, |\, f(x) \ge 0 \ \mbox{$\mu$-a.e.}\}$.
\end{Prop}
\begin{Proof}
First, we recall a  useful fact:  Let $\mathcal{R}$ be a convex cone in $\mathcal{X}$. Then the {\it dual cone} of $\mathcal{R}$ is defined by 
$\mathcal{R}^{\dagger}=\{\phi\in \mathcal{X}\, |\, \langle \phi, \psi\rangle \ge 0\ \forall \psi\in \mathcal{R}\}$.  
We say that $\mathcal{R}$ is {\it self-dual},  if $\mathcal{R}=\mathcal{R}^{\dagger}$. 
Note that  $\mathcal{R}$ is a self-dual cone,  if and only if, $\mathcal{R}$ is a Hilbert cone \cite{Bs1976,Bratteli1987}. 

We denote by $\mathcal{A}_0$ the right hand side of (\ref{P=Q}).
Let $\phi\in \mathcal{C} $ and $f\in L_+^2(M)$. Trivially, $\phi\otimes f\in \mathcal{A}_0$.
Because 
 $f(x)\phi\in \mathcal{C}$ $\mu$-a.e.,
   we have $\phi\otimes f=\int_M^\oplus f(x)\phi \, d\mu(x)\in\mathcal{A}$, which implies  $\mathcal{A}_0 \subseteq \mathcal{A}$. Therefore,  $\mathcal{A}_0^\dagger\supseteq \mathcal{A}^\dagger=\mathcal{A}$ holds, where we have used the above fact. 
   
It suffices to prove $\mathcal{A}_0^\dagger\subseteq \mathcal{A}$. Let  $\psi\in\mathcal{A}_0^\dagger$.
For any $\phi\in \mathcal{C}$ and $f\in L^2_+(M)$, we have
  $\i<\psi,\phi\otimes f>=\int_M\i<\psi(x),\phi>f(x)\,d\mu(x)\geq 0$. Since $\int_M \mathrm{Im}\i<\psi(x),\phi>f(x)\, d\mu(x)=0$ for any $f\in L_+^2(M)$, we conclude $\mathrm{Im}\i<\psi(x),\phi>=0\ \mu$-a.e..
  Next, we claim that $\mathrm{Re}\i<\psi(x),\phi>\geq 0$. To this end, 
    suppose $\mu(\{x\in M\,|\,\mathrm{Re}\i<\psi(x),\phi><0\})>0$.  Because $M$ is $\sigma$-finite, there exists a subset $D\subset\{x\in M\,|\,\mathrm{Re}\i<\psi(x),\phi><0\}$ with $0<\mu(D)<\infty$. 
    Let $\chi_D$ be the indicator function of the set $D$.
      Because $\chi_D\in L_+^2(M)$, we have $\i<\psi,\phi\otimes\chi_D>=\int_D\mathrm{Re}\i<\psi(x),\phi>\,d\mu(x)<0$. This contradicts with the property  $\i<\psi,\phi\otimes\chi_D>\geq0$, which follows from the fact  that $\phi\otimes \chi_D\in \mathcal{A}_0$.  Hence,   $\mathrm{Re}\i<\psi(x),\phi>\geq 0
      $ holds for $ \mu$-a.e. $x$. Therefore,  we finally conclude  that $\psi(x)\in \mathcal{C}\ \mu$-a.e.  and $\mathcal{A}_0^\dagger\subseteq\mathcal{A}$.\qed
\end{Proof}

\subsubsection*{Proof of Proposition \ref{QCone}}
Apply Proposition \ref{usefulDCONE} with 
 $\mathcal X=Q_0 \mathcal{L}_N, \mathcal{C}=Q_0\mathcal{L}_{N, +}, M=\mathbb R^{|\Lambda|}$ and $\mu$ the  Lebesgue measure on $\mathbb{R}^{|\Lambda|}$.\qed

\bibliographystyle{abbrv}

\end{document}